\documentclass[twocolumn,published,twocolappendix]{aastex631}
\usepackage[T1]{fontenc}
\shorttitle{SO emission in CQ~Tau and MWC~758}
\shortauthors{Zagaria, Jiang et al.}

\graphicspath{{./}{figures/}}
\usepackage{multirow}
\usepackage{xcolor}
\definecolor{emerald}{RGB}{0,155,155}

\newcommand{\hjadd}[1]{\textcolor{emerald}{#1}}

\begin{document}

\title{SO emission in the dynamically perturbed protoplanetary disks around CQ Tau and MWC 758}%: hints of non-equilibrium chemistry?

\correspondingauthor{Francesco Zagaria and Haochang Jiang contributed equally to this work. For enquiries, please email:}
\email{frzagaria@mpia.de, h-jiang@mpia.de}

\author[0000-0001-6417-7380]{Francesco Zagaria}
\affiliation{Max-Planck Institute for Astronomy (MPIA), Königstuhl 17, 69117 Heidelberg, Germany}

\author[0000-0003-2948-5614]{Haochang Jiang}
\affiliation{Max-Planck Institute for Astronomy (MPIA), Königstuhl 17, 69117 Heidelberg, Germany}

\author[0000-0002-2700-9676]{Gianni Cataldi}
\affiliation{National Astronomical Observatory of Japan, 2-21-1 Osawa, Mitaka, Tokyo 181-8588, Japan}
%\affiliation{Department of Astronomy, Graduate School of Science, University of Tokyo, Tokyo 113-0033, Japan}

\author[0000-0003-4689-2684]{Stefano Facchini}
\affiliation{Dipartimento di Fisica, Università degli Studi di Milano, Via Celoria 16, 20133 Milano, Italy}

\author[0000-0002-7695-7605]{Myriam Benisty}
\affiliation{Max-Planck Institute for Astronomy (MPIA), Königstuhl 17, 69117 Heidelberg, Germany}

\author[0000-0003-3283-6884]{Yuri Aikawa}
\affiliation{Department of Astronomy, Graduate School of Science, University of Tokyo, Tokyo 113-0033, Japan}

\author[0000-0003-2253-2270]{Sean Andrews}
\affiliation{Center for Astrophysics | Harvard \& Smithsonian, 60 Garden St., Cambridge, MA 02138, USA}

\author[0000-0001-7258-770X]{Jaehan Bae}
\affiliation{Department of Astronomy, University of Florida, Gainesville, FL 32611, USA}

\author[0000-0001-6378-7873]{Marcelo Barraza-Alfaro}
\affiliation{Department of Earth, Atmospheric and Planetary Sciences, Massachusetts Institute of Technology, Cambridge, MA 02139, USA}

\author[0000-0003-2045-2154]{Pietro Curone}
\affiliation{Departamento de Astronomía, Universidad de Chile, Camino El Observatorio 1515, Las Condes, Santiago, Chile}

\author[0000-0002-1483-8811]{Ian Czekala}
\affiliation{School of Physics \& Astronomy, University of St. Andrews, North Haugh, St. Andrews KY16 9SS, UK}

\author[0000-0003-4679-4072]{Daniele Fasano}
\affiliation{Université Côte d’Azur, Observatoire de la Côte d’Azur, CNRS, Laboratoire Lagrange, 06300 Nice, France}
\affiliation{Max-Planck Institute for Astronomy (MPIA), Königstuhl 17, 69117 Heidelberg, Germany}

\author[0000-0002-8138-0425]{Cassandra Hall} 
\affiliation{Department of Physics and Astronomy, The University of Georgia, Athens, GA 30602, USA}
\affiliation{Center for Simulational Physics, The University of Georgia, Athens, GA 30602, USA}

\author[0000-0003-1502-4315]{Iain Hammond}
\affiliation{School of Physics and Astronomy, Monash University, Clayton, VIC 3800, Australia}
\affiliation{Max-Planck Institute for Astronomy (MPIA), Königstuhl 17, 69117 Heidelberg, Germany}

\author[0000-0001-6947-6072]{Jane Huang}
\affiliation{Department of Astronomy, Columbia University, 538 W. 120th Street, Pupin Hall, New York, NY 10027, USA}

\author[0000-0003-1008-1142]{John D. Ilee}
\affiliation{School of Physics and Astronomy, University of Leeds, Leeds, LS2 9JT, UK}

\author[0000-0001-8446-3026]{Andrés F. Izquierdo}
\altaffiliation{NASA Hubble Fellowship Program Sagan Fellow}
\affiliation{Department of Astronomy, University of Florida, Gainesville, FL 32611, USA}

\author[0009-0007-5371-3548]{Jensen Lawrence}
\affiliation{Department of Earth, Atmospheric and Planetary Sciences, Massachusetts Institute of Technology, Cambridge, MA 02139, USA}

\author[0000-0002-2357-7692]{Giuseppe Lodato}
\affiliation{Dipartimento di Fisica, Università degli Studi di Milano, Via Celoria 16, 20133 Milano, Italy}

\author[0000-0002-1637-7393]{François Ménard} 
\affiliation{Univ. Grenoble Alpes, CNRS, IPAG, F-38000 Grenoble, France}

\author[0000-0001-5907-5179]{Christophe Pinte}
\affiliation{School of Physics and Astronomy, Monash University, Clayton, VIC 3800, Australia}
\affiliation{Univ. Grenoble Alpes, CNRS, IPAG, F-38000 Grenoble, France}

\author[0000-0003-4853-5736]{Giovanni Rosotti} 
\affiliation{Dipartimento di Fisica, Università degli Studi di Milano, Via Celoria 16, 20133 Milano, Italy}

\author[0000-0002-0491-143X]{Jochen Stadler} 
%\affiliation{Univ. Grenoble Alpes, CNRS, IPAG, F-38000 Grenoble, France}
\affiliation{Université Côte d’Azur, Observatoire de la Côte d’Azur, CNRS, Laboratoire Lagrange, 06300 Nice, France}

\author[0000-0003-1534-5186]{Richard Teague}
\affiliation{Department of Earth, Atmospheric and Planetary Sciences, Massachusetts Institute of Technology, Cambridge, MA 02139, USA}

\author[0000-0003-1859-3070]{Leonardo Testi}
\affiliation{Alma Mater Studiorum Università di Bologna, Dipartimento di Fisica e Astronomia (DIFA), Via Gobetti 93/2, 40129 Bologna, Italy}

\author[0000-0003-1526-7587]{David Wilner}
\affiliation{Center for Astrophysics | Harvard \& Smithsonian, 60 Garden St., Cambridge, MA 02138, USA}

\author[0000-0002-7501-9801]{Andrew Winter} 
\affiliation{Université Côte d’Azur, Observatoire de la Côte d’Azur, CNRS, Laboratoire Lagrange, 06300 Nice, France}
\affiliation{Max-Planck Institute for Astronomy (MPIA), Königstuhl 17, 69117 Heidelberg, Germany}

\author[0000-0001-8002-8473]{Tomohiro Yoshida} 
\affiliation{National Astronomical Observatory of Japan, 2-21-1 Osawa, Mitaka, Tokyo 181-8588, Japan}
\affiliation{Department of Astronomical Science, The Graduate University for Advanced Studies, SOKENDAI, 2-21-1 Osawa, Mitaka, Tokyo 181-8588, Japan}

%\collaboration{20}{(AAS Journals Data Editors)}
%
%\author{F.X Timmes}
%\affiliation{Arizona State University}
%\affiliation{AAS Journals Associate Editor-in-Chief}
%\altaffiliation{AASTeX v6+ programmer}
%\affiliation{AAS Director of Publishing}
%\affiliation{American Astronomical Society \\
%1667 K Street NW, Suite 800 \\



\begin{abstract}
We report the serendipitous detection of the SO~$J_N=6_5-5_4$ (219.949~GHz) rotational transition in archival Atacama Large Millimeter/submillimeter Array (ALMA) observations of the spiral hosting protoplanetary disks around CQ~Tau (with $\approx4.9\sigma$ significance) and MWC~758 (with $\approx3.4\sigma$~significance). In the former, the SO emission comes in the shape of a ring, arises from the edge of the~continuum cavity, and is qualitatively consistent, at the currently available spectral resolution,~with being in Keplerian rotation. %, and is tentatively co-located with the bases of the spirals detected in scattered light observations. 
In the latter, instead, while arising primarily from inside the continuum cavity% and showing tentative connections with the scattered-light spirals
, the SO emission also extends to the continuum ring(s) and its morphology and kinematics are less~clear. We put these sources in the context of the other protoplanetary disks where SO detections have been previously reported in the literature and discuss the possible origins of SO in terms of (thermal) desorption or formation in the gas phase. We argue that these processes might be fostered by dynamical perturbations caused by unseen embedded massive companions, shadows, or late-time infall,~thus suggesting %that there might exist a strong link between dynamical perturbations and SO detections in protoplanetary disks. 
a possible link between perturbed dynamics and SO emission in (these)~protoplanetary~disks. If confirmed, our interpretation %our results 
would imply that chemical evolution timescales could be significantly shorter in these systems than is commonly assumed, %chemistry may be out of equilibrium in these systems. 
%\red{Yet, due to biases in the current sample, we cannot exclude ``passive'' thermal desorption as the origin of SO emission in some of these sources.} 
%Our results indicate 
indicating that dynamical perturbations might influence the composition of newborn (proto-)planets by altering the volatile makeup of their formation environment.%As a consequence, by altering the volatile makeup of their forming environment, our results hint that those perturbations might influence the composition of newborn (proto-)planets. 
%Our results indicate that dynamical perturbations can significantly influence the distribution of disk volatiles, challenging the traditional view that planets inherit their composition solely from unperturbed disks. %, with potentially significant impacts on the composition of planets forming therein.
\end{abstract}%shall add gas-phse formation



\section{Introduction}\label{sec:1_introduction}
High angular resolution observations of protoplanetary disks have revealed a wide variety of structures, offering indirect constraints on the physical mechanisms shaping them, such as planet-disk interactions \citep[e.g.,][]{BaeEtal2023} or (magneto-)hydrodynamical instabilities \citep[e.g.,][]{LesurEtal2023}. At (sub-)mm wavelengths,~the most common features in the dust continuum are concentric gaps and rings \citep[e.g.,][]{AndrewsEtal2018b,LongEtal2018f,Andrews2020}. Molecular line observations often show analogous annular morphologies, reflecting a combination of temperature and abundance gradients that influence line intensities \citep[e.g.,][]{OebergBergin2021,OebergEtal2021,FacchiniEtal2021}. Sophisticated thermochemical models can reproduce many observed molecular line features, providing %stringent 
constraints on molecular abundances and the formation/destruction pathways of volatile species (see \citealt{OebergEtal2023} and references therein). However, besides a few notable exceptions~(e.g. \citealt{IleeEtal2011,YonedaEtal2016}), these models most often considered disks to be dynamically unperturbed environments in a quasi-steady state, where chemical evolution occurs on secular timescales much longer than the Keplerian orbital timescale (cf. \citealt{OebergEtal2023}). %typically assume a static disk (i.e. dynamically quiet) reached a quasi-steady state after long enough run, where chemical evolution occurs on secular timescales much longer than the Keplerian orbital timescale \citep[yet, see e.g.,][]{IleeEtal2011,YonedaEtal2016}. 

In reality, it is increasingly evident that even in the Class~II stage protoplanetary disks might experience significant dynamical perturbations %throughout their evolution 
\citep[e.g.,][]{BaeEtal2023,PinteEtal2023a} that can take place on short-enough timescales to give rise to (locally) active~chemical processing. %bring chemistry out of equilibrium. 
Additionally, even if chemical evolution were able to respond fast enough to such dynamical perturbations, thus relaxing back to a quasi-steady~state, the local physical and chemical conditions under which chemical evolution would take place might still be substantially different from those typically assumed to forward-model the data \citep[see e.g.,][]{CleevesEtal2015,EvansEtal2015,EvansEtal2019,JiangEtal2023}.

Observationally, some of the best examples of dynamically perturbed disks are those with spiral structures.~In Class~II protoplanetary disks, spirals have~been~detected in dust thermal continuum, scattered-light emission, and %dust emission observed from $\mu$m- to mm-wavelengths, and 
in bright CO rotational line emission~\citep[see e.g.,][]{PerezEtal2016,HuangEtal2018c,TeagueEtal2022,WoelferEtal2023,BenistyEtal2023}. Interestingly,~spirals~in scattered-light are often associated with %high near-IR excess \citep{GarufiEtal2018}, suggestive of dynamically perturbed inner disk regions, and with 
azimuthally asymmetric structures in (sub-)mm continuum emission \citep[see e.g.,][]{TangEtal2017,PinedaEtal2019,YangEtal2023h}, possibly linked to long-lived anticyclonic vortices or horseshoes %generated in the disk and 
able to trap dust particles \citep{vanderMarelEtal2021a}.

Several hypotheses have been proposed to explain the possible origins of these spirals. A class of young~sources, including IM~Lup, Elias~2-27 \citep{Paneque-CarrenoEtal2021,HuangEtal2018c}, and AB~Aur \citep{SpeedieEtal2024}, are often regarded as self-gravitating or undergoing gravitational instability (GI, \citealt{KratterLodato2016}), as is supported by indirect kinematic evidence, such as (1) high dynamical masses \citep{VeronesiEtal2021}, (2) GI-wiggles \citep{HallEtal2020,LongariniEtal2024,SpeedieEtal2024}, in addition to, when present, (3) prominent continuum spirals, expected to be a telltale of self-gravity \citep{KratterLodato2016}. 

Another popular interpretation involves (sub-)stellar companions launching Lindblad spirals that propagate at sonic velocities within the disk \citep[see e.g.,][]{DongEtal2015a,BaeZhu2018}. However, while clear connections between the presence of close companions and spiral features can be made in some cases (e.g., HD~100543 and HD~142527, \citealt{GonzalezEtAl2020,GargEtal2021}), most of the planet-mass companions proposed to explain prominent spiral structures (such as those in MWC~758 and HD~169142, e.g., \citealt{DongEtal2018a,RenEtal2020,HammondEtal2023}) remain debated. Conversely, PDS~70, the only disk with confirmed planet-mass companions \citep{KepplerEtal2018,HaffertEtal2019}, does not show any clear spiral features. %However, while in some cases clear connections could be made between the presence of close companions and spiral features (see e.g., HD~100543 or HD~142527, \citealt{GonzalezEtAl2020,GargEtal2021}), even in disks showing prominent spiral structures, most of the proposed planet-mass companions (such as those in MWC~758 and HD~169142, e.g., \citealt{DongEtal2018a,RenEtal2020,HammondEtal2023}) remain debated, except for PDS~70 \citep{KepplerEtal2018,HaffertEtal2019}, that, however, does not show any clear spiral features. 

Alternative explanations involve azimuthal disk temperature variations, such as those caused by shadows induced by % This scenario does not necessarily require companions but often depends on 
a misaligned inner component or a warped~disk \citep[e.g.,][]{MontesinosEtal2016,SuBai2024,ZiamprasEtal2025,ZhuEtal2025}, % casting shadows on the outer disk.
%A final compelling interpretation is 
or infalling material from the surrounding environment %, as emission from infalling material is seen at both %. With an increasing number of detected infall events in at both
%NIR and (sub-)mm wavelengths 
\citep[e.g.,][]{GinskiEtal2021,MesaEtal2022,HuangEtal2023}, %it has also been shown that infalling material 
that can generate spiral-like structures within~disks~\citep[e.g.,][]{LesurEtal2015,HennebelleEtal2017,KuffmeierEtal2020,KuznetsovaEtal2022,CalcinoEtal2025}. %Additionally, infall is hypothesized to induce misalignment between the inner and outer disks \citep{KuffmeierEtal2021,BohnEtal2022}, as the infalling material may carry angular momentum vectors misaligned with the primary disk. Such misalignment could further facilitate shadow-driven spirals.

All these mechanisms %coexist with 
are expected to generate strong dynamical perturbations that can lead to (1) local temperature enhancement via viscous heating or the generation of shocks/compression \citep[e.g.,][]{OnoEtal2025}, (2) an increase in the relative dust particle velocities, which makes ice-coated grains more susceptible to (potentially destructive) collisions \citep{ErikssonEtal2025}, and (3) enhanced local turbulence that stirs solids into the warm, UV-illuminated upper disk layers \citep{BiEtal2021,BinkertEtal2023}. These processes create favorable conditions for the release of volatiles, such as H$_2$CO,~SO,~and~other oxygen-bearing species, into the gas phase through~direct thermal desorption or dust shattering, and their~formation in the gas-phase via chemical reactions driven by energetic radiation \citep[e.g.,][]{AotaEtal2015}. More generally, these perturbations would break the quasi-steady state of the disk, fostering active chemical processing. In this sense, it is no surprise that tracers like SO have been robustly identified in several spiral- and candidate-planet-hosting protoplanetary disks, such as HD~100546 \citep{BoothEtal2021b,BoothEtal2024a}, AB~Aur \citep{DutreyEtal2024,SpeedieEtal2025}, HD~169142 \citep{LawEtal2023,BoothEtal2023b}, and TW~Hya \citep{YoshidaEtal2024}.

In this work, we %aim to 
explore the possible link between the dynamical state of the disk and 
%Given the rarity of SO detections in protoplanetary disks, the consistent association with dynamically perturbed systems points to a plausible link between complex gas dynamics and the 
SO chemistry. %This correlation highlights the critical role of disk-companion and disk-environment interactions in shaping the chemical composition of planet-forming disks. 
%In this work, 
We report the serendipitous detection of SO in archival Atacama Large Millimeter/submillimeter Array (ALMA) observations of the dynamically perturbed protoplanetary disks around CQ~Tau and MWC~758. We~put~them in the context of the other Class~II sources where SO detections were previously published, discussing different possible origins for SO emission. In \autoref{sec:2_data} we introduce these two systems, describe the available datasets that cover the SO~$J_N=6_5-5_4$ transition, and their self-calibration. Our new SO detections, their morphology, kinematics, and connections with scattered light observations are presented in \autoref{sec:3_results}. Our results are discussed in the broader context of the other SO-bearing Class~II protoplanetary disks in \autoref{sec:4_discussion}. In \autoref{sec:5_summary} we summarize our results and draw our conclusions.

\section{Sources, data, self-calibration}\label{sec:2_data}

\subsection{The perturbed CQ~Tau and MWC~758 disks}\label{subsec:2.1_sources}
CQ~Tau -- $M_\star=1.47\pm0.18\ M_\odot$ \citep{VioqueEtal2018}, $d=149.4\pm1.3$~pc \citep{GaiaCollaboration2020}, SpT~F2 \citep{GarciaLopezEtal2006} -- and MWC~758 -- $M_\star=1.56\pm0.11\ M_\odot$ \citep{VioqueEtal2018}, $d=155.9\pm0.8$~pc \citep{GaiaCollaboration2020}, SpT A8 \citep{VieiraEtal2003} -- are two Herbig stars located in the nearby~HD~35187~association,
%, with an estimated age of $18\pm4$~Myr, situated
near the Taurus star-forming region \citep{LuhmanEtal2023a}. %Based on Br$\gamma$ line observations, \citet{GrantEtal2022} estimated the stellar accretion rate to be $<4.7\times 10^{-9}\ M_\odot\ {\rm yr}^{-1}$ for CQ~Tau and 0.2 to 1.2$\times 10^{-7}\ M_\odot\ {\rm yr}^{-1}$ for MWC~758.

Thanks to their close proximity and high~luminosity, the planet-forming disks around CQ~Tau and MWC~758 are among the best studied ones across a broad range of frequencies. Observations at (sub-)millimeter wavelengths, first with the Submillimeter Array (SMA, \citealt{IsellaEtal2010,TripathiEtal2017}) and later with the Atacama Large Millimeter/submillimeter Array (ALMA), revealed the presence of (1) extended ($R_{\rm 1.3mm}\approx66$~au and 92~au, respectively) continuum emission with a ring-like morphology around wide ($R_{\rm cav,mm}\approx50$ to 60~au) and dust-depleted cavities \citep{BoehlerEtal2018,DongEtal2018a,UbeiraGabelliniEtal2019,CuroneEtal2025}, and (2) larger-scale ($R_{\rm CO}\approx200$~au and 300~au, respectively, \citealt{Galloway-SprietsmaEtal2025}) CO emission with smaller but still well resolved ($R_{\rm cav,CO}\approx30$ to 50~au) cavities in $^{13}$CO and C$^{18}$O emission \citep{BoehlerEtal2018,UbeiraGabelliniEtal2019,WoelferEtal2021,WoelferEtal2023}.

Both sources show telltale signatures of ongoing dynamical perturbations, as evidenced by their prominent scattered-light spiral structures, characterized by a dominant two-arm pattern with additional smaller-scale features \citep[e.g.,][]{BenistyEtal2015,HammondEtal2022,RenEtal2023b}, and their highly asymmetric (sub-)mm continuum emission with two prominent bright arcs on top of a highly substructured wide ring on opposite sides of their cavity rims \citep{BoehlerEtal2018,DongEtal2018a,UbeiraGabelliniEtal2019}, roughly aligning with the bases of the scattered-light spirals. Moreover, the cavity of MWC~758 shows a significant eccentricity of $e\approx0.1$ \citep{DongEtal2018a,KuoEtal2022}, further suggesting active dynamical sculpting.

Additionally, $^{12}$CO observations revealed strong spiral structures in the kinematics of CQ~Tau \citep{WoelferEtal2021,WoelferEtal2023} and MWC~758 \citep{WoelferEtal2025}. These spirals are evident in the velocity residuals obtained after subtracting a Keplerian model from the line centroid maps (as done by \citealt{IzquierdoEtal2022}, see \citealt{WoelferEtal2025}; Benisty et al., in prep.). Their magnitude is comparable to the local sound speed, consistently with the expectations for propagating sound waves, albeit likely a lower limit, due to the limited angular and spectral resolution of the available data. %\hjc{Depending on whether we cite exoALMA, consider deleting the following sentence.} The high quality of the kinematic data also enables the detection of gas spirals in the line width maps, tracing enhanced local turbulent motions inside the spirals. These features together strongly indicate that both disks are undergoing active dynamical perturbations.
%J=2-1 for CQ~Tau, 3-2 and 2-1 for MWC~758.

\subsection{Data and self-calibration}\label{subsec:2.2_data&self-cal}
We collected from the ALMA archive all the publicly available data of CQ~Tau and MWC~758~that~cover the SO~$J_N=6_5-5_4$ rotational transition~at~219.949~GHz (Band 6). CQ~Tau observations were conducted in Cycles~2, 4, and 5 between Aug.~2015 and Nov.~2017, as part of the programs 2013.1.00498.S~(PI: L.~Pérez), 2016.A.00026.S (PI: L.~Testi), and 2017.1.01404.S (PI: L.~Testi). MWC~758 was observed in Cycle~5 as part of the program 2017.1.00940.S (PI: L.~Ricci) between Oct. and Dec. 2017. The observational setup details~are summarized in \autoref{app:data}.

The data were first pipeline-calibrated by the ESO node of the European ALMA Regional Centre (ARC), and then self-calibrated using the software CASA (Common Astronomy Software Applications, see \citealt{CASATeamEtal2022}) \texttt{v6.6.5-31}, following the procedure adopted by the \href{https://almascience.eso.org/almadata/lp/DSHARP/}{DSHARP} \citep{AndrewsEtal2018b}, \href{https://alma-maps.info/overview.html}{MAPS} \citep{CzekalaEtal2021}, and \href{https://www.exoalma.com/}{exoALMA} \citep{LoomisEtal2025} collaborations. %Here we briefly summarize the key steps of the process and we refer to \autoref{app:self-cal} for a more thorough description.
A detailed description of the self-calibration procedure can be found in \autoref{app:self-cal}. We measured a peak signal-to-noise ratio (SNR) of $\approx90$ and~$\approx170$~in our fiducial continuum images of CQ~Tau and MWC~758. Their root mean square (RMS)~noise is within 24.3\% and 91.6\% of the expected thermal noise (as estimated using the \href{https://casadocs.readthedocs.io/en/stable/api/tt/casatasks.imaging.apparentsens.html}{\texttt{apparentsens}} CASA task), indicative of some residual phase noise at high angular resolution. The self-calibrated and continuum-subtracted datasets were used to generate SO emission cubes (at the highest achievable spectral resolution of 0.7~km~s$^{-1}$ for CQ~Tau, and 1.4~km~s$^{-1}$ for MWC~758) and moment maps, as is detailed in \autoref{app:imaging}.

\section{Results}\label{sec:3_results}

% \begin{figure*}[t!]
%     \centering
%     % \includegraphics[width=0.95\textwidth]{figures/CQ_Tau_figure1_paper.pdf}
%     % \includegraphics[width=0.95\textwidth]{figures/MWC_758_figure1_paper.pdf}
%     \includegraphics[width=\textwidth]{figures/summary_continuum.pdf}
%     \caption{From left to right: 1.3~mm continuum emission map, highpass filtered continuum, and polarized intensity for CQ~Tau (top row) and MWC~758 (bottom row). The dotted gray line display the $[5,65]\times\sigma$ (CQ~Tau) and the $[5,40]\times\sigma$ (MWC~758) emission contours. The synthesized CLEAN beam is shown as an ellipse in the bottom left corner of each continuum panel. The regions within 0\farcs1 of the scattered light images cannot be accessed due to the coronagraph and are masked out.}
%     \label{fig:1_continuum}
% \end{figure*}

\paragraph{Continuum emission} The continuum emission maps reconstructed with our fiducial imaging parameters (\autoref{app:imaging} for the details) are shown in panels a and b of \autoref{fig:1_continuum}. Here and in the following figures, the brightness temperature is computed in the Rayleigh-Jeans approximation. The dotted gray lines display the $[5,65]\times\sigma$ and the $[5,40]\times\sigma$ emission contours. These values were chosen to highlight the location of the cavity rim and the continuum crescents. As previously discussed by \citet[][]{UbeiraGabelliniEtal2019,BoehlerEtal2018,DongEtal2018a,CuroneEtal2025}, in both sources the continuum morphology is characterized by (1) a deep, almost empty (and, in MWC~758, also clearly eccentric) cavity with a faint inner disk, and (2) bright crescent-like asymmetries on top of a wide, potentially substructured, continuum ring, suggestive of strong dynamical perturbations (as is highlighted in the highpass continuum and scattered-light images shown in \autoref{app:imaging}). %CQ~Tau shows evidence of ongoing dynamical perturbations in the form of tightly wound spirals that span the full ring width and azimuth, connecting the two crescents (see~e.g., the high-pass filter image in \autoref{app:imaging}). Similarly, an arc-like feature departing from the inner west-side crescent towards the east side of the disk can be seen in MWC~758, as reported by \citet{DongEtal2018a}. 
%Such asymmetries show up more distinctly in the central panels (1b and 1e) of \autoref{fig:1_continuum}, that display the highpass-filtered continuum emission from both sources. This was obtained by subtracting to the images reconstructed with \texttt{robust=0.5} and \texttt{1.5} their convolution with Gaussian kernels with a standard deviation of 5 and 8 pixels, respectively. Three arcs veering from the continuum ring towards the central cavity are clearly visible in CQ~Tau. One of them (labeled ``Arc 1'') is connected with one the spirals detected in scattered light (cf. \autoref{fig:1_continuum}, panel 1c, that shows the $K_{\rm s}$ band $Q_\phi$ polarized differential image of \citealt{RenEtal2023b}), suggesting that they might be tracing the same disk structure. In MWC~758, an azimuthally-wide arc, similarly co-located with one of the scattered-light spirals (see panel 1f in \autoref{fig:1_continuum}, that also shows the $K_{\rm s}$ band $Q_\phi$ polarized differential image of \citealt{RenEtal2023b}), connects the cavity rim with a bright annular structure (as was previously reported by \citealt{DongEtal2018a} and \citealt{CuroneEtal2025}).

\begin{figure}[t!]
    \centering
    \includegraphics[width=\linewidth]{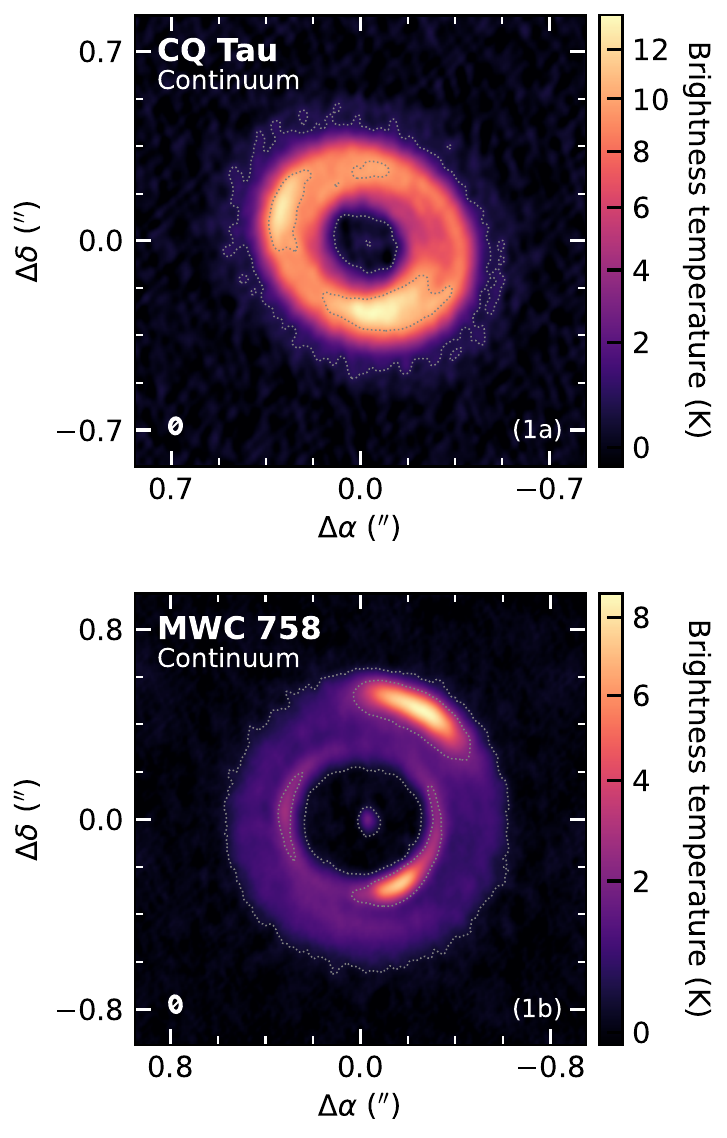}
    \caption{1.3~mm continuum emission map of CQ~Tau (top) and MWC~758 (bottom). The dotted gray lines highlight the $[5,65]\times\sigma$ (CQ~Tau) and the $[5,40]\times\sigma$ (MWC~758) emission contours. The ellipse in the bottom left corner of each panel displays the synthesized CLEAN beam.}
    \label{fig:1_continuum}
\end{figure}

\begin{figure*}[t!]
    \centering
    \includegraphics[width=\textwidth]{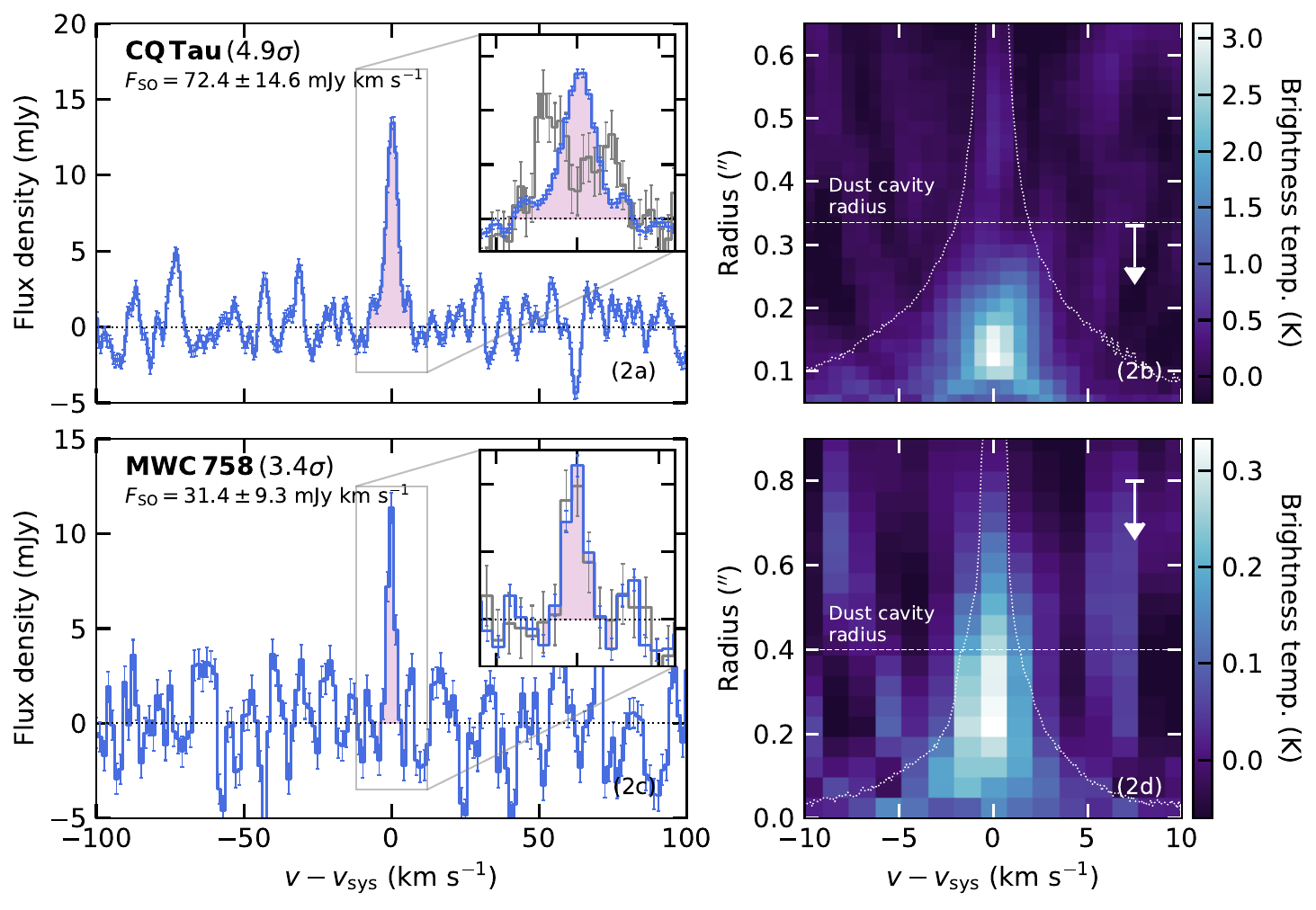}
    \caption{Left: Shifted and stacked SO spectra (blue). %extracted from our fiducial images over an area of 0\farcs33 and 0\farcs80$$. 
    The zoom-in inserts around the systematic velocity also display the native spectra (gray) for comparison. The plum areas highlight the velocity ranges adopted to measure the SO flux and generate moment maps (see \autoref{app:imaging}). Right: Teardrop plots. The dotted white lines mark the region where Keplerian emission~from the disk is expected. Those correspond to the location where the CO emission (from the fiducial exoALMA cubes, \citealt{TeagueEtal2025}) drops below 3~K. The white arrows indicate the apertures over which the spectra were extracted and integrated over~to measure the SO flux. The dust cavity radius is indicated with a white dashed line. SO emission is clearly detected and primarily originates within the cavity.}
    \label{fig:2_spectra}
\end{figure*}

\paragraph{SO emission} SO emission is detected both in CQ~Tau and in MWC~758 as is clear from the shifted~and~stacked spectra (blue lines, panels 2a and 2c) and teardrop~plots (panels 2b and 2d) displayed in \autoref{fig:2_spectra},~both~generated using \texttt{gofish} \citep{Teague2019}, from our fiducial~cubes (\autoref{app:imaging} for details), with fluxes of~$72.4\pm14.6$~($4.9\sigma$ detection) and $31.4\pm9.3$ ($3.4\sigma$ detection)~mJy~km~s$^{-1}$, respectively. These line intensities were measured~by~integrating the shifted and stacked spectra extracted from a circular mask with a radius of $0\farcs33$ (CQ~Tau)~and $0\farcs80$ (MWC~758). These apertures correspond to the disk radius where the cumulative distribution of the SO emission, obtained integrating the shifted and stacked spectra extracted from progressively larger circular masks, flattens out (cf. the arrows in the teardrop plots in panels b and d of \autoref{fig:2_spectra}). The flux uncertainties, instead, were estimated as the standard deviation of the fluxes measured by integrating the shifted and stacked spectra in non-overlapping spectral ranges as wide as those used to measure the line intensities (to take into account the spectral correlation across channels induced by line shifting and stacking), but not covering the SO transition, extracted from the same disk region.% 360 off-source copies of the mask used to measure the SO flux \citep[e.g.,][]{RampinelliEtal2024}. %This method provides systematically larger uncertainties by a factor of 1.5 to 2.0 than the traditionally adopted one by \citet{CarneyEtal2019}. The channel maps, integrated intensity, peak intensity, velocity and teardrop maps for other cleaning parameters are available at this \red{link}.

%SO emission is detected both in CQ~Tau and in MWC~758, as is clear from the shifted and stacked spectra (blue lines, panels 2a and 2c) and teardrop plots (panels 2b and 2d) displayed in \autoref{fig:2_spectra}, both generated using \texttt{gofish}, \citep{Teague2019} from our fiducial cubes (see \autoref{app:imaging}), with fluxes of $72.4\pm14.6$ ($4.9\sigma$ detection) and $31.4\pm9.3$ ($3.4\sigma$ detection)~mJy~km~s$^{-1}$, respectively. These line intensities were measured by integrating the shifted and stacked spectra extracted from a circular mask with radius of $0\farcs33$ (CQ~Tau) and $0\farcs80$ (MWC~758). These apertures correspond to the locations where the cumulative distribution of the SO emission, obtained integrating the shifted and stacked spectra extracted from progressively larger circular masks, flattens out. Flux uncertainties, instead, were estimated as the standard deviation of the fluxes measured by integrating the shifted and stacked spectra in non-overlapping spectral ranges as wide as those used to measure the line intensities but not covering the SO transition, extracted from the same disk region.% 360 off-source copies of the mask used to measure the SO flux \citep[e.g.,][]{RampinelliEtal2024}. %This method provides systematically larger uncertainties by a factor of 1.5 to 2.0 than the traditionally adopted one by \citet{CarneyEtal2019}. The channel maps, integrated intensity, peak intensity, velocity and teardrop maps for other cleaning parameters are available at this \red{link}.

\begin{figure*}[t!]
    \centering
    \includegraphics[width=\textwidth]{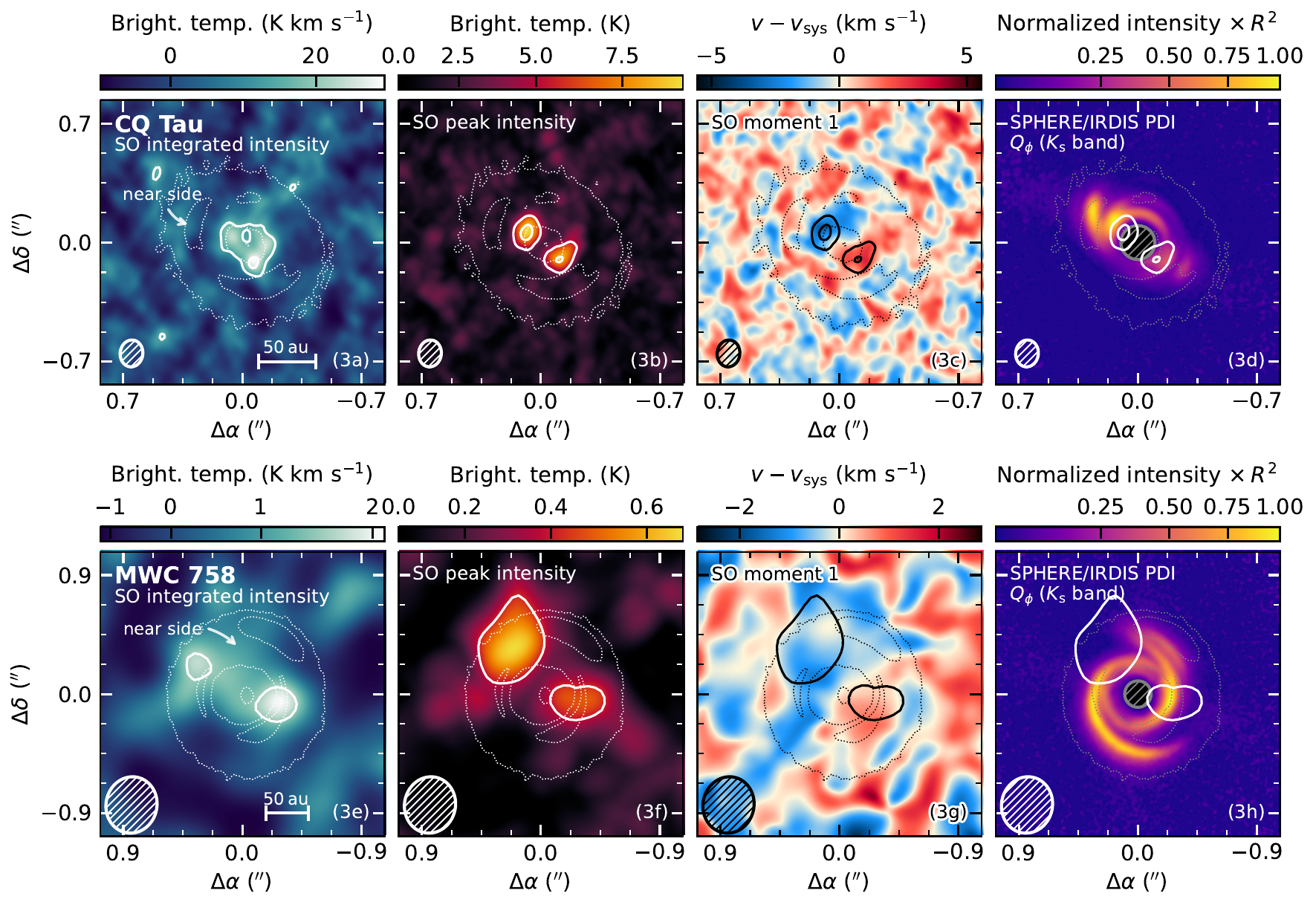}
    \caption{From left to right: SO integrated intensity (``moment 0''), peak intensity (``moment 8''), velocity (``moment 1'') maps, and comparison between scattered-light images and SO peak intensity for CQ~Tau (top row) and MWC~758 (bottom row). The dotted contours display the $[5,65]\times\sigma$ (CQ~Tau) and the $[5,40]\times\sigma$ (MWC~758) emission levels. The white (black for panels 3c and 3g) solid contours, instead, mark the $[3,5]\times\sigma$ SO detection levels. The synthesized CLEAN beam is shown as an ellipse in the bottom left corner of each panel. The regions within 0\farcs1 of the scattered light images cannot be accessed because of the coronagraph and are masked out.}
    \label{fig:3_summary}
\end{figure*}

%The central and right panels of \autoref{fig:1_continuum} display the SO peak intensity and velocity maps (generated using \texttt{bettermoments}, \citealt{TeagueForeman-Mackey2018}) for our fiducial CLEANing parameters (\texttt{natural} weighting and a Gaussian $uv$-taper with $0\farcs05$ full width at half maximum, see \autoref{app:imaging} for more details). The channel maps, integrated intensity map, shifted and stacked spectrum, and a teardrop plot (generated using \texttt{gofish}, \citealt{Teague2019}) for the same fiducial parameters are shown in \autoref{fig:A1_CQTau_summary} for CQ~Tau and \autoref{fig:A2_MWC_758_summary} for MWC~758.

\autoref{fig:3_summary} displays the SO integrated intensity (panels 3a and 3e), peak intensity (panels 3b and 3f), and velocity (``moment 1'', panels 3c and 3g) maps, generated using \texttt{bettermoments} \citep{TeagueForeman-Mackey2018}, without applying any channel or threshold clipping, for our fiducial CLEANing parameters. %(\texttt{natural} weighting and a Gaussian $uv$-taper with $0\farcs05$ full width at half maximum for CQ~Tau and $0\farcs25$ for MWC~758, see \autoref{app:imaging} for more details). 
In addition, a comparison between SO peak intensity and scattered-light morphology is displayed in panels 3d and 3h. Some relevant channel maps, reconstructed with the same fiducial parameters, are shown in \autoref{app:imaging}.%\autoref{fig:A2_channels}.

%In CQ~Tau, SO is detected with high significance (i.e., $>5\sigma$) across several independent channels (\autoref{fig:A2_channels}). 
In CQ~Tau, SO is detected with high (5 to $10\sigma$)~significance across several independent channels~(see~\autoref{app:imaging}). The integrated intensity map indicates that SO emission primarily arises from the edge of the continuum cavity (as is suggested by the teardrop plot in \autoref{fig:2_spectra}) in the form of a ring, with a small inner gap and a~bright spot in the direction of the south-east crescent (on the far side of the disc). The SO peak intensity map is also brighter at the cavity rim. However, its morphology is not ring-like, but asymmetric~and brighter along the disk major axis and azimuthally closer to the continuum crescents. The comparison between the SO peak intensity emission morphology and the $K_{\rm s}$ band $Q_\phi$ polarized differential image of \citet{RenEtal2023b}, displayed in panel 3d, suggests that the SO emission is preferentially co-located with the innermost regions of the scattered light spiral wakes. While, at a first glance, this alignment might suggest a common dynamical origin for the scattered-light spirals and the SO emission, we caution that their co-location is not necessarily physical, as is indicated by the morphological differences between integrated and peak intensity maps, and could be a consequence of limb brightening. In fact, due to a combination of the potentially low optical depth of SO (cf., e.g., the C$^{18}$O integrated intensity map in PDS~70, \citealt{LawEtal2024}) and the limited resolution and sensitivity of our dataset (e.g., \citealt{WoelferEtal2021}), emission from the higher velocity channels could be artificially enhanced along the disk minor axis. %Moreover, the SO and scattered light observations have different resolution elements, preventing a robust comparison. 
Finally, the SO velocity map (\autoref{fig:3_summary}, panel 3c), indicates that, at the currently-available spectral resolution (of $\approx0.7$~km~s$^{-1}$) the SO emission is qualitatively compatible with Keplerian rotation. However, we cannot exclude lower-amplitude velocity perturbations (e.g., due to the presence of spirals; cf. e.g., the CO velocity residuals published by \citealt{IzquierdoEtal2023}, typically of order $\leq0.4$~km~s$^{-1}$). % (cf. e.g., the CO velocity maps published by \citealt{WoelferEtal2021,WoelferEtal2023}).

In MWC~758, SO is detected only in few independent channels with $>3\sigma$ significance (\autoref{app:imaging}), not only because of the lower source inclination, but also possibly because SO emission is more extended but overall fainter. %Consequently, both the peak intensity and velocity maps that can be drawn from the available data (central and right panels of \autoref{fig:1_continuum}) are relatively uninformative and would require deeper and higher spectral resolution observations to provide better insights into the SO emission morphology. However, it is noteworthy that the teardrop plot in \autoref{fig:A2_MWC_758_summary} indicates that, as in the case of CQ~Tau, also in MWC~758 SO is emitted primarily within the continuum cavity. 
Both the SO integrated intensity and peak intensity maps (panels e and f of \autoref{fig:3_summary}) display two bright spots along the disk major axis and along the main ring, close to the continuum cavity edge (as was hinted by the teardrop plot in \autoref{fig:2_spectra}). %, although in this case limb brightening seems unlikely due to the low(er) inclination. 
Similarly to CQ~Tau, the SO emission velocity of MWC~758 (panel 3g) is qualitatively consistent with being Keplerian. Finally, the comparison between SO peak intensity and the $K_{\rm s}$ band $Q_\phi$ polarized differential image of \citet{RenEtal2023b}, displayed in panel 3h, does not show any connections between SO emission and scattered-light spirals.%, but similar caveats as for CQ~Tau apply.

\section{Discussion}\label{sec:4_discussion}

\subsection{Comparison with the other SO-bearing Class~II disks}
We detected emission from the SO~$J_N=6_5-5_4$ transition in CQ~Tau and MWC~758. %with fluxes of $72.4\pm14.6$ and  $44.3\pm12.9$~mJy~km~s$^{-1}$. 
With these~two~new sources, SO detections have been so far reported in nine Class~II disks: UY~Aur \citep{TangEtal2014}, HD~100546 \citep{BoothEtal2018,BoothEtal2023,BoothEtal2024a}, IRS~48 \citep{BoothEtal2021b,BoothEtal2024b}, HD~169142 \citep{LawEtal2023,BoothEtal2023b}, AB~Aur \citep{FuenteEtal2010,Pacheco-VazquezEtal2016,Riviere-MarichalarEtal2020,DutreyEtal2024,SpeedieEtal2025}, DR~Tau~\citep{HuangEtal2023,HuangEtal2024}, TW~Hya \citep{YoshidaEtal2024}, and CQ~Tau and MWC~758. %\citep[][]{TangEtal2014,BoothEtal2021b,BoothEtal2023,BoothEtal2023b,BoothEtal2024a,BoothEtal2024b,HuangEtal2023,HuangEtal2024,LawEtal2023,DutreyEtal2024,YoshidaEtal2024,SpeedieEtal2025}. 
%\citep[][]{BoothEtal2021b,BoothEtal2023,BoothEtal2023b,BoothEtal2024a,BoothEtal2024b}. 
Remarkably, six of these systems, including CQ~Tau and MWC~758, (1) are around Herbig stars~and (2) show large inner cavities in their ALMA continuum, $^{13}$CO, and C$^{18}$O emission. Moreover, their detected SO emission is consistently located inside or near the edge of these cavities.%, hinting at the presence of a potential common origin tied to the formation of the cavities themselves, see discussion in \autoref{sec:origins}.
%, \hjadd{e.g., SO desorption due to the direct exposure to the stellar irradiation inside the cavities, where the temperature could steeply increase \citep[e.g.,][]{LeemkerEtal2022}}.

We first checked for any correlations between SO emission and stellar properties in our sample of Class~II disks with SO detections. For disks lacking SO~$J_N=6_5-5_4$ observations, we estimated its flux based on the line~flux of the detected transition with the closest upper state energy to the $J_N=6_5-5_4$ one, and rescaled~the~fluxes to a common distance of 140~pc. Both the optically~thin and thick regimes were considered, with the conversion process detailed in \autoref{app:SO_flux}. We note that~different assumptions on the excitation temperature do~not~qualitatively change our results, as the derived flux differences are relatively small. The stellar and disk parameters of interest and the converted SO fluxes are summarized in \autoref{tab:1_parameters} and \autoref{app:SO_flux}.

\begin{table*}
    \caption{Stellar and disk properties of the Class~II sources with SO detections.}
    \movetableright=-1.0in
    % \centering
    \renewcommand{\arraystretch}{1.25}
    \begin{tabular}{llllllllll}
    \hline
    \multirow{2}{*}{Source} & $M_\star$ & $\log_{10}L_\star$ & $t_\star$ & \multirow{2}{*}{SpT} & $M_{\rm dust}$ & $\log_{10}\dot{M}_{\rm acc}$ & \multirow{2}{*}{SO tran.s} & $F_{\rm SO}$ & \multirow{2}{*}{Ref.s} \\
     & ($M_\odot$) & ($L_\odot$) & (Myr) & & ($M_\oplus$) & ($M_\odot$ yr$^{-1}$) & & (mJy km s$^{-1}$) & \\
    (1) & (2) & (3) & (4) & (5) & (6) & (7) & (8) & (9) & (10) \\
    \hline
    \hline
    TW~Hya & $0.9^{+0.1}_{-0.2}$ & $-0.54^{+0.13}_{-0.10}$ & $9.8^{+6.2}_{-1.4}$ & M0.5 & $57.1\pm0.1$ & $-8.65\pm0.22$ & $7_8-6_7$ & $13\pm3$ & \footnotesize{$1,1,1,2,3,1,4$} \\ %Hercseg et al, 2023, Hercseg et al, 2023, Hercseg et al, 2023, Hercseg & Hillenbrand, 2014, Macias et al, 2021, Hercseg et al, 2023, Yoshida et al, 2024. Flux is 575.4 ± 0.9, uclear where uncertainties come from.
    \hline
    HD~100546 & $2.06^{+0.10}_{-0.12}$ & $1.37^{+0.07}_{-0.05}$ & $5.5^{+1.4}_{-0.8}$ & A0 & $38.0\pm3.9$ & $-6.95\pm0.36$ & $7_8-6_7$ & $343\pm26$ & \footnotesize{$5,5,5,6,7,8,9$} \\ %Vioque et al, 2018, Vioque et al, 2018, Vioque et al, 2018, Gray et al, 2017, Stapper et al, 2022, Grant et al., 2022, \citet{BoothEtal2024a}
    \hline
    AB~Aur & $2.15^{+0.36}_{-0.21}$ & $1.61^{+0.19}_{-0.21}$ & $4.0^{+1.4}_{-1.5}$ & A0 & $11.8\pm1.2$ & $-6.13\pm0.27$ & $6_5-5_4$ & $330\pm10$ & \footnotesize{$5,5,5,10,7,11,12$} \\ %Vioque et al, 2018, Vioque et al, 2018, Vioque et al, 2018, Mooley et al, 2013, Stapper et al, 2022, Wichittanakom et al, 2020, \citet{DutreyEtal2024}
    \hline 
    IRS~48 & 2.5 & 1.62 & 4.1 & A0 & 7.8 & $-8.40\pm0.50$ & $7_8-6_7$ & $1063\pm23$ & \footnotesize{$13,13,13,14,15,16,17$} \\ % Brown et al, 2012, Brown et al, 2012, Brown et al, 2012, Brown et al, 2012, , Salyck et al, 2013 (cosiderazioni spicciole su uncr.), Booth et al, 2024b
    \hline
    HD~169142 & $2.00\pm0.13$ & $1.31^{+0.12}_{-0.22}$ & $9.0^{+11.0}_{-3.9}$ & F1 & $22.9\pm2.4$ & $-7.09\pm0.21$ & $8_8-7_7$ & $120\pm10$ & \footnotesize{$5,5,5,6,7,11,18$} \\ %Vioque et al, 2018, Vioque et al, 2018, Vioque et al, 2018, Gray et al, 2017, Stapper et al, 2022, Wichittanakom et al, 2020, \citet{BoothEtal2023}
    \hline
    UY~Aur~A & 0.60 & $-0.49\pm0.2$ & 5.13 & M0   & $11.2\pm0.6$ & $-7.64$ & $6_5-5_4$ & N/A & \footnotesize{$19,19,19,19,20,19,21$} \\ %Hartigan&Kenyon, 2003, Hartigan&Kenyon, 2003, Hartigan&Kenyon, 2003, Manara et al, 2023, Hartigan&Kenyon, 2003, Tang et al, 2014
    UY~Aur~B & 0.34 & $-0.77\pm0.2$ & 3.89 & M2.5 & $ 4.4\pm0.1$ & $-7.70$ & $6_5-5_4$ & N/A & \footnotesize{$19,19,19,19,20,19,21$} \\ %Hartigan&Kenyon, 2003, Hartigan&Kenyon, 2003, Hartigan&Kenyon, 2003, Manara et al, 2023, Hartigan&Kenyon, 2003, Tang et al, 2014
%    UY~Aur~A & $0.65^{+0.17}_{-0.13}$ & $0.38\pm0.14$ & $0.9\pm0.4$ & K7 & $11.2\pm0.6$ & $-7.26\pm0.35$ & $6_5-5_4$ & N/A & abcdefg \\ %Manara et al, 2019, Manara et al, 2023, Long et al, 2019, Manara et al, 2023, Manara et al, 2023, Manara et al, 2023, Tang et al, 2014. Assumed a conservative 28% luminosity uncertainty (Herczeg&Hillenbrand2014) 
%    UY~Aur~B & $0.32\pm0.20$ & -- & -- & M2.5 & $ 4.4\pm0.1$ & -- & $6_5-5_4$ & N/A & abcdefg\\
    \hline
%    DR~Tau, 6   &  &  &  & K6 & $70.4\pm0.1$ &  & $6_5-5_4$ & $195\pm9$ & \\ %Manara et al, 2023, Manara et al, 2023
    DR~Tau & $0.83^{+0.34}_{-0.24}$ & $-0.54\pm0.20$ & $3.2^{+2.7}_{-1.4}$ & K4 & $70.4\pm0.1$ & $-7.93\pm0.40$ & $6_5-5_4$ & $195\pm9$ & \footnotesize{$22,22,22,22,20,22,23$} \\ %Gangi et al, 2022, Gangi et al, 2022, Long et al, 2019 (but not self-consistent), Gangi et al, 2022, Manara et al., 2023, Gangi et al, 2022, Huang et al, 2023
    \hline
    CQ~Tau & $1.47^{+0.19}_{-0.11}$ & $0.87^{+0.18}_{-0.12}$ & $8.9^{+2.8}_{-2.5}$ & F2 & $44.2\pm4.8$ & $<-8.33$ & $6_5-5_4$ & $72.4\pm14.6$ & \footnotesize{$5,5,5,24,7,8,25$} \\ %Vioque et al, 2018, Vioque et al, 2018, Vioque et al, 2018, Garcia Lopez et al, 2006, Stapper et al, 2022, Grant et al, 2022, this work
    \hline
    MWC~758 & $1.56^{+0.11}_{-0.08}$ & $1.04^{+0.12}_{-0.08}$ & $8.3^{+0.4}_{-1.4}$ & A8 & $18.8\pm2.0$ & $-7.32\pm0.40$ & $6_5-5_4$ & $31.4\pm9.3$ & \footnotesize{$5,5,5,26,7,8,25$} \\ %Vioque et al, 2018, Vioque et al, 2018, Vioque et al, 2018, Vieira et al., 2003, Stapper et al, 2022, Grant et al, 2022, this work
    \hline
    \end{tabular}
    \tablecomments{Column 1: Source name. Columns 2 to 5: Stellar mass, luminosity, age, and spectral type. Column 6: Disk dust mass (conversion from tabulated 1.3~mm - 0.9~mm for IRS~48 - continuum fluxes in the optically thin approximation, \citealt{Hildebrand1983}, the assumptions on temperatures and opacities are as by \citealt{ManaraEtal2023} for T~Tauri stars and by \citet{StapperEtal2022} for Herbig disks). Column 7: Mass accretion rate. Column 8: SO $J_N$ rotational transition. Column 9: SO flux. Column 10: References (ordered by column). Source distances and line transition corrected flux comparisons are available in \autoref{app:correlations}.\\
    The stellar parameters of IRS~48 are based on radiative transfer modeling of near-infrared direct imaging and spectroscopy \citep{BrownEtal2012,SchworerEtal2017}. Its mass accretion rate was estimated form the Pf$\beta$ accretion line luminosity by adopting the stellar mass ($2.0M_\odot$) and radius ($1.4R_\odot$) of \citet{BrownEtal2012}. For UY~Aur~AB the stellar masses of \citet{HartiganKenyon2003} are consistent within their errors with those reported by \citet{LongEtal2019,ManaraEtal2019,ManaraEtal2023}, who, however, estimated a stellar luminosity of $1.94L_\odot$, SpT K7, and a much smaller system age of $0.9\pm0.4$~Myr. For DR~Tau, \citet{GangiEtal2022} did not report a stellar age, we thus adopted the estimate of \citet{LongEtal2019} from slightly different measurements of the stellar luminosity and effective temperature, still consistent within their uncertainties with those of \citet{GangiEtal2022}. }
    %\\ \textsc{References---1:}
    %For UY~Aur~AB \citet{HartiganKenyon2003} measured stellar properties for both components but do not report uncertainties.
    %For UY~Aur and DR~Tau, \citet{GangiEtal2022} and \citet{ManaraEtal2023} did not report stellar ages, we thus adopted those estimated by \citet{LongEtal2019} from slightly different, but still consistent within their uncertainties, measurements of the stellar luminosity and effective temperature.
    \tablerefs{1: \citet{HerczegEtal2023}, 2: \citet{Herczeg&Hillenbrand2014}, 3: \citet{MaciasEtal2021}, 4: \citet{YoshidaEtal2024}, 5: \citet{VioqueEtal2018}, 6: \citet{Gray2017}, 7: \citet{StapperEtal2022}, 8: \citet{GrantEtal2022}, 9: \citet{BoothEtal2024a}, 10: \citet{Mooley2013}, 11: \citet{WichittanakomEtal2020}, 12: \citet{DutreyEtal2024}, 13: \citet{SchworerEtal2017}, 14: \citet{BrownEtal2012}, 15: \citet{OhashiEtal2020}, 16: \citet{SalykEtal2013}, 17: \citet{BoothEtal2024b}, 18: \citet{BoothEtal2023}, 19: \citet{HartiganKenyon2003}, 20: \citet{ManaraEtal2023}, 21: \citet{TangEtal2014}, 22: \citet{GangiEtal2022}, 23: \citet{HuangEtal2023}, 24: \citet{GarciaLopezEtal2006}, 25: this work, 26: \citet{VieiraEtal2003}.}
    \label{tab:1_parameters}
\end{table*}

With the caveats of %a strong detection bias 
a higher detection occurrence towards Herbig stars and our small sample size, we did~not detect any correlations between SO luminosity and stellar mass, luminosity, accretion rate, or (sub-)mm continuum luminosity. However, we identified a tentative anti-correlation between SO luminosity and individual stellar age (cf., \autoref{app:correlations} for a more quantitative discussion on the correlations), suggesting that younger~systems~in our sample have stronger SO emission, perhaps reflecting its intrinsic origin. This tentative trend is qualitatively consistent with the evidence that SO is commonly detected in young embedded protostars. In fact, 76\% of~the~21 Class~0 sources part of the CALYPSO survey \citep{PodioEtal2021}, 86\% of the Class~0 and 57\% of the Class~I sources targeted by the PEACHES survey \citep{ArturdelaVillarmoisEtal2023}, as well as several protostars observed by the eDisk \citep[e.g.,][]{van'tHoffEtal2023,YamatoEtal2023,FloresEtal2023} and FAUST \citep[e.g.,][]{Codella2024,ZhangEtal2024_SO,OyaEtal2025} ALMA Large Programs show clear evidence of SO emission. Still, in many of these young sources SO is primarily tracing stellar outflows or disk winds, rather than the disk itself or shocks at the interface between the disk and its surrounding infalling envelope. Moreover, we acknowledge that age estimates of young stellar objects are highly uncertain \citep{SoderblomEtal2014} and our sample size is too small to draw a definitive conclusion. In fact, the lack of dependence of the SO flux on the other stellar and disk parameters might simply indicate that multiple SO origins coexist.

%Class 0 detection rate of  and Class I is. In both cases it's either extended and trace outflow cavity walls/ jet (mainly class 0s), or more compact and trace accretion shocks between disk and envelope (mainly class Is).

\subsection{Potential origins of SO emission}\label{sec:origins}
The production of SO in the gas-phase comes from two groups of processes: (1) (thermal) desorption or sputtering \citep[e.g.,][]{AotaEtal2015}, and (2) gas-phase~SO formation chemistry \citep[e.g.,][]{vanGelderEtal2021}, driven by the production of (2.1) the OH radical, either with endothermic reactions between H$_2$ and atomic oxygen, or by photodissociation of H$_2$O \citep[e.g.,][]{HartquistEtal1980}, or (2.2) the SH radical, by desorption of H$_2$S,~which is expected to be the major sulfur carrier in the ices \citep[e.g.,][]{VidalEtal2017}. For both these processes to work, an excess heating or UV source is required, thus explaining why SO is often suggested as a shock~tracer.~However, this might also explain %our current detection bias 
the higher occurrence~of~SO towards Herbig stars in our sample as a consequence of %, due to 
their stronger radiation field, especially in the presence of continuum cavities, where direct irradiation of the cavity walls can create favorable condition, both in terms of temperature and UV irradiation, for SO desorption or gas-phase formation. We discuss the possible origins of the detected~gas-phase SO as follows.

\subsubsection{Direct desorption of SO at the inner cavity edge}
The morphology of some sources in our sample, characterized by deep continuum cavities and SO emission arising primarily from or within their rims, suggests that the detected SO could simply be SO vapor sublimated from the icy pebbles trapped at the edge of these cavities. %From the disk morphology, the most straightforward interpretation is that the detected SO is simply SO vapor sublimated from the icy pebbles trapped at the edge of the cavity. 
Indeed, detailed modeling has shown that the gas temperature can rise steeply inside continuum cavities, where the gas is exposed to the direct heating and UV irradiation from the host star (e.g., \citealt{LeemkerEtal2022}, and \citealt{UbeiraGabelliniEtal2019} in the case of CQ~Tau). This ``passive'' desorption mechanism has been proposed as an explanation for SO emission in IRS~48 \citep{BoothEtal2021b} and HD~100546 \citep{BoothEtal2023}, %where the SO emission shows good overlap with the inner edge of the continuum cavities, 
and may also account for the SO emission morphology of CQ~Tau. % However, this is less clear for other sources,  this interpretation cannot explain the diversity of SO emission morphologies seen in our sample.

However, whether ``direct'' thermal desorption can be an explanation for SO emission also in other sources is less straightforward. Additionally, this mechanism cannot explain the diversity of SO emission morphologies~in our sample. For example, while SO in MWC~758 (\autoref{fig:3_summary}) and HD~169142 \citep{BoothEtal2023b} still arises close to the inner edge of the continuum cavity, it is too patchy to yield robust conclusions on its morphology.~In %For example, in MWC~758 and HD~169142 the SO emission, although still arising close to the inner edge of the cavity, is too patchy to yield robust conclusions on its morphology. 
AB~Aur, instead, the SO~emission~peaks slightly beyond, and not inside, the continuum ring~\citep[e.g.,][]{DutreyEtal2024,SpeedieEtal2025}. %Finally the two T~Tauri stars, TW~Hya and DR~Tau, do not show a clear large cavity. Thus, although it remains a possibility, the elevated temperature inside the cavity could not fully explain all the SO emission detected. Meanwhile, the cavity itself would still require an origin.
Finally, for the T~Tauri stars in our sample, they are either not known to host continuum cavities (DR~Tau and UY~Aur), or their SO emission is not co-located with such cavities when present (TW~Hya, \citealt{YoshidaEtal2024}).

Similar considerations apply to the desorption of other S-bearing volatiles, such as H$_2$S, that are needed to explain the observed SO emission via gas-phase formation.

\subsubsection{Spirals and companions}
Similar to CQ~Tau and MWC~758, almost all Class~II disks with SO detection in our sample show prominent spirals in near-infrared scattered light images: DR~Tau \citep{MesaEtal2022}, AB~Aur \citep{HashimotoEtal2011,BoccalettiEtal2020}, UY~Aur \citep{TangEtal2014}, IRS~48 \citep{FolletteEtal2015}, HD~100546 \citep{FolletteEtal2017}, and HD~169142 (possibly triggered by the protoplanet candidate HD~169142~b, \citealt{HammondEtal2023}). Furthermore, these disks exhibit significantly non-Keplerian gas kinematics \citep[e.g.,][]{CalcinoEtal2019,GargEtal2022,CasassusEtal2022,SpeedieEtal2024,WoelferEtal2023,WoelferEtal2025,TeagueEtal2025}, where spiral-like features have been identified in $^{12}$CO observations through residuals in line width, centroid velocity, and brightness temperature. Spirals are also present in the CO line emission~of TW~Hya \citep{TeagueEtal2022}, which is otherwise seemingly unperturbed in (sub-)mm continuum \citep[e.g.,][]{AndrewsEtal2016} and near-infrared imaging (\citealt{vanBoekelEtal2017}, but notice that \citealt{DebesEtal2017} identified clear asymmetries moving with non-Keplerian velocity in their scattered-light observations).

This consistent association between disk spirals and SO emission suggests a plausible link between complex gas dynamics and SO chemistry. Indeed, spiral structures in protoplanetary disks are potential drivers of dynamical and chemical activity \citep[e.g.,][]{IleeEtal2011}: as they propagate through the disk, spirals can compress gas, creating shocks that locally increase temperature and density \citep{OnoEtal2025}, favoring the direct thermal desorption of SO, or its gas-phase formation. %The presence of spirals in CQ~Tau and MWC~758 and other SO-bearing Class~II disks, therefore, provides potential evidence that spiral awakening can lead to SO formation \hjadd{\citep[though simulations designed for Class~0/I disks, see ][]{EvansEtal2015,EvansEtal2019}}.

While the origin of the spirals detected in CQ~Tau~and MWC~758 remains uncertain, (sub-)stellar companions are promising candidates to explain their origins~\citep[e.g.,][]{GoldreichTremaine1979,ZhuEtal2015b}.~In~fact, several companions have been proposed based on their disk morphology or claimed through high-contrast imaging \citep[e.g.,][]{DongEtal2015a,ReggianiEtal2018,UbeiraGabelliniEtal2019,HammondEtal2022,WagnerEtal2019b,WagnerEtal2023}. If these companions were responsible for exciting the spirals by launching Lindblad waves, their Keplerian orbital frequencies should reflect in the spiral motion.~However,~the~current~measurements~of~the spiral motion in these two sources suggest significantly lower angular velocities than those expected~from companions inside their cavities \citep[][]{RenEtal2018,SafonovEtal2022}, and at the time of writing, no such companions have been confirmed \citep[see e.g.,][]{UyamaEtal2020b,WagnerEtal2019b,WagnerEtal2024}. %However, it is noteworthy that an M2V star has been identified as a co-moving object $30\farcs$ away from CQ~Tau, confirmed by HST/STIS coronagraphic imaging and inspection of POSS (The Palomar Sky Survey) data \citep{GradyEtal2005}. Indeed, statistically, both close and wide binary fractions increase with stellar mass \citep{OffnerEtal2023}, but there is no clear dependence of SO flux on stellar mass as discussed. 
%The companion scenario contradicts the tentative trend of weaker SO in older systems, since the perturbation should be stronger if the companion is more massive, while the growth of the companion may take time, making it more likely to occur in older systems. Overall, we could not rule out unseen companions causing the spiral and the SO emission, which is left for future companion searches.

%An alternative explanation could be that these spirals are excited by the shadow of a misaligned inner disk \citep{ZhangEtal2024,ZiamprasEtal2024}.
As an alternative explanation, these spirals might have been excited by the shadow of a misaligned inner disk \citep[e.g.,][]{MontesinosEtal2016,ZiamprasEtal2025}. In this scenario, the spiral motion would follow the precession of the inner disk rather than the Keplerian velocity at the spiral's radial location \citep[see e.g.,][]{ZhuEtal2025}, thus leaving open the possibility that unseen inner disk companions could indirectly generate the spiral features detected in the outer disk by misaligning the inner disk through gravitational interactions \citep[e.g.,][]{FacchiniEtal2018a,NealonEtal2018}. %This leaves open the possibility that the misalignment of the inner disk is due to gravitational interactions with unseen companions \hjadd{in the inner disk} \citep{FoucartLai2014,FacchiniEtal2018a}, and indirectly leads to the spiral in the outer disk. 

Finally, we note that accreting companions themselves could also act as an excess heating source, either through accretion heating \citep{CleevesEtal2015,JiangEtal2023} or viscous heating around the circum-companion disks \citep{CridlandEtal2025}. %, leading to the desorption and/or gas-phase formation of SO. This has been proposed to be the origin of SO in HD~100546 and HD~169142 \citep{BoothEtal2023,LawEtal2023}.
However,~while~this~process~could lead to the desorption and/or gas-phase formation of SO and was proposed to be at the origin of SO emission in HD~100546 and HD~169142 \citep{BoothEtal2023,LawEtal2023}, its morphology is expected to be potentially more azimuthally localized than what is observed e.g., in CQ~Tau and MWC~758.

All in all, we argue that by locally~heating~up~a disk, dynamical perturbations, made evident by the presence of spiral-like features in scattered light observations or gas kinematics, potentially driven by the presence of (yet unseen) companions, could foster the production of gas-phase SO by thermal desorption or gas-phase chemistry. However, we stress that our hypothesis is based on the concurrent presence of SO emission and dynamical perturbations in a small sample with a significant fraction~of bright early type stars, thus making it more difficult to disentangle whether the formation of SO is driven by direct stellar irradiation or excess heating fostered by dynamical perturbations. %. Moreover, the evidence that spiral structures are more frequently detected in disks around early-type stars \citep[e.g.,][]{BenistyEtal2023}, makes it difficult to disentangle whether thermal desorption is driven by direct stellar irradiation of excess heating due to dynamical perturbations. %``passive'' (driven by direct stellar irradiation) and ``active'' (caused by excess heating due to dynamical perturbations) desorption.

\subsubsection{Environmental interactions and infall}
%Since there is no dependence of the SO on the isolated stellar and disk properties, 
Another possibility is that the observed SO emission is due to environmental interactions, i.e., slow shocks occurring at the interface between the disk and the infalling material \citep[][]{AotaEtal2015,vanGelderEtal2021}. This interpretation has been proposed to explain the SO emission detected in the Class~I disks around DG~Tau and HL~Tau \citep{GarufiEtal2022a}, IRS~63 \citep{FloresEtal2023}, and the Class~II disks around DR~Tau \citep{HuangEtal2024} and AB~Aur \citep{Riviere-MarichalarEtal2020,DutreyEtal2024,SpeedieEtal2025}, where prominent gas streamers traced by CO emission on scales of hundreds of au %that deviate from the Keplerian disk structure 
were detected in Northern Extended Millimeter Array (NOEMA) and ALMA observations \citep[cf.,][]{HuangEtal2023,SpeedieEtal2025}. 

Curiously, both sources with newly-reported SO detection introduced in the paper, CQ~Tau and MWC~758, are located near the Taurus star-forming region in close proximity to four other disks, HL~Tau, DG~Tau, DR~Tau, and AB~Aur, where SO emission was also observed. %The two sources with SO detections reported in this paper, CQ~Tau and MWC~758, are both located near the Taurus star-forming region, in close proximity to four other sources (HL~Tau, DG~Tau, DR~Tau, and AB~Aur) that also exhibit SO detections. 
Interestingly, the VLT/SPHERE GTO and DESTINYS surveys detected signs of ambient material in about a third of the sources in Taurus %the Taurus region 
\citep{GarufiEtal2024}, highlighting the importance of environmental interactions in disk evolution as suggested by recent theoretical studies \citep{WinterEtal2024,PadoanEtal2024}. 

While there is no direct evidence for ongoing infall in CQ~Tau nor MWC~758, a kilo-au-scale optical reflection nebulosity was detected by HST around CQ~Tau \citep{GradyEtal2005}. In fact, by searching~the~vicinity~of disks hosting large-scale gaseous structures~in~ALMA~data, \citet{GuptaEtal2023} found that all such sources have near-infrared and/or optical reflection nebulae in their vicinity, suggesting a strong correlation between large-scale nebulae and late-infall events. This raises the possibility that environmental interactions, although not directly visible in the current kinematic data, may have played a critical role in shaping the chemical and morphological properties of these disks in the past.

We note that companion-, spiral-, and infall-induced SO formation are not mutually exclusive. For instance, the infall of material from the surrounding environment has been proposed as a promising mechanism to generate spirals \citep[e.g.,][]{HennebelleEtal2017,KuffmeierEtal2019,KuffmeierEtal2020}. In this scenario, the observed spirals could be remnants of past infall events, which can persist on viscous timescales up to $>10^4$~yr in the outer region of disks \citep{CalcinoEtal2025}. If Bondi-Hoyle accretion onto disks is frequent \citep{WinterEtal2024,PadoanEtal2024}, disks may never fully evolve in isolation, providing a continuous source of perturbations to sustain SO production. %Once formed, the timescale for SO abundance decay depends on several parameters, \citep{GarufiEtal2022a} suggests that it takes $>10^3$~yr for the abundance to decrease by an order of magnitude \hjadd{\citep[but also see][who find shorter timescale when higher densities]{EvansEtal2015}}. Compared to the free fall timescale of $\sim10^4-10^5$~yr typically associated with observed streamers at $\sim10^2-10^3$au, this is not critically short, especially given the observed frequency of ongoing infall events. This alleviates concerns about SO detection in the absence of ongoing infall.

Meanwhile, infall-driven SO formation can explain the tentative anti-correlation between SO flux and system age. In a word, younger disks with more active dynamical environments, higher infall frequency and pre-shock gas density will have stronger detectable SO emission, and the decaying trend of SO could serve as a signal of cloud density dissipation.
%On the other hand, by assuming that the SO flux is proportional to the rate of decay of turbulent energy in the disk, late-infall models predict a correlation between SO emission and a combination of disk and stellar parameters, $F_{\rm SO}\propto(M_\star^2\,\dot{M}_{\rm acc}^2\,M_{\rm disk})^{1/3}$ (cf. \citealt{WinterEtal2024}), that we did not detect. 
However, this could well be a consequence of our very small sample sizes and inherent higher occurrence rate towards more massive stars and brighter disks. Indeed, the anti-correlation between SO flux and system age might simply reflect the different origins of SO emission in our sample: infall-driven in younger sources and excited by companions or shadows in more evolved systems.

\subsection{Implications for disk evolution and planet formation}
The possible link between SO emission and dynamical perturbations underscores the critical role of disk dynamics and environmental interactions in shaping disk chemistry. In this picture, SO serves as a valuable tracer of excess heating induced by spiral density waves, thus providing insights into the mechanisms driving disk evolution and potentially supporting the hypothesis that this process does not take place in isolation. 

The presence of potentially frequent dynamical perturbations raises two concerns. On the one hand, these disks may not always reach (quasi-)equilibrium chemistry, since the relatively long chemical timescales can be comparable to or even longer than the local dynamical timescale \citep{SemenovWiebe2011,OebergEtal2023}. On the other hand, these perturbations lead to significant azimuthal asymmetries, which will change the local material and energy sources for chemical reactions. Both could have major implications for the interpretation of the disk chemistry, as most works assume a static disk over the simulation time and set up the thermo-chemical model as azimuthally symmetric. For example, the CS/SO ratio may not be a robust indicator of the C/O ratio when CS and SO are forward modeled with static and passively irradiated disk models \citep[e.g.,][]{LeGalEtal2019a,KeyteEtal2023,HuangEtal2024}.

%The co-spatial distribution of the SO emission, and thus the potential shock front, with the continuum cavities also provides clues to the formation of disk substructures. If an inner companion is responsible for the spiral awakening and SO formation, the co-location would naturally manifest after the companion(s) open the cavities \citep{ArtymowiczLubow1996,ZhuEtal2011}. Alternatively, it has been shown in models that infalling material can also lead to ring formation \citep{KuznetsovaEtal2022}. Together with the fact that some of these transition disks show misalignment between the inner sub-au and the outer ALMA continuum disks \citep{BohmEtal2022}, it is therefore possible that some transition disks are built through an infall event \citep{KuffmeierEtal2021}. 

If confirmed, this link between SO formation and dynamical perturbations might %Our results 
have profound implications for planet formation. First, it shows that dynamical perturbations can significantly influence the distribution of volatiles, challenging the traditional view that planets inherit their composition solely from unperturbed disks \citep[e.g.,][]{OebergEtal2011, EistrupEtal2016}. In addition, the presence of SO near the rims of continuum disk cavities highlights these regions as chemically active zones, potentially influencing the composition of protoplanets forming within the rings \citep[e.g.,][]{vanderMarelEtal2021b,RampinelliEtal2024}. Future observations mapping the chemical inventory of CQ~Tau and MWC~758, such as those performed for IRS~48 and HD~100546 \citep{BoothEtal2024a,BoothEtal2024b}, will be crucial to assess the relationship between SO chemistry and dynamical perturbations and to better understand the composition of the planet-forming material.

\section{Summary and conclusions}\label{sec:5_summary}
We report the serendipitous detection of the SO~$J_N=6_5-5_4$ rotational line transition in archival ALMA observations of the planet-forming disks around CQ~Tau and MWC~758, two dynamically perturbed systems known for their prominent spiral structures in scattered-light observations. SO emission is detected with $4.9\sigma$ significance in CQ~Tau and $3.4\sigma$ significance in MWC~758. In both cases, it arises preferentially from the edge of the continuum cavity, but while in CQ~Tau it comes in the shape of a ring and is qualitatively compatible with Keplerian rotation, %might be associated is tentatively co-located with the bases of the scattered-light spirals, 
the SO emission is more extended, with a less clear morphology and kinematics %association with scattered light features 
in MWC~758.

These results add to the growing sample of SO-bearing Class~II disks, which now includes nine systems, six of which are Herbig stars with large inner cavities. We argue that the detection of SO in these dynamically perturbed disks suggests a possible link between disk dynamical perturbations and sulfur chemistry. In this picture, strong additional heating and/or turbulence-driven dust collisions, %sources, whether driven by spirals, companions, shadows, or infall, can enhance the production of SO in the gas phase. 
whether caused by yet unseen massive companions, spirals (due to companions themselves or shadows), or infall, can foster the thermal desorption~of SO or its production in the gas phase.~However,~we~stress that, since our sample is small and made up by a significant fraction of bright and massive Herbig stars, ``direct'' thermal desorption due to passive stellar irradiation remains a possible explanation for the origin of SO emission in at least some of these systems.

We found no statistically significant correlations between SO flux and stellar or disk parameters other than a tentative anti-correlation between SO flux and system age, either suggesting that younger disks, with more active dynamical environments, are likely to have stronger SO emission, or different origins for SO emission. This could be infall driven in younger sources and due to ``passive'' irradiation or ``active'' spiral-driven thermal desorption/dust shattering in older perturbed systems. %Comparing the SO flux with stellar and disk parameters in the SO-bearing disks, we found that the SO flux has no statistically significant correlation with stellar mass, luminosity, accretion rate, and dust disk mass. However, there is a tentative anti-correlation between SO flux and system age, suggesting that younger disks, with more active dynamical environments, are likely to have stronger detectable SO emission. In the context of infall-driven SO emission, this trend could therefore reflect higher infall rates and hence stronger shocks in the earlier evolutionary stages of protoplanetary disks. In addition, the infall could simultaneously explain the spiral structures and continuum cavity. Alternatively, we could see a dichotomy of young infall-driven SO formation and old companion/spiral-driven SO formation. 

The detection of SO in such dynamically active systems raises the possibility that chemical evolution might take place on significantly shorter timescales than those commonly assumed in disk evolution and planet formation models. Our results have important implications for interpreting disk chemistry and predicting the composition of forming planets, since tracers such as the CS/SO ratio might not be reliably used to infer the C/O ratio in dynamically perturbed and chemically active disks if forward models are not representative of the out-of-equilibrium environment conditions, and emphasize the need for future studies to consider chemical and dynamical evolution simultaneously. %The presence of SO near the cavity rims of transition disks highlights these regions as chemically active zones, potentially influencing the composition of forming planets \citep[e.g.,][]{JiangEtal2023,RampinelliEtal2024,CridlandEtal2025}. Meanwhile, the detection of SO in dynamically perturbed disks raises the possibility that disks may not always reach chemical equilibrium, challenging the traditional assumption of steady-state chemistry in disk evolution and planet formation models. This has important implications for interpreting disk chemistry and predicting the composition of forming planets, since tracers such as the CS/SO ratio may not be reliably used to infer the C/O ratio in dynamically active disks if the forward model is not representative of the out-of-equilibrium environment conditions, and emphasizes the need for future studies to consider chemical and dynamical evolution simultaneously.

Future higher resolution and sensitivity observations will be crucial to unravel the detailed mechanisms linking disk dynamics, chemistry, and planet formation in these systems. In particular, a larger sample of SO bearing disks, combined with detailed kinematic studies on both large and small scales, will help to clarify the role of environmental interactions and dynamical perturbations in shaping disk chemistry.

%% Also note that the akcnowlodgment environment does not support long amounts of text. If you have a lot of people and institutions to acknowledge, do not use this command. Instead, create a new \section{Acknowledgments}.
\begin{acknowledgments}
This paper makes use of the following ALMA data: 

\noindent ADS/JAO.ALMA\#2013.1.00498.S,

\noindent ADS/JAO.ALMA\#2016.A.00026.S,

\noindent ADS/JAO.ALMA\#2017.1.01404.S, and

\noindent ADS/JAO.ALMA\#2017.1.00940.S.

\noindent ALMA is a partnership of ESO (representing its member states), NSF (USA) and NINS (Japan), together with NRC (Canada), NSTC and ASIAA (Taiwan), and KASI (Republic of Korea), in cooperation with the Republic of Chile. The Joint ALMA Observatory is operated by ESO, AUI/NRAO and NAOJ. We are grateful to the anonymous referee for their insightful comments. F.Z. is grateful to Marta De Simone, Carlo F. Manara, and Anna Miotello for insightful discussions. This project has received funding from the European Research Council (ERC) under the European Union’s Horizon 2020 research and innovation programme (PROTOPLANETS, grant agreement No. 101002188). S.F. is funded by the European Union (ERC, UNVEIL, 101076613), and acknowledges financial contribution from PRIN-MUR 2022YP5ACE. Y.A. acknowledges the support by JSPS KAKENHI grant No. 24K00674. P.C. acknowledges support by the ANID BASAL project FB210003. J.D.I. acknowledges support from an STFC Ernest Rutherford Fellowship (ST/W004119/1) and a University Academic Fellowship from the University of Leeds. G.L. received funding from the European Union, Next Generation EU, CUP: G53D23000870006. F.Me. has received funding from the European Research Council (ERC) under the European Union's Horizon Europe research and innovation program (grant agreement No. 101053020, project Dust2Planets). G.R. is funded by the European Union under the European Union's Horizon Europe Research \& Innovation Programme No.101039651 (DiscEvol) and by the Fondazione Cariplo, grant No. 2022-1217. Views and opinions expressed are however those of the author(s) only and do not necessarily reflect those of the European Union or the European Research Council Executive Agency. Neither the European Union nor the granting authority can be held responsible for them. 
\end{acknowledgments}


\vspace{5mm}
\facilities{ALMA, VLT/SPHERE}


\software{
    \texttt{CASA} \citep{CASATeamEtal2022},
    \texttt{gofish} \citep{Teague2019}
    \texttt{bettermoments} \citep{TeagueForeman-Mackey2018}
    \texttt{numpy} \citep{HarrisEtal2020},
    \texttt{matplotlib} \citep{Hunter2007},
    \texttt{scipy} \citep{VirtanenEtal2020},
    \texttt{astropy} \citep{AstropyCollaborationEtal2013}
}



\appendix

\section{Data}\label{app:data}
We introduce the archival observations underlying our analysis. Key information is summarized in \autoref{tab:A1}.

\begin{table*}
    \caption{Summary of the archival ALMA observations analyzed in this letter.}
    \movetableright=-1.05in
    % \centering
    % \renewcommand{\arraystretch}{1.25}
    \begin{tabular}{ccccccccccc}
    \hline
    \multirow{2}{*}{Source} & \multirow{2}{*}{Project code} & \multirow{2}{*}{PI} & Obs. date & ToS & MBL & \multirow{2}{*}{\# ant.} & \multirow{2}{*}{Phase cal.} & \multirow{2}{*}{Flux cal.} & PWV & MRS \\
     & & & (dd/mm/yy) & (min) & (km) & & & & (mm) & ($''$) \\
    (1) & (2) & (3) & (4) & (5) & (6) & (7) & (8) & (9) & (10) & (11) \\
    \hline
    \hline
    \multirow{4}{*}{CQ~Tau} & 2013.1.00498.S & Pérez & 30/08/15 & 15.817 & 1.5 & 35 & J0550+2326 & J0510+1800 & 1.18 & 2.9 \\
    & 2016.A.00026.S & Testi & 07/08/17 & 19.656 & 3.7 & 41 & J0547+2721 & J0510+1800 & 0.93 & 1.3 \\
    & \multirow{2}{*}{2017.1.01404.S} & \multirow{2}{*}{Testi} & 20/11/17 & 29.083 & 8.5 & 44 & \multirow{2}{*}{J0521+2112} & \multirow{2}{*}{J0510+1800} & 0.53 & 0.8 \\
    & & & 23/11/17 & 29.050 & 8.5 & 48 & & & 0.46 & 1.0 \\
    \hline
    \multirow{9}{*}{MWC~758} & \multirow{9}{*}{2017.1.00940.S} & \multirow{9}{*}{Ricci} & 10/10/17 & 44.633 & 16.2 & 47 & \multirow{9}{*}{J0521+2112} & \multirow{9}{*}{J0510+1800} & 0.40 & 0.3 \\
    & & & 10/10/17 & 44.617 & 16.2 & 47 & & & 0.41 & 0.3 \\
    & & & 11/10/17 & 44:600 & 16.2 & 47 & & & 1.31 & 0.3 \\
    & & & 15/10/17 & 44:667 & 16.2 & 52 & & & 2.45 & 0.4 \\
    & & & 16/10/17 & 44.717 & 16.2 & 52 & & & 2.74 & 0.4 \\
    & & & 09/10/17 & 38.500 &  3.1 & 44 & & & 2.24 & 1.6 \\
    & & & 17/10/17 & 38.517 &  2.5 & 45 & & & 2.18 & 2.4 \\
    & & & 27/10/17 & 38.550 &  2.5 & 46 & & & 2.38 & 2.3 \\
    & & & 28/10/17 & 38.517 &  2.5 & 45 & & & 1.04 & 2.5 \\
    \hline
    \end{tabular}
    \tablecomments{Column 1: Source. Column 2: Project code. Column 3: PI. Column 4: Observation date. Column 5: Integration time on the science target (time on source, ToS). Column 6: Maximum baseline length (MBL). Column 7: Number of antennas. Column 8 and 9: Phase and flux calibrators. Column 10: Precipitable water vapor (PWV). Column 11: Maximum recoverable scale (MRS).}
    \label{tab:A1}
\end{table*}

\paragraph{CQ~Tau} The observations were conducted in Cycles~2, 4, and 5 between Aug.~2015 and Nov.~2017, as part of the programs 2013.1.00498.S (PI: L.~Pérez), 2016.A.00026.S (PI: L.~Testi), and 2017.1.01404.S (PI: L.~Testi). These programs used only Band~6 receivers and the data were correlated from eight spectral windows (SPWs) in dual polarization mode. Six SPWs were set in frequency division mode (FDM), one with 1920 channels 244.1~kHz-wide centered at 230.7~GHz, two with 960 channels 448.3~kHz-wide centered at 219.7 and 220.2~GHz, two with 60 channels 7.8~MHz-wide centered at 218.8 and 219.3~GHz, one with 120 channels 3.9~MHz-wide centered at 231.2~GHz, all spanning a total bandwidth of 468.8~MHz. The remaining two SPWs, centered at 217.0 (2013.1.00498.S), 217.6 (2016.A.00026.S and 2017.1.01404.S) and 232.4~GHz, were set in time division mode (TDM), with 128 channels 15.6~MHz-wide, and spanned a total bandwidth of 2.0~GHz.

For the program 2013.1.00498.S, the observations used an array with baseline lengths from 15.1~m to 1.5~km (resolution of $\approx0\farcs260$) and 35 antennas in the C34-7 configuration. The total integration time on the science target was $\approx15$~min. The bandpass calibrator J0423-0120 and the flux calibrator J0510+1800 were observed at the beginning of the observing block, the phase calibrator J0550+2326 and an additional delay calibrator J0547+2721 were observed in an alternating sequence with the science target, every $\approx10$ and 15~min.

For the program 2016.A.00026.S the observations used an array with baseline lengths from 81.0~m to 3.7~km (resolution of $\approx0\farcs110$) and 41 antennas in the C40-7 configuration. The total integration time on the science target was $\approx20$~min. The bandpass and flux calibrator J0510+1800 was observed at the beginning of the observing block, the phase calibrator J0547+2721 was observed in an alternating sequence with the science target, every $\approx1$ to 2~min. An additional ``check'' calibrator, J0521+2112, was observed every $\approx7$~min to assess the quality of the phase transfer.

For the program 2017.1.01404.S, the observations used an array with baseline lengths from 92.1~m to 8.5~km (resolution of $\approx0\farcs060$). Two execution blocks (EBs) were scheduled, with 44 and 48 antennas, respectively, in the C43-8 configuration. The total integration time on the science target was $\approx30$~min for each EB. The bandpass, flux, and ``check'' calibrator J0510+1800 was observed at the beginning of the observing block, then every $\approx10$~min to assess the quality of phase transfer, the phase calibrator J0521+2112 was observed in an alternating sequence with the science target, every $\approx1$ to 2~min.

\paragraph{MWC~758} The observations were conducted in Cycle~5 between Oct. and Dec. 2017 as part of the program 2017.1.00940.S (PI: L.~Ricci). The data were correlated from four SPWs in dual polarization mode. Two SPWs were set in FDM each with 1920 channels 976.6~kHz-wide centered at 219.5 and 231.5~GHz and spanning a total bandwidth of 1.9~GHz. The remaining SPWs, centered at 217.5 and 234.2~GHz, was set in TDM, with 128 channels 15.6~MHz-wide, and spanning a total bandwidth of 2.0~GHz. The target was observed in the more compact C43-6 configuration (short baselines, four scheduled EBs), and the more extended C43-10 configuration (long baselines, five scheduled EBs).

The observations at short baselines used an array with baseline lengths from 15.1~m to 2.5 or 3.1~km (resolution of $\approx0\farcs130$) and between 44 and 46 antennas. The total integration time on the science target was $\approx40$~min for each EB. The bandpass and flux calibrator J0510+1800 was observed at the beginning of each EB, while the phase calibrator J0521+2112 was observed in an alternating sequence with the science target, every $\approx10$~min. An additional ``check'' calibrator, J0519+2744, was observed every $\approx30$~min to assess the quality of the phase transfer.

The observations at long baselines used an array with baseline lengths from 41.4 or 233.0~m to 16.2 km (resolution of $\approx0\farcs020$) and between 47 and 52 antennas. Five EBs were scheduled. The total integration time on the science target was $\approx45$~min for each of them. The bandpass and flux calibrator J0510+1800 was observed at the beginning of each EB, while the phase calibrator J0521+2112 was observed in an alternating sequence with the science target, every $\approx1$ to 2~min. An additional ``check'' calibrator, J0518+2054, was observed every $\approx15$~min to assess the quality of the phase transfer.

\section{Self-calibration}\label{app:self-cal}
To begin with, we created pseudo-continuum datasets flagging channels within $\pm15$~km~s$^{-1}$ of the systematic velocity around the rest frequencies of the $^{12}$CO ($J=2-1$ at 230.5380~GHz), $^{13}$CO ($J=2-1$ at 220.3986~GHz), C$^{18}$O ($J=2-1$ at 219.5603~GHz), DCN ($J=3-2$ at 217.2385~GHz), SO ($J_N=6_5-5_4$ at 219.9494~GHz),~and H$_2$CO ($N_{K_a,K_c}=3_{0,3}-2_{0,2}$, $3_{2,1}-2_{2,0}$, $3_{2,2}-2_{2,1}$, and $9_{1,8}-9_{1,9}$ at 218.2221, 218.7601, 218.4756, and 216.5686~GHz) transitions. 

These pseudo-continuum EBs were channel-averaged ensuring a $<1\%$ reduction in peak response to a point source at the edge of the primary beam to avoid bandwidth smearing, according to the criterion of \citet{BridleSchwab1999}, and imaged (separately) over the entire primary beam, using the task \texttt{tclean}. CLEANing was performed over a user-defined elliptical mask (of $0\farcs85\times0\farcs72$, ${\rm PA}=55$~deg for CQ~Tau and $0\farcs80\times0\farcs80$, ${\rm PA}=62$~deg for MWC~758 that covers all the continuum emission, with a threshold of $\approx6\sigma$, where~$\sigma$ is the root mean squared (RMS) noise calculated over an emission-free annular region between 4$\farcs$ and 6$\farcs$ centered on the source phase center and larger than the continuum emission extent, and $\approx8$ pixels along the synthesized beam minor axis. We adopted the \texttt{multiscale} deconvolver \citep{Cornwell2008}, with scales including a point source and some (sub-)multiples of the synthesized beam that could be accommodated within the user-defined mask ([0, 2, 4, 8, 12] for short baseline and [0, 2, 4, 8, 16, 24] for long baseline observations), with preference for smaller ones (\texttt{smallscalebias=0.6}), \texttt{cycleniter=300}, \texttt{gain=0.3}, and a \texttt{briggs} weighting scheme with \texttt{robust=0.5} \citep{Briggs1995}. For CQ~Tau, since some of the data were taken in early ALMA cycles, we recomputed the weights using the task \texttt{statwt} combining different polarisations, scans, and SPWs. %actually for the first CLEANing I used [0,2,4,8] for the short-baseline observations... also the mask was 0.75\times0.65 for CQ~Tau long-baseline data for the first CLEANing... 

Subsequently, a round of phase-only self-calibration was performed on each EB (separately) to increase their signal-to-noise ratio (SNR) prior to alignment. The gain solutions were computed using the task \texttt{gaincal} over an infinite solution interval, for both polarisations together, combining all the available SPWs and time scans, and rejecting those gain solutions corresponding to antennas with $<3$ baselines and ${\rm SNR}<4$. We then shifted the phase centers of these self-calibrated EBs to a common direction chosen to be that of the phase center of the long baseline EB with the highest peak SNR. The alignment was performed in visibility space, minimizing the phase differences between EBs in the widest continuum SPW, and checked for by comparing the ratio images (reconstructed using the same parameters described above) of each dataset to the reference EB (see \citealt{CasassusCarcamo2022,LoomisEtal2025}). While in the case of CQ~Tau the peak SNR of each self-calibrated EB was high-enough ($\gtrsim65$) to provide a significant reduction of the amount of structures in the ratio images, indicative of an accurate alignment, for MWC~758, the much lower peak SNR, especially for the long baseline EBs ($\lesssim15$), often led to a (marginal) worsening of the EB alignment. Thus, since the direct inspection of the images suggested that the EBs were already almost perfectly aligned, for MWC~758 we chose not to shift the EBs.

For both sources, we inspected the EBs for amplitude-scale offsets, plotting the ratio of their projected visibilities to those of a reference EB. For CQ~Tau, the EB with highest peak SNR shows flux offsets $>30\%$ compared to the others. Thus, we adopted as a reference EB the one taken on Nov. 23, 2017, because of its lower phase RMS (22.58 deg) and higher peak SNR compared to the remaining ones. Instead, for MWC~758, we selected as a reference EB the one taken on Dec. 28, 2017, because its peak SNR is the highest among the available EBs, and its phase RMS (16.72 deg) is lower than for the other SBs. For both sources, most EBs show a (sometimes significant) dependence of the visibility amplitude ratio on the deprojected baselines, which we interpreted as an indication of phase-decoherence. For this reason, although for some EBs flux-scale offsets $>4\%$ were estimated, at this stage we chose not to rescale their amplitudes. Instead, we performed a first iteration of phase-only self-calibration, first on the concatenated short-baseline observations and then combining short- and long-baseline data to correct for decoherence and identify more accurate flux-scale offsets. %Any residual amplitude mismatches $>4\%$ were then corrected for, and a second self-calibration iteration was run (Loomis et al., subm.).

Self-calibration was performed iteratively, first on the short-baseline observations and then on the combined short- and long-baseline EBs (e.g., \citealt{AndrewsEtal2018b}). The images were reconstructed after each step adopting the same fiducial parameters described above, except for the long-baseline observations of MWC~758, whose best trade-off between angular resolution and peak SNR was obtained for \texttt{robust=1.0}. For the short-baseline data, the gain solutions were first computed separately for each polarization on an infinite solution interval, combining different SPWs and time scans. Then, on progressively shorter solution intervals of 360, 120, 60, and 20~s, for both polarisations together, combining different SPWs and time scans (but no scans were combined for solution intervals shorter than the average scan length). The gain solutions for antennas with $<4$ baselines and ${\rm SNR}<3$ were flagged. Visual inspection led to further manual flagging of too large phase-gain solutions. For the long-baseline data, we followed a similar procedure, with solution intervals decreasing from the full EB, to 360, 120, 60, 30, and 18~s\footnote{For MWC~758 we only considered \texttt{solint} parameters $\geq60$~s.~For shorter solution intervals the peak SNR leveled off and the phase gain solutions obtained for the EBs taken on Dec. 15 and 16,~2017 became very extreme ($>150$ or $<-150$~deg) and heavily time-dependent for most antennas, leading to substantial flux-scale offset variations among the long-baseline EBs.}, rejecting gain solutions corresponding to antennas with $<4$ baselines and ${\rm SNR}<2$ (for CQ~Tau) or 3 (for MWC~758). For CQ~Tau, the peak SNR increased from 112.77 to 151.32 ($\approx34\%$) after self-calibration of the short-baseline data and from 138.88 to 181.98 ($\approx31\%$) after self-calibration of the combined short and long baseline EBs. For MWC~758, the peak SNR increased from 317.43 to 442.55 ($\approx39\%$) after self-calibration of the short-baseline observations~and from 171.13 to 186.57 ($\approx9\%$) after self-calibration of the combined short and long baseline EBs.%\footnote{In this case we chose to stop at solution intervals of 30~s. For shorter intervals, the gain solutions obtained for the EBs taken on Dec. 15 and 16, 2017, become very high and heavily time-dependent, leading to flux-scale offset variations $>6\%$ and a marginal decrease of peak SNR to 87.81.}. %For MWC~758 you also checked the phase and amplitude gains and rescaled all EB offsets regardless of their significance...

We re-inspected the amplitude ratios between different EBs and the reference ones after each self-calibration step, finding that the flux offsets were progressively reduced and decoherence decreased in most cases. Then, we re-scaled those EBs that showed flux ratios $>4\%$ and performed a new self-calibration iteration. For CQ~Tau the peak SNR increased from 148.65 to 201.44 ($\approx36\%$) after self-calibration of the short-baseline data and from 164.87 to 222.13 ($\approx35\%$) after self-calibration of the combined short and long baseline EBs. For MWC~758, the peak SNR increased from 322.03 to 450.81 ($\approx40\%$) after self-calibration of the short-baseline observations and from 167.33 to 176.44 ($\approx5\%$) after self-calibration of the combined short and long baseline EBs. %\footnote{Similarly to the previous iteration, we chose to stop at solution intervals of 60~s. For shorter intervals, the gain solutions obtained for the EBs taken on Dec. 16, 2017, become very high and heavily time dependent, leading to flux-scale offset variations $\approx10\%$.}. 

We then performed two cycles of amplitude and phase self-calibration. The solutions were obtained with an infinite solution interval, first combining all the available SPWs and scans, then only the SPWs, for both polarisations together. Images were reconstructed with the same fiducial parameters described above and a CLEANing threshold of $\approx1\sigma$. In all cases, gain solutions for antennas with $<4$ baselines and ${\rm SNR}<5$ were flagged. Amplitude gains $>20\%$ were also flagged, and phase-gain solutions were further visually inspected and manually flagged when too large. For CQ~Tau both steps were successful and led to an improvement in the peak SNR from 222.35 to 224.17 ($\lesssim1\%$), while for MWC~758 the peak SNR improved from 167.98 to 179.98 ($\approx7\%$) after the first step and no second iteration was performed\footnote{Because the too short solution intervals (scan length $\approx30$ to 60~s) led to too high gain solutions for most of the antennas.}. %After amplitude and phase self-calibration, the flux offsets among EBs were $\lesssim0.01\%$ for CQ~Tau and $\lesssim3\%$ for MWC~758.

Finally, we applied the phase shifts, flux scaling, and gain solutions to the original datasets (i.e., those including the previously flagged line emission channels) and time-binned the self-calibrated measurement sets by 30~s to reduce the data volume. Then, using the CASA \texttt{v6.2.1-7} implementation of the \texttt{uvcontsub} task, with \texttt{solint=1}, \texttt{fitorder=1}, and excluding the same channels used to create the pseudo-continuum datasets, these measurement sets were continuum-subtracted and independent datasets for each relevant rotational transition were created.%, spanning a range within $\pm30$~km~s$^{-1}$ of their systematic velocities.

The calibration and imaging scripts are available \href{https://github.com/fzagaria/selfcal_CQTau_and_MWC758}{here}.

\section{Imaging and summary plots}\label{app:imaging}
The continuum images were reconstructed from the measurement sets obtained after the last step of phase and amplitude self-calibration, time-averaged by 30~s. These (line-flagged) datasets were CLEANed with the same fiducial parameters described in \autoref{app:self-cal}, \texttt{robust=0.0} for CQ~Tau and \texttt{robust=1.0} for MWC~758, a noise threshold of $\approx1\sigma$ (the RMS was measured on an emission-free circular annulus between $4\farcs$ and $6\farcs$ centered on the source phase center using the task \texttt{imstat}), and \texttt{gain=0.02} (to minimize any artifacts in the continuum cavities). Our fiducial images are displayed in \autoref{fig:1_continuum}, while \autoref{tab:A2} summarizes their imaging parameters and continuum properties. %As is commented on in \autoref{sec:3_results}, the continuum emission from both sources shows significant evidence of ongoing dynamical perturbations in the form of high-contrast asymmetries and arc-like features. To make this clearer, the left panel of \autoref{fig:A0_CQTau_highpass} shows the continuum emission from CQ~Tau through a high-pass filter (obtained subtracting to the \texttt{robust=0.5} continuum image its convolution with a 2D Gaussian kernel). Three arcs veering from the continuum ring towards the central cavity are clearly visible. One of them (labeled ``Arc 1'' in the plot) aligns with one the spirals detected in scattered-light images, as is shown in the left panel, suggesting that they might be tracing the same disk structure.

\begin{table*}
    \caption{Continuum emission imaging parameters and key properties.}
    % \movetableright=-1.05in
    \movetableright=-0.65in
    % \centering
    % \renewcommand{\arraystretch}{1.25}
    \begin{tabular}{ccccccccc}%ccc
    \hline
    \multirow{2}{*}{Source} & Wav. & \multirow{2}{*}{Rob.} & Beam size & PA & RMS noise & Peak intensity & \multirow{2}{*}{Peak SNR} & Integrated flux \\%& $R_{\rm 68}$ & $R_{\rm 95}$ & $R_{\rm cav}$ \\
     & (mm) & & ($''^2$) & (deg) & (mJy beam$^{-1}$) & (mJy beam$^{-1}$) & & (mJy) \\%& ($''$) & ($''$) & ($''$) \\
    (1) & (2) & (3) & (4) & (5) & (6) & (7) & (8) & (9) \\%& (10) & (11) & (12) \\
    \hline
    \hline
    CQ~Tau  & 1.33 & 0.0 & $0.055\times0.039$ & $-4.1$ & $1.43\times10^{-3}$ & 1.29 & 90.04 & $146.80\pm0.37$ \\%& & & \\
    %MWC~758 & 1.33 & 0.5 & $0.045\times0.027$ & $ 8.0$ & $6.62\times10^{-3}$ & 1.11 & 172.33 & $58.01\pm0.30$ \\%& & & \\
    MWC~758 & 1.33 & 1.0 & $0.066\times0.044$ & $ 8.5$ & $6.46\times10^{-3}$ & 1.11 & 172.33 & $58.01\pm0.30$ \\%& & & \\
    \hline
    \end{tabular}
    \tablecomments{Column 1: Source. Column 2: Average observation wavelength. Column 3: Briggs weighting \texttt{robust} parameter. Column 4 and 5: Synthesized beam size and PA. Column 6: RMS noise. Column 7: Peak intensity. Column 8: Peak SNR. Column 9: Integrated flux. Uncertainties were estimated as the standard deviation of the integrated fluxes measured in 24 off-source non-overlapping copies of the mask used to image the continuum over.}% Column 10 and 11: Dust disk size, defined as the disk radius enclosing 68\% and 95\% of the integrated flux. Column 12: Cavity radius, defined as .}
    \label{tab:A2}
\end{table*}

\begin{figure*}[t!]
    \centering
    \includegraphics[width=\textwidth]{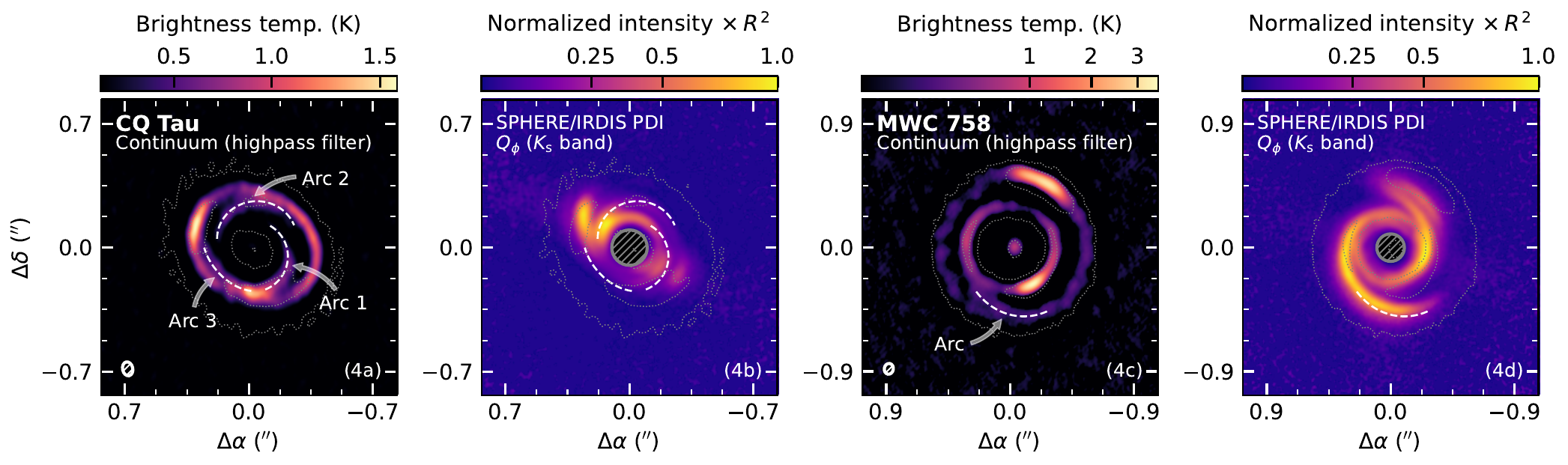}
    \caption{Highpass-filtered 1.3 mm continuum emission (panels 4a, 4c), and polarized intensity (panels 4b, 4d), for CQ Tau and MWC 758. The dotted gray lines display the $[5,65]\times\sigma$ (CQ~Tau) and the $[5,40]\times\sigma$ (MWC~758) emission contours. The synthesized CLEAN beam is shown as an ellipse in the bottom left corner of each continuum panel. The regions within 0\farcs1 of the scattered light images cannot be accessed because of the coronagraph and are masked out.}
    \label{fig:A1_continuum}
\end{figure*}

As is commented on in \autoref{sec:3_results}, the continuum emission from both sources shows significant evidence of ongoing dynamical perturbations in the form of high-contrast asymmetries. Such asymmetries show up more distinctly in panels a and c of \autoref{fig:A1_continuum}, that display the highpass-filtered continuum emission from both sources. These highpass-filtered maps were obtained by subtracting to the images reconstructed with \texttt{robust=0.5} and \texttt{1.5} their convolution with Gaussian kernels with a standard deviation of 5 and 8 pixels, respectively. In the case of MWC~758, \citet{DongEtal2018a} had already reported the presence of an azimuthally-wide arc, co-located with one of the scattered-light spirals (cf. panel d, that shows the $K_{\rm s}$ band $Q_\phi$ polarized differential image published by \citealt{RenEtal2023b}), that connects the cavity rim with a bright annular structure. For CQ~Tau, the higher SNR and angular resolution of our combined dataset allowed detecting, for the first time, three arcs veering from the continuum ring towards the central cavity. One of them (labeled ``Arc 1'' in panel a) is connected with one of the spirals detected in scattered light (cf. panel b, that also shows the $K_{\rm s}$ band $Q_\phi$ polarized differential image published by \citealt{RenEtal2023b}), suggesting that they might be tracing the same physical disk structure.

% \begin{figure*}
%     \centering
%     \includegraphics[width=0.8\textwidth]{figures/CQ_Tau_highpass.pdf}
%     \caption{Left: the high-pass filtered continuum emission from CQ~Tau shows three arc-like features connecting the continuum cavity and the ring outer edge. Right: ``Arc 1'' in the continuum is well aligned with one of the spirals detected in scattered-light observations.}
%     \label{fig:A0_CQTau_highpass}
% \end{figure*}

For SO, the continuum-subtracted measurement sets were imaged using \texttt{tclean} with \texttt{specmode=`cube'}, 23 channels of 0.7~km~s$^{-1}$ (corresponding to absolute velocities from $-1.3$ to 13.9~km~s$^{-1}$ in the LSRK reference frame) for CQ~Tau and 7 channels of 1.4~km~s$^{-1}$ (corresponding to absolute velocities from 1.7 to 10.1~km~s$^{-1}$ in the LSRK reference frame) for MWC~758 (plum area in the spectra in \autoref{fig:2_spectra}), a \texttt{multiscale} deconvolver and scales including a point source and sub-multiples of the synthesized beam (i.e., [0, 2, 4, 8]). CLEANing was performed with a common restoring beam across different channels, $\approx8$ pixels along the synthesized beam minor axis, a $1\sigma$ threshold (the RMS was estimated as for the continuum), and a conservative \texttt{gain=0.01}, over a user-defined mask, either the same one used for the continuum (equally across all the channels), or a Keplerian mask (generated using the code of \citealt{Teague2020} and the geometric parameters of \citealt{CuroneEtal2025} and \citealt{IzquierdoEtal2025}, listed in \autoref{tab:A3}). For MWC~758 some differences between the images reconstructed with the continuum and Keplerian masks can be seen due to clear non-Keplerian emission features (see e.g., teardrop plot in \autoref{fig:2_spectra}). Different weighting schemes, \texttt{robust} parameters, and Gaussian $uv$-tapers were tested. %; summary plots for each of these parameters are available at this \red{link}. 

\begin{figure*}
    \centering
    \includegraphics[width=\textwidth]{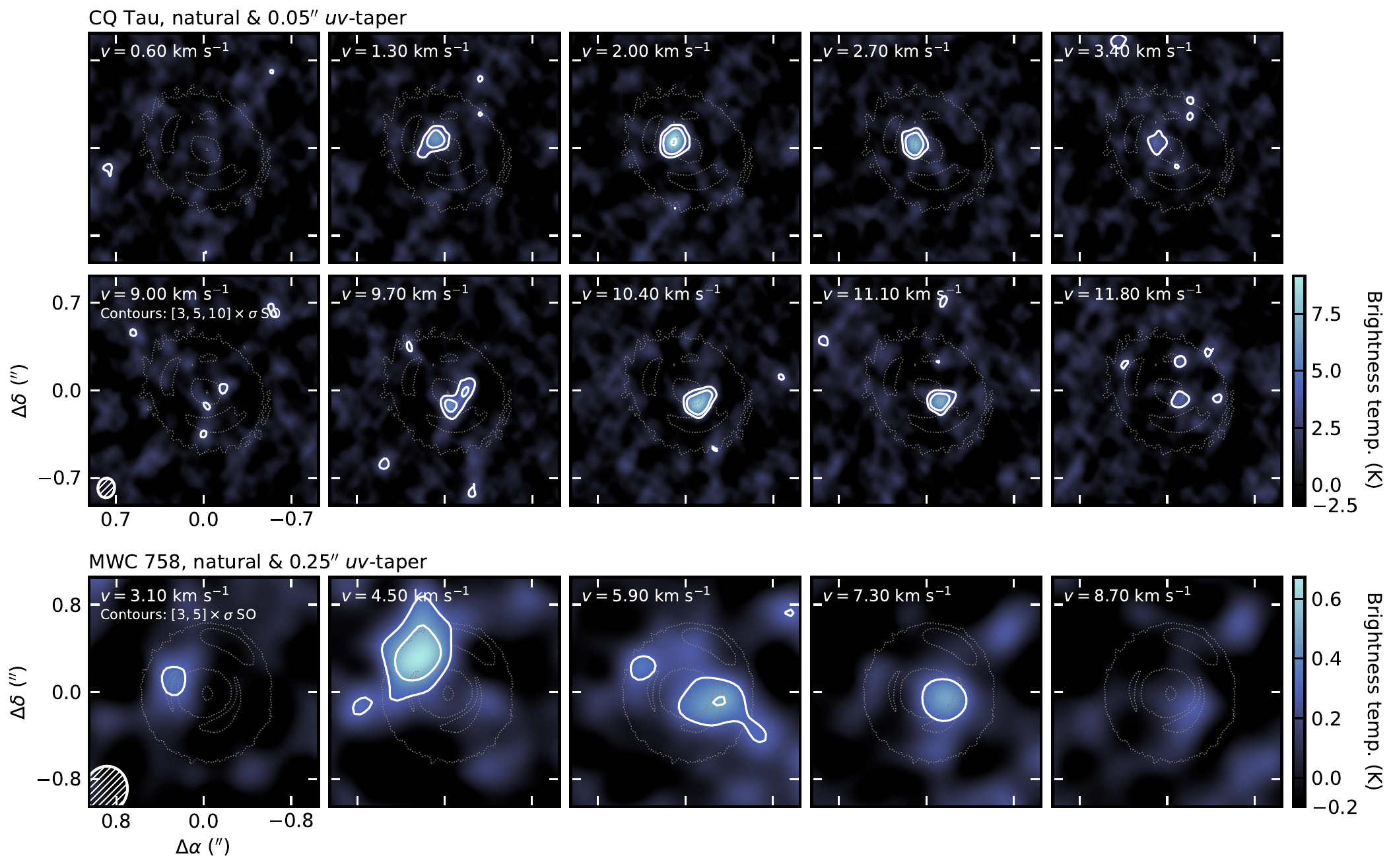}
    %\caption{CQ~Tau. Top panel: SO channel maps. Bottom panels (from left to right): SO integrated intensity map, shifted and stacked spectrum (blue for the SO transition), teardrop plot (where the white dotted line marks the location where the $^{12}$CO emission drops below 3~K, and the dashed line the location of the continuum cavity). The integrated SO flux and its uncertainty are annotated in the central panel.}
    \caption{SO channel maps for CQ~Tau ($v_{\rm sys}=6.2$~km~s$^{-1}$, top rows) and MWC~758 ($v_{\rm sys}=5.9$~km~s$^{-1}$, bottom row). The dotted gray contours display the $[5,65]\times\sigma$ (CQ~Tau) and the $[5,40]\times\sigma$ (MWC~758) continuum emission levels. The synthesized CLEAN beam is shown as an ellipse in the bottom left corner of the bottom left panels.}
    \label{fig:A2_channels}
\end{figure*}

Our fiducial images were reconstructed with \texttt{natural} weighting scheme and a Gaussian $uv$-taper with $0\farcs05$ (CQ~Tau) and $0\farcs25$ (MWC~758) full width at half maximum. Some relevant channel maps are displayed in~\autoref{fig:A2_channels} and their imaging parameters and emission properties are summarized in \autoref{tab:A4}. These~same~fiducial~images were also used to generate the integrated~intensity, peak intensity, and velocity maps displayed in~\autoref{fig:3_summary}, using \href{https://bettermoments.readthedocs.io/en/latest/index.html}{\texttt{bettermoments}}, a python package that collapses (smoothed and spatially or spectrally masked) image cubes along the velocity axis to generate moment maps \citep{TeagueForeman-Mackey2018}. Because of the moderate detection significance of SO, we did not apply any channel or noise clipping. No clear signs of spacial noise correlations can be seen in any of the moment maps, nor in any channel map, for all weighting schemes. %%Our fiducial images, reconstructed with \texttt{natural} weighting scheme and a Gaussian $uv$-taper with $0\farcs05$ (CQ~Tau) and $0\farcs25$ (MWC~758) full width at half maximum, were used to generate the peak intensity and velocity maps displayed in \autoref{fig:3_summary}. Some relevant channel maps are shown in \autoref{fig:A2_channels}, \autoref{tab:A4} summarizes their imaging parameters. \red{In the integrated and peak intensity maps, as well as separately in each channel map, no sign of small or large spacial scale noise correlations can be seen for all weighting schemes.}

\begin{table*}
    \caption{Deprojection and stacking parameters (adopted from \citealt{IzquierdoEtal2025}).}
    \movetableright=1.325in
    \begin{tabular}{cccccc}
    \hline
    \multirow{2}{*}{Source} & $i$ & PA & $M_\star$ & $d$ & $v_{\rm sys}$ \\
     & (deg) & (deg) & ($M_\odot$) & (pc) & (km s$^{-1}$) \\
    (1) & (2) & (3) & (4) & (5) & (6) \\
    \hline
    \hline
    CQ~Tau  & $-36.3$ & 235.1 & 1.40 & 149.4 & 6.19 \\
    MWC~758 & $ 19.4$ & 240.3 & 1.40 & 155.9 & 5.89 \\
    \hline
    \end{tabular}
    \tablecomments{Column 1: Source. Column 2: Inclination. Column 3: Position angle. Column 4: Stellar mass. Column 5: Distance. Column 6: Systematic velocity. (The fiducial stellar masses adopted in our analysis were inferred dynamically by modeling CO emission \citep{IzquierdoEtal2025} and are consistent, within the uncertainties, with the spectroscopically-based estimates reported in \autoref{subsec:2.1_sources} and \autoref{tab:1_parameters}. We prefer such dynamical estimates because they are not affected by the systematics in the pre-main-sequence evolutionary tracks (cf., \citealt{GarciaLopezEtal2006,VioqueEtal2018}) and for consistency with exoALMA.}
    \label{tab:A3}
\end{table*}

\begin{table*}
    \caption{SO emission imaging parameters and key properties.}
    \movetableright=-0.50in
    % \centering
    % \renewcommand{\arraystretch}{1.25}
    \begin{tabular}{cccccccc}
    \hline
    \multirow{2}{*}{Source} & \multirow{2}{*}{Weighting} & $uv$-tapering & Beam size & PA & Avg. RMS noise & Peak intensity & \multirow{2}{*}{Peak SNR} \\
     & & ($''$) & ($''^2$) & (deg) & (mJy beam$^{-1}$) & (mJy beam$^{-1}$) & \\
    (1) & (2) & (3) & (4) & (5) & (6) & (7) & (8) \\
    \hline
    \hline
    CQ~Tau  & \texttt{natural} & 0.05 & $0.159\times0.133$ & $-6.76$ & $7.44\times10^{-1}$ & 7.67 & 10.31 \\
    MWC~758 & \texttt{natural} & 0.25 & $0.428\times0.326$ & $-11.73$ & $6.61\times10^{-1}$ & 4.39 & 6.65 \\
    \hline
    \end{tabular}
    \tablecomments{Column 1: Source. Column 2: Weighting. Column 3: $uv$-tapering. Column 4 and 5: Synthesized beam size and PA. Column 6: Average RMS noise across the line-emission channels. Column 7: Peak intensity. Column 8: Peak SNR.}
    \label{tab:A4}
\end{table*}

%Hereafter, we show some relevant channel maps, the integrated intensity map, shifted and stacked spectra, and teardrop plots generated from the same fiducial images for CQ~Tau in \autoref{fig:A1_CQTau_summary} and MWC~758 in \autoref{fig:A2_MWC_758_summary}.

% \begin{figure*}
%     \centering
%     \includegraphics[width=\textwidth]{figures/MWC_758_SOchannels_natural_1.0sigma_uvtaper0.05arcsec_0.01gain_8.0ppb_goodchannels.png}
%     \includegraphics[width=\textwidth]{figures/MWC_758_SOsummaryplot_integrated_natural_1.0sigma_uvtaper0.05arcsec_0.01gain_8.0ppb_bestchans_spectra.pdf}
%     \caption{Same as in \autoref{fig:A1_CQTau_summary} for MWC~758.}
%     \label{fig:A2_MWC_758_summary}
% \end{figure*}

To assess the SO detection significance, a set of images were reconstructed with these same fiducial parameters, but across a much wider velocity range ($\gtrsim200$~km~s$^{-1}$), essential to accurately determine the background noise in line free channels. We used \href{https://fishing.readthedocs.io/en/latest/index.html}{\texttt{gofish}} \citep{Teague2019}, a python package that extracts and averages spectra from an image cube across a given disk region and, under the assumption of a Keplerian velocity field, corrects for the known disk rotation by shifting such spectra to a common line center (cf., \citealt{YenEtal2016}), to stack our SO emission lines and increase their peak SNR, adopting the physical and geometrical parameters in \autoref{tab:A3}. Such an improvement can be seen by comparing the gray and blue spectra in the insert in panel a and c of \autoref{fig:2_spectra}.

\section{Flux conversion between SO transitions}\label{app:SO_flux}
To fairly compare the SO emission detected in different transitions, we estimated the SO $J_N=6_5-5_4$ flux for disks that do not have SO $J_N=6_5-5_4$ observations, based on the line flux from the detected transition. We considered two regimes: optically thin and optically thick. This estimation depends on the excitation conditions, energy levels, and line properties. Our conversion results are summarized in \autoref{tab:SO_fluxes}.

% \subsection{Optically thin}
In the optically thin regime, assuming local thermodynamic equilibrium (LTE) and a constant excitation temperature ($T=30$, 50, and 70~K), we calculated the flux ratio of the known line $F_1$ to the desired line flux $F_2$ by \citep{Goldsmith&Langer1999}
\begin{equation}
    \frac{F_2}{F_1} = \frac{\nu_2}{\nu_1} \cdot \frac{A_{ul}^{(2)} g_{u}^{(2)} e^{-E_{u}^{(2)}/kT}}{A_{ul}^{(1)} g_{u}^{(1)} e^{-E_{u}^{(1)}/kT}},
\end{equation}
where $k$ is the Boltzmann constant, $\nu_1$, $\nu_2$ are the line frequencies, $A_{ul}^{(1)}$, $A_{ul}^{(2)}$ the Einstein $A_{ul}$ coefficients, $g_{u}^{(1)}$, $g_{u}^{(2)}$ the statistical weights of the upper levels, and $E_{u}^{(1)}$, $E_{u}^{(2)}$ the upper-level energies of the two transitions. %The line properties of relevant transition are listed in \autoref{tab:SO_transitions}. 
In the optically thick case, the line becomes saturated, and the intensity depends primarily on the excitation temperature and source size, $F \propto B_\nu(T) \cdot \Omega$,
% \begin{equation}
%     F \propto B_\nu(T_{\rm ex}) \cdot \Omega
% \end{equation}
where $B_{\nu}(T)$ is the Planck function at the excitation temperature $T$ and frequency $\nu$, and $\Omega$ is the solid angle of the emitting region. Assuming the same excitation temperature and emitting area for both transitions, we have
\begin{equation}
    \frac{F_2}{F_1} \approx \frac{B_{\nu_2}(T)}{B_{\nu_1}(T)}\approx \frac{\nu_2^2}{\nu_1^2}
\end{equation}
where the last step follows from the Rayleigh-Jeans approximation.
% \begin{equation}
%     \frac{F_2}{F_1} \approx \frac{B_{\nu_2}(T_{\text{ex}})}{B_{\nu_1}(T_{\text{ex}})}.
% \end{equation}
% In the Rayleigh-Jeans approximation, this expression can be approximated as
% \begin{equation}
%     \frac{F_2}{F_1} \approx \frac{\nu_2^2}{\nu_1^2},
% \end{equation}
% which is independent of the assumed excitation temperature.

\begin{table*}
    \caption{SO flux conversion.}
    % \movetableright=-0.65in
    % \centering
    % \renewcommand{\arraystretch}{1.25}
    \begin{tabular}{lllllll}
    \hline
    \multirow{2}{*}{Source} & $d$ & \multirow{2}{*}{SO transition} & $F_{\rm SO}$ obsv. & \multirow{2}{*}{Optical depth} & $T$ & $F_{\rm SO}$ at 140~pc \\
    & (pc) & & (mJy km s$^{-1}$) & & (K) & (mJy km s$^{-1}$) \\
    (1) & (2) & (3) & (4) & (5) & (6) & (7) \\
    \hline
    \hline
    \multirow{5}{*}{TW~Hya} & \multirow{5}{*}{$60.1\pm0.1$} & $7_8-6_7$ & $13\pm3$ & -- & -- & 2.4 \\
    & & $6_5-5_4$ & -- & Thin  & 30 & 1.7 \\
    & & $6_5-5_4$ & -- & Thin  & 50 & 0.9 \\
    & & $6_5-5_4$ & -- & Thin  & 70 & 0.7 \\
    & & $6_5-5_4$ & -- & Thick & -- & 1.0 \\
    \hline
    \multirow{5}{*}{HD~100546} & \multirow{5}{*}{$108.1\pm0.4$} & $7_8-6_7$ & $343\pm26$ & -- & -- & 204.6 \\
    & & $6_5-5_4$ & -- & Thin  & 30 & 143.0 \\
    & & $6_5-5_4$ & -- & Thin  & 50 &  77.2 \\
    & & $6_5-5_4$ & -- & Thin  & 70 &  59.3 \\
    & & $6_5-5_4$ & -- & Thick & -- &  85.3 \\
    \hline
    AB~Aur     & $155.9\pm0.9$ & $6_5-5_4$ & $330\pm10$ & -- & -- & 409.4 \\
    \hline
    \multirow{5}{*}{IRS~48} & \multirow{5}{*}{$136.3\pm1.9$} & $7_8-6_7$ & $1063\pm23$ & -- & --  & 1007.9 \\
    & & $6_5-5_4$ & -- & Thin  & 30 & 704.4 \\
    & & $6_5-5_4$ & -- & Thin  & 50 & 380.5 \\
    & & $6_5-5_4$ & -- & Thin  & 70 & 292.2 \\
    & & $6_5-5_4$ & -- & Thick & -- & 420.0 \\
    \hline
    \multirow{5}{*}{HD~169142} & \multirow{5}{*}{$114.9\pm0.3$} & $8_8-7_7$ & $120\pm10$ & -- & -- & 80.8 \\
    & & $6_5-5_4$ & -- & Thin  & 30 & 58.5 \\
    & & $6_5-5_4$ & -- & Thin  & 50 & 29.0 \\
    & & $6_5-5_4$ & -- & Thin  & 70 & 21.5 \\
    & & $6_5-5_4$ & -- & Thick & -- & 33.0 \\
    \hline
    UY~Aur~AB & $152.3\pm0.9$ & $6_5-5_4$ & N/A & -- & -- & N/A \\
    \hline
    DR~Tau    & $193.0\pm1.2$ & $6_5-5_4$ & $195\pm9$ & -- & -- & 370.5 \\
    \hline
    CQ~Tau    & $149.4\pm1.3$ & $6_5-5_4$ & $72.4\pm14.6$ & -- & -- & 82.4 \\
    \hline
    MWC~758   & $155.9\pm0.8$ & $6_5-5_4$ & $31.4\pm9.3$ & -- & -- & 38.9 \\
    \hline
    \end{tabular}
    \tablecomments{Column 1: Source name. Column 2: Source distance. Column 3: SO $J_N$ transition. Upper state energies are 35.0, 81.2, and 87.5~K for the $J_N=6_5-5_4$, $7_8-6_7$, and $8_8-7_7$ transitions, respectively cf. the Leiden Atomic and Molecular Database (\href{https://home.strw.leidenuniv.nl/~moldata/SO.html}{LAMDA}), and \citet{Clark&DeLucia1976}. Column 4: SO flux. Column 5: Optical depth assumption. Column 6: Excitation temperature assumption. Column 7: SO $J_N=6_5-5_4$ flux rescaled to 140~pc.}
    \label{tab:SO_fluxes}
\end{table*}

\section{Correlation test}\label{app:correlations}
\begin{figure*}
    \centering
    \includegraphics[width=\textwidth]{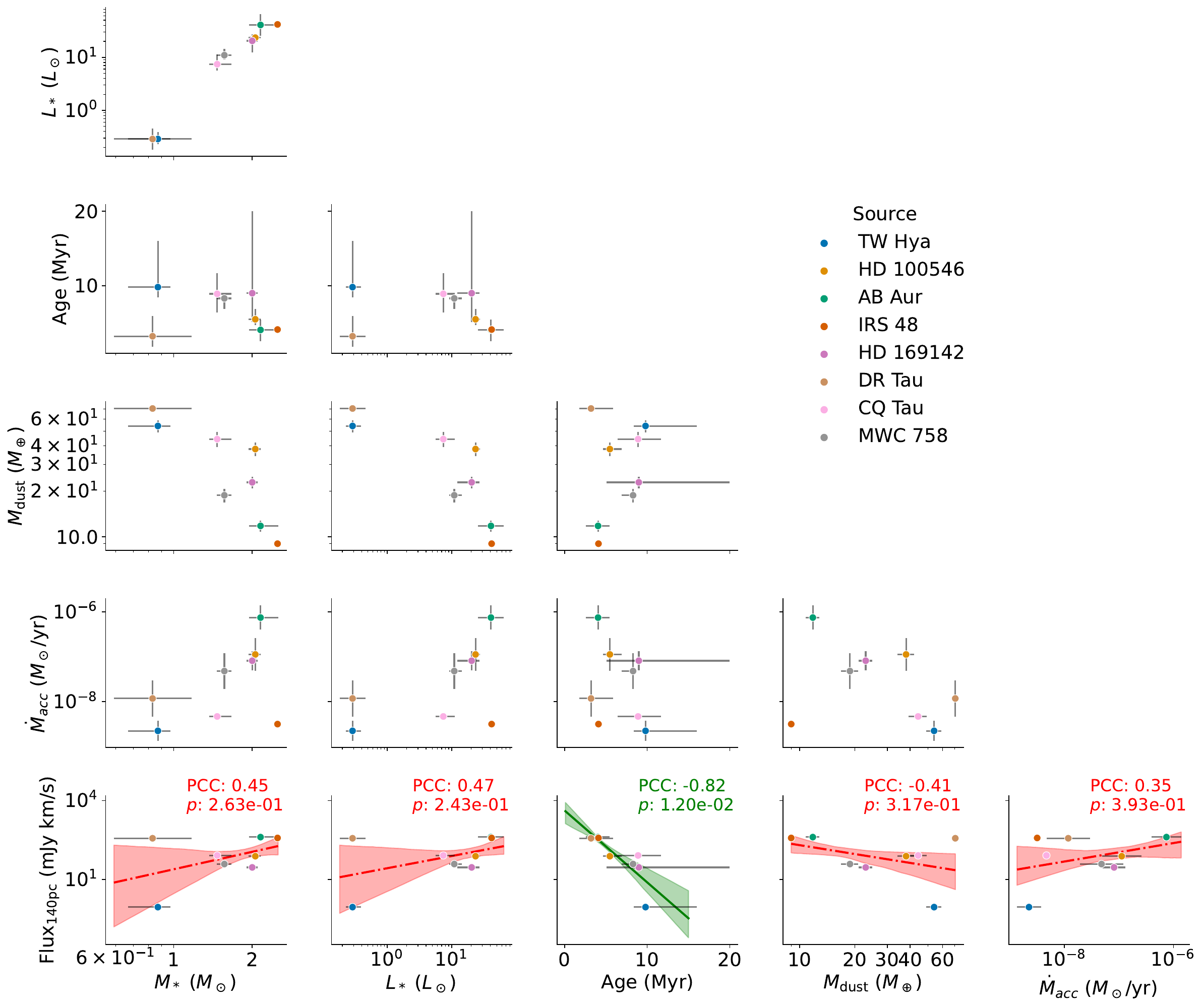}
    \caption{Pair plot of stellar and disk parameters for all known SO-bearing Class~II protoplanetary disks. The bottom row lists the calculated PCC and the corresponding $p$-values between the scaled SO flux and other parameters. A linear regression analysis is performed for each parameter pair, with the best fits shown as solid lines and the 16$^{\rm th}$ and 84$^{\rm th}$ percentiles of the bootstrap distribution shown as collar dashes, indicating $1\sigma$ confidence interval. The linear regression is shown in green when $p<0.05$.}
    \label{fig:pairplot_thin}
\end{figure*}

We visualize the stellar and disk parameters of interest and the converted SO fluxes summarized from \autoref{tab:1_parameters} and \autoref{tab:SO_fluxes} in the pair plot in \autoref{fig:pairplot_thin}. We adopted the Pearson Correlation Coefficient (PCC) and the corresponding $p$-values to quantitatively assess how the SO fluxes correlate with other parameters, which are labeled in the pair plots. While no statistically significant ($p<0.05$) correlations are found with stellar mass, luminosity, accretion rate, and disk mass, the SO fluxes show a tentative anti-correlation with system age, with a PCC of $-0.82$ when SO flux is in logarithmic space and age is in linear space.
%\begin{equation}
%    F_\mathrm{SO,140\,au} \simeq 1400\,\mathrm{mJy\,km\,s^{-1}} \text{exp}\left(-\frac{t}{2\,{\rm Myr}}\right).
%\end{equation}

\bibliography{sample631}{}

\begin{thebibliography}{}
\expandafter\ifx\csname natexlab\endcsname\relax\def\natexlab#1{#1}\fi
\providecommand{\url}[1]{\href{#1}{#1}}
\providecommand{\dodoi}[1]{doi:~\href{http://doi.org/#1}{\nolinkurl{#1}}}
\providecommand{\doeprint}[1]{\href{http://ascl.net/#1}{\nolinkurl{http://ascl.net/#1}}}
\providecommand{\doarXiv}[1]{\href{https://arxiv.org/abs/#1}{\nolinkurl{https://arxiv.org/abs/#1}}}

\bibitem[{{Andrews}(2020)}]{Andrews2020}
{Andrews}, S.~M. 2020, \araa, 58, 483, \dodoi{10.1146/annurev-astro-031220-010302}

\bibitem[{{Andrews} {et~al.}(2016){Andrews}, {Wilner}, {Zhu}, {Birnstiel}, {Carpenter}, {P{\'e}rez}, {Bai}, {{\"O}berg}, {Hughes}, {Isella}, \& {Ricci}}]{AndrewsEtal2016}
{Andrews}, S.~M., {Wilner}, D.~J., {Zhu}, Z., {et~al.} 2016, \apjl, 820, L40, \dodoi{10.3847/2041-8205/820/2/L40}

\bibitem[{{Andrews} {et~al.}(2018){Andrews}, {Huang}, {P{\'e}rez}, {Isella}, {Dullemond}, {Kurtovic}, {Guzm{\'a}n}, {Carpenter}, {Wilner}, {Zhang}, {Zhu}, {Birnstiel}, {Bai}, {Benisty}, {Hughes}, {{\"O}berg}, \& {Ricci}}]{AndrewsEtal2018b}
{Andrews}, S.~M., {Huang}, J., {P{\'e}rez}, L.~M., {et~al.} 2018, \apjl, 869, L41, \dodoi{10.3847/2041-8213/aaf741}

\bibitem[{{Aota} {et~al.}(2015){Aota}, {Inoue}, \& {Aikawa}}]{AotaEtal2015}
{Aota}, T., {Inoue}, T., \& {Aikawa}, Y. 2015, \apj, 799, 141, \dodoi{10.1088/0004-637X/799/2/141}

\bibitem[{{Artur de la Villarmois} {et~al.}(2023){Artur de la Villarmois}, {Guzm{\'a}n}, {Yang}, {Zhang}, \& {Sakai}}]{ArturdelaVillarmoisEtal2023}
{Artur de la Villarmois}, E., {Guzm{\'a}n}, V.~V., {Yang}, Y.~L., {Zhang}, Y., \& {Sakai}, N. 2023, \aap, 678, A124, \dodoi{10.1051/0004-6361/202346728}

\bibitem[{{Astropy Collaboration} {et~al.}(2013){Astropy Collaboration}, {Robitaille}, {Tollerud}, {Greenfield}, {Droettboom}, {Bray}, {Aldcroft}, {Davis}, {Ginsburg}, {Price-Whelan}, {Kerzendorf}, {Conley}, {Crighton}, {Barbary}, {Muna}, {Ferguson}, {Grollier}, {Parikh}, {Nair}, {Unther}, {Deil}, {Woillez}, {Conseil}, {Kramer}, {Turner}, {Singer}, {Fox}, {Weaver}, {Zabalza}, {Edwards}, {Azalee Bostroem}, {Burke}, {Casey}, {Crawford}, {Dencheva}, {Ely}, {Jenness}, {Labrie}, {Lim}, {Pierfederici}, {Pontzen}, {Ptak}, {Refsdal}, {Servillat}, \& {Streicher}}]{AstropyCollaborationEtal2013}
{Astropy Collaboration}, {Robitaille}, T.~P., {Tollerud}, E.~J., {et~al.} 2013, \aap, 558, A33, \dodoi{10.1051/0004-6361/201322068}

\bibitem[{{Bae} {et~al.}(2023){Bae}, {Isella}, {Zhu}, {Martin}, {Okuzumi}, \& {Suriano}}]{BaeEtal2023}
{Bae}, J., {Isella}, A., {Zhu}, Z., {et~al.} 2023, in Astronomical Society of the Pacific Conference Series, Vol. 534, Astronomical Society of the Pacific Conference Series, ed. S.~{Inutsuka}, Y.~{Aikawa}, T.~{Muto}, K.~{Tomida}, \& M.~{Tamura}, 423

\bibitem[{{Bae} \& {Zhu}(2018)}]{BaeZhu2018}
{Bae}, J., \& {Zhu}, Z. 2018, \apj, 859, 118, \dodoi{10.3847/1538-4357/aabf8c}

\bibitem[{{Benisty} {et~al.}(2015){Benisty}, {Juhasz}, {Boccaletti}, {Avenhaus}, {Milli}, {Thalmann}, {Dominik}, {Pinilla}, {Buenzli}, {Pohl}, {Beuzit}, {Birnstiel}, {de Boer}, {Bonnefoy}, {Chauvin}, {Christiaens}, {Garufi}, {Grady}, {Henning}, {Huelamo}, {Isella}, {Langlois}, {M{\'e}nard}, {Mouillet}, {Olofsson}, {Pantin}, {Pinte}, \& {Pueyo}}]{BenistyEtal2015}
{Benisty}, M., {Juhasz}, A., {Boccaletti}, A., {et~al.} 2015, \aap, 578, L6, \dodoi{10.1051/0004-6361/201526011}

\bibitem[{{Benisty} {et~al.}(2023){Benisty}, {Dominik}, {Follette}, {Garufi}, {Ginski}, {Hashimoto}, {Keppler}, {Kley}, \& {Monnier}}]{BenistyEtal2023}
{Benisty}, M., {Dominik}, C., {Follette}, K., {et~al.} 2023, in Astronomical Society of the Pacific Conference Series, Vol. 534, Protostars and Planets VII, ed. S.~{Inutsuka}, Y.~{Aikawa}, T.~{Muto}, K.~{Tomida}, \& M.~{Tamura}, 605, \dodoi{10.48550/arXiv.2203.09991}

\bibitem[{{Bi} {et~al.}(2021){Bi}, {Lin}, \& {Dong}}]{BiEtal2021}
{Bi}, J., {Lin}, M.-K., \& {Dong}, R. 2021, \apj, 912, 107, \dodoi{10.3847/1538-4357/abef6b}

\bibitem[{{Binkert} {et~al.}(2023){Binkert}, {Szul{\'a}gyi}, \& {Birnstiel}}]{BinkertEtal2023}
{Binkert}, F., {Szul{\'a}gyi}, J., \& {Birnstiel}, T. 2023, \mnras, 523, 55, \dodoi{10.1093/mnras/stad1405}

\bibitem[{{Boccaletti} {et~al.}(2020){Boccaletti}, {Di Folco}, {Pantin}, {Dutrey}, {Guilloteau}, {Tang}, {Pi{\'e}tu}, {Habart}, {Milli}, {Beck}, \& {Maire}}]{BoccalettiEtal2020}
{Boccaletti}, A., {Di Folco}, E., {Pantin}, E., {et~al.} 2020, \aap, 637, L5, \dodoi{10.1051/0004-6361/202038008}

\bibitem[{{Boehler} {et~al.}(2018){Boehler}, {Ricci}, {Weaver}, {Isella}, {Benisty}, {Carpenter}, {Grady}, {Shen}, {Tang}, \& {Perez}}]{BoehlerEtal2018}
{Boehler}, Y., {Ricci}, L., {Weaver}, E., {et~al.} 2018, \apj, 853, 162, \dodoi{10.3847/1538-4357/aaa19c}

\bibitem[{{Booth} {et~al.}(2023{\natexlab{a}}){Booth}, {Ilee}, {Walsh}, {Kama}, {Keyte}, {van Dishoeck}, \& {Nomura}}]{BoothEtal2023}
{Booth}, A.~S., {Ilee}, J.~D., {Walsh}, C., {et~al.} 2023{\natexlab{a}}, \aap, 669, A53, \dodoi{10.1051/0004-6361/202244472}

\bibitem[{{Booth} {et~al.}(2023{\natexlab{b}}){Booth}, {Law}, {Temmink}, {Leemker}, \& {Mac{\'\i}as}}]{BoothEtal2023b}
{Booth}, A.~S., {Law}, C.~J., {Temmink}, M., {Leemker}, M., \& {Mac{\'\i}as}, E. 2023{\natexlab{b}}, \aap, 678, A146, \dodoi{10.1051/0004-6361/202346974}

\bibitem[{{Booth} {et~al.}(2021){Booth}, {van der Marel}, {Leemker}, {van Dishoeck}, \& {Ohashi}}]{BoothEtal2021b}
{Booth}, A.~S., {van der Marel}, N., {Leemker}, M., {van Dishoeck}, E.~F., \& {Ohashi}, S. 2021, \aap, 651, L6, \dodoi{10.1051/0004-6361/202141057}

\bibitem[{{Booth} {et~al.}(2018){Booth}, {Walsh}, {Kama}, {Loomis}, {Maud}, \& {Juh{\'a}sz}}]{BoothEtal2018}
{Booth}, A.~S., {Walsh}, C., {Kama}, M., {et~al.} 2018, \aap, 611, A16, \dodoi{10.1051/0004-6361/201731347}

\bibitem[{{Booth} {et~al.}(2024{\natexlab{a}}){Booth}, {Leemker}, {van Dishoeck}, {Evans}, {Ilee}, {Kama}, {Keyte}, {Law}, {van der Marel}, {Nomura}, {Notsu}, {{\"O}berg}, {Temmink}, \& {Walsh}}]{BoothEtal2024a}
{Booth}, A.~S., {Leemker}, M., {van Dishoeck}, E.~F., {et~al.} 2024{\natexlab{a}}, \aj, 167, 164, \dodoi{10.3847/1538-3881/ad2700}

\bibitem[{{Booth} {et~al.}(2024{\natexlab{b}}){Booth}, {Temmink}, {van Dishoeck}, {Evans}, {Ilee}, {Kama}, {Keyte}, {Law}, {Leemker}, {van der Marel}, {Nomura}, {Notsu}, {{\"O}berg}, \& {Walsh}}]{BoothEtal2024b}
{Booth}, A.~S., {Temmink}, M., {van Dishoeck}, E.~F., {et~al.} 2024{\natexlab{b}}, \aj, 167, 165, \dodoi{10.3847/1538-3881/ad26ff}

\bibitem[{{Bridle} \& {Schwab}(1999)}]{BridleSchwab1999}
{Bridle}, A.~H., \& {Schwab}, F.~R. 1999, in Astronomical Society of the Pacific Conference Series, Vol. 180, Synthesis Imaging in Radio Astronomy II, ed. G.~B. {Taylor}, C.~L. {Carilli}, \& R.~A. {Perley}, 371

\bibitem[{{Briggs}(1995)}]{Briggs1995}
{Briggs}, D.~S. 1995, PhD thesis, New Mexico Institute of Mining and Technology

\bibitem[{{Brown} {et~al.}(2012){Brown}, {Herczeg}, {Pontoppidan}, \& {van Dishoeck}}]{BrownEtal2012}
{Brown}, J.~M., {Herczeg}, G.~J., {Pontoppidan}, K.~M., \& {van Dishoeck}, E.~F. 2012, \apj, 744, 116, \dodoi{10.1088/0004-637X/744/2/116}

\bibitem[{{Calcino} {et~al.}(2025){Calcino}, {Price}, {Hilder}, {Christiaens}, {Speedie}, \& {Ormel}}]{CalcinoEtal2025}
{Calcino}, J., {Price}, D.~J., {Hilder}, T., {et~al.} 2025, \mnras, 537, 2695, \dodoi{10.1093/mnras/staf135}

\bibitem[{{Calcino} {et~al.}(2019){Calcino}, {Price}, {Pinte}, {van der Marel}, {Ragusa}, {Dipierro}, {Cuello}, \& {Christiaens}}]{CalcinoEtal2019}
{Calcino}, J., {Price}, D.~J., {Pinte}, C., {et~al.} 2019, \mnras, 490, 2579, \dodoi{10.1093/mnras/stz2770}

\bibitem[{{CASA Team} {et~al.}(2022){CASA Team}, {Bean}, {Bhatnagar}, {Castro}, {Donovan Meyer}, {Emonts}, {Garcia}, {Garwood}, {Golap}, {Gonzalez Villalba}, {Harris}, {Hayashi}, {Hoskins}, {Hsieh}, {Jagannathan}, {Kawasaki}, {Keimpema}, {Kettenis}, {Lopez}, {Marvil}, {Masters}, {McNichols}, {Mehringer}, {Miel}, {Moellenbrock}, {Montesino}, {Nakazato}, {Ott}, {Petry}, {Pokorny}, {Raba}, {Rau}, {Schiebel}, {Schweighart}, {Sekhar}, {Shimada}, {Small}, {Steeb}, {Sugimoto}, {Suoranta}, {Tsutsumi}, {van Bemmel}, {Verkouter}, {Wells}, {Xiong}, {Szomoru}, {Griffith}, {Glendenning}, \& {Kern}}]{CASATeamEtal2022}
{CASA Team}, {Bean}, B., {Bhatnagar}, S., {et~al.} 2022, \pasp, 134, 114501, \dodoi{10.1088/1538-3873/ac9642}

\bibitem[{{Casassus} \& {C{\'a}rcamo}(2022)}]{CasassusCarcamo2022}
{Casassus}, S., \& {C{\'a}rcamo}, M. 2022, \mnras, 513, 5790, \dodoi{10.1093/mnras/stac1285}

\bibitem[{{Casassus} {et~al.}(2022){Casassus}, {C{\'a}rcamo}, {Hales}, {Weber}, \& {Dent}}]{CasassusEtal2022}
{Casassus}, S., {C{\'a}rcamo}, M., {Hales}, A., {Weber}, P., \& {Dent}, B. 2022, \apjl, 933, L4, \dodoi{10.3847/2041-8213/ac75e8}

\bibitem[{{Clark} \& {De Lucia}(1976)}]{Clark&DeLucia1976}
{Clark}, W.~W., \& {De Lucia}, F.~C. 1976, Journal of Molecular Spectroscopy, 60, 332, \dodoi{10.1016/0022-2852(76)90136-3}

\bibitem[{{Cleeves} {et~al.}(2015){Cleeves}, {Bergin}, \& {Harries}}]{CleevesEtal2015}
{Cleeves}, L.~I., {Bergin}, E.~A., \& {Harries}, T.~J. 2015, \apj, 807, 2, \dodoi{10.1088/0004-637X/807/1/2}

\bibitem[{{Codella} {et~al.}(2024){Codella}, {Podio}, {De Simone}, {Ceccarelli}, {Ohashi}, {Chandler}, {Sakai}, {Pineda}, {Segura-Cox}, {Bianchi}, {Cuello}, {L{\'o}pez-Sepulcre}, {Fedele}, {Caselli}, {Charnley}, {Johnstone}, {Zhang}, {Maureira}, {Zhang}, {Sabatini}, {Svoboda}, {Jim{\'e}nez-Serra}, {Loinard}, {Mercimek}, {Murillo}, \& {Yamamoto}}]{Codella2024}
{Codella}, C., {Podio}, L., {De Simone}, M., {et~al.} 2024, \mnras, 528, 7383, \dodoi{10.1093/mnras/stae472}

\bibitem[{{Cornwell}(2008)}]{Cornwell2008}
{Cornwell}, T.~J. 2008, IEEE Journal of Selected Topics in Signal Processing, 2, 793, \dodoi{10.1109/JSTSP.2008.2006388}

\bibitem[{{Cridland} {et~al.}(2025){Cridland}, {Lega}, \& {Benisty}}]{CridlandEtal2025}
{Cridland}, A.~J., {Lega}, E., \& {Benisty}, M. 2025, \aap, 693, A86, \dodoi{10.1051/0004-6361/202451140}

\bibitem[{{Curone} {et~al.}(2025){Curone}, {Facchini}, {Andrews}, {Testi}, {Benisty}, {Czekala}, {Huang}, {Ilee}, {Isella}, {Lodato}, {Loomis}, {Stadler}, {Winter}, {Bae}, {Barraza-Alfaro}, {Cataldi}, {Cuello}, {Fasano}, {Flock}, {Fukagawa}, {Galloway-Sprietsma}, {Garg}, {Hall}, {Izquierdo}, {Kanagawa}, {Lesur}, {Longarini}, {Menard}, {Orihara}, {Pinte}, {Price}, {Rosotti}, {Teague}, {Wafflard-Fernandez}, {Wilner}, {W{\"o}lfer}, {Yen}, {Yoshida}, \& {Zawadzki}}]{CuroneEtal2025}
{Curone}, P., {Facchini}, S., {Andrews}, S.~M., {et~al.} 2025, \apjl, 984, L9, \dodoi{10.3847/2041-8213/adc438}

\bibitem[{{Czekala} {et~al.}(2021){Czekala}, {Loomis}, {Teague}, {Booth}, {Huang}, {Cataldi}, {Ilee}, {Law}, {Walsh}, {Bosman}, {Guzm{\'a}n}, {Le Gal}, {{\"O}berg}, {Yamato}, {Aikawa}, {Andrews}, {Bae}, {Bergin}, {Bergner}, {Cleeves}, {Kurtovic}, {M{\'e}nard}, {Nomura}, {P{\'e}rez}, {Qi}, {Schwarz}, {Tsukagoshi}, {Waggoner}, {Wilner}, \& {Zhang}}]{CzekalaEtal2021}
{Czekala}, I., {Loomis}, R.~A., {Teague}, R., {et~al.} 2021, \apjs, 257, 2, \dodoi{10.3847/1538-4365/ac1430}

\bibitem[{{Debes} {et~al.}(2017){Debes}, {Poteet}, {Jang-Condell}, {Gaspar}, {Hines}, {Kastner}, {Pueyo}, {Rapson}, {Roberge}, {Schneider}, \& {Weinberger}}]{DebesEtal2017}
{Debes}, J.~H., {Poteet}, C.~A., {Jang-Condell}, H., {et~al.} 2017, \apj, 835, 205, \dodoi{10.3847/1538-4357/835/2/205}

\bibitem[{{Dong} {et~al.}(2015){Dong}, {Zhu}, {Rafikov}, \& {Stone}}]{DongEtal2015a}
{Dong}, R., {Zhu}, Z., {Rafikov}, R.~R., \& {Stone}, J.~M. 2015, \apjl, 809, L5, \dodoi{10.1088/2041-8205/809/1/L5}

\bibitem[{{Dong} {et~al.}(2018){Dong}, {Liu}, {Eisner}, {Andrews}, {Fung}, {Zhu}, {Chiang}, {Hashimoto}, {Liu}, {Casassus}, {Esposito}, {Hasegawa}, {Muto}, {Pavlyuchenkov}, {Wilner}, {Akiyama}, {Tamura}, \& {Wisniewski}}]{DongEtal2018a}
{Dong}, R., {Liu}, S.-y., {Eisner}, J., {et~al.} 2018, \apj, 860, 124, \dodoi{10.3847/1538-4357/aac6cb}

\bibitem[{{Dutrey} {et~al.}(2024){Dutrey}, {Chapillon}, {Guilloteau}, {Tang}, {Boccaletti}, {Bouscasse}, {Collin-Dufresne}, {Di Folco}, {Fuente}, {Pi{\'e}tu}, {Rivi{\`e}re-Marichalar}, \& {Semenov}}]{DutreyEtal2024}
{Dutrey}, A., {Chapillon}, E., {Guilloteau}, S., {et~al.} 2024, \aap, 689, L7, \dodoi{10.1051/0004-6361/202451299}

\bibitem[{{Eistrup} {et~al.}(2016){Eistrup}, {Walsh}, \& {van Dishoeck}}]{EistrupEtal2016}
{Eistrup}, C., {Walsh}, C., \& {van Dishoeck}, E.~F. 2016, \aap, 595, A83, \dodoi{10.1051/0004-6361/201628509}

\bibitem[{{Eriksson} {et~al.}(2025){Eriksson}, {Yang}, \& {Armitage}}]{ErikssonEtal2025}
{Eriksson}, L. E.~J., {Yang}, C.-C., \& {Armitage}, P.~J. 2025, \mnras, 537, L26, \dodoi{10.1093/mnrasl/slae110}

\bibitem[{{Evans} {et~al.}(2019){Evans}, {Hartquist}, {Caselli}, {Boley}, {Ilee}, \& {Rawlings}}]{EvansEtal2019}
{Evans}, M.~G., {Hartquist}, T.~W., {Caselli}, P., {et~al.} 2019, \mnras, 483, 1266, \dodoi{10.1093/mnras/sty2765}

\bibitem[{{Evans} {et~al.}(2015){Evans}, {Ilee}, {Boley}, {Caselli}, {Durisen}, {Hartquist}, \& {Rawlings}}]{EvansEtal2015}
{Evans}, M.~G., {Ilee}, J.~D., {Boley}, A.~C., {et~al.} 2015, \mnras, 453, 1147, \dodoi{10.1093/mnras/stv1698}

\bibitem[{{Facchini} {et~al.}(2018){Facchini}, {Juh{\'a}sz}, \& {Lodato}}]{FacchiniEtal2018a}
{Facchini}, S., {Juh{\'a}sz}, A., \& {Lodato}, G. 2018, \mnras, 473, 4459, \dodoi{10.1093/mnras/stx2523}

\bibitem[{{Facchini} {et~al.}(2021){Facchini}, {Teague}, {Bae}, {Benisty}, {Keppler}, \& {Isella}}]{FacchiniEtal2021}
{Facchini}, S., {Teague}, R., {Bae}, J., {et~al.} 2021, \aj, 162, 99, \dodoi{10.3847/1538-3881/abf0a4}

\bibitem[{{Flores} {et~al.}(2023){Flores}, {Ohashi}, {Tobin}, {J{\o}rgensen}, {Takakuwa}, {Li}, {Lin}, {van't Hoff}, {Plunkett}, {Yamato}, {Sai (Insa Choi)}, {Koch}, {Yen}, {Aikawa}, {Aso}, {de Gregorio-Monsalvo}, {Kido}, {Kwon}, {Lee}, {Lee}, {Looney}, {Santamar{\'\i}a-Miranda}, {Sharma}, {Thieme}, {Williams}, {Han}, {Narayanan}, \& {Lai}}]{FloresEtal2023}
{Flores}, C., {Ohashi}, N., {Tobin}, J.~J., {et~al.} 2023, \apj, 958, 98, \dodoi{10.3847/1538-4357/acf7c1}

\bibitem[{{Follette} {et~al.}(2015){Follette}, {Grady}, {Swearingen}, {Sitko}, {Champney}, {van der Marel}, {Takami}, {Kuchner}, {Close}, {Muto}, {Mayama}, {McElwain}, {Fukagawa}, {Maaskant}, {Min}, {Russell}, {Kudo}, {Kusakabe}, {Hashimoto}, {Abe}, {Akiyama}, {Brandner}, {Brandt}, {Carson}, {Currie}, {Egner}, {Feldt}, {Goto}, {Guyon}, {Hayano}, {Hayashi}, {Hayashi}, {Henning}, {Hodapp}, {Ishii}, {Iye}, {Janson}, {Kandori}, {Knapp}, {Kuzuhara}, {Kwon}, {Matsuo}, {Miyama}, {Morino}, {Moro-Martin}, {Nishimura}, {Pyo}, {Serabyn}, {Suenaga}, {Suto}, {Suzuki}, {Takahashi}, {Takato}, {Terada}, {Thalmann}, {Tomono}, {Turner}, {Watanabe}, {Wisniewski}, {Yamada}, {Takami}, {Usuda}, \& {Tamura}}]{FolletteEtal2015}
{Follette}, K.~B., {Grady}, C.~A., {Swearingen}, J.~R., {et~al.} 2015, \apj, 798, 132, \dodoi{10.1088/0004-637X/798/2/132}

\bibitem[{{Follette} {et~al.}(2017){Follette}, {Rameau}, {Dong}, {Pueyo}, {Close}, {Duch{\^e}ne}, {Fung}, {Leonard}, {Macintosh}, {Males}, {Marois}, {Millar-Blanchaer}, {Morzinski}, {Mullen}, {Perrin}, {Spiro}, {Wang}, {Ammons}, {Bailey}, {Barman}, {Bulger}, {Chilcote}, {Cotten}, {De Rosa}, {Doyon}, {Fitzgerald}, {Goodsell}, {Graham}, {Greenbaum}, {Hibon}, {Hung}, {Ingraham}, {Kalas}, {Konopacky}, {Larkin}, {Maire}, {Marchis}, {Metchev}, {Nielsen}, {Oppenheimer}, {Palmer}, {Patience}, {Poyneer}, {Rajan}, {Rantakyr{\"o}}, {Savransky}, {Schneider}, {Sivaramakrishnan}, {Song}, {Soummer}, {Thomas}, {Vega}, {Wallace}, {Ward-Duong}, {Wiktorowicz}, \& {Wolff}}]{FolletteEtal2017}
{Follette}, K.~B., {Rameau}, J., {Dong}, R., {et~al.} 2017, \aj, 153, 264, \dodoi{10.3847/1538-3881/aa6d85}

\bibitem[{{Fuente} {et~al.}(2010){Fuente}, {Cernicharo}, {Ag{\'u}ndez}, {Bern{\'e}}, {Goicoechea}, {Alonso-Albi}, \& {Marcelino}}]{FuenteEtal2010}
{Fuente}, A., {Cernicharo}, J., {Ag{\'u}ndez}, M., {et~al.} 2010, \aap, 524, A19, \dodoi{10.1051/0004-6361/201014905}

\bibitem[{{Gaia Collaboration}(2020)}]{GaiaCollaboration2020}
{Gaia Collaboration}. 2020, {VizieR Online Data Catalog: Gaia EDR3 (Gaia Collaboration, 2020)}, VizieR On-line Data Catalog: I/350. Originally published in: 2021A\&amp;A...649A...1G, \dodoi{10.26093/cds/vizier.1350}

\bibitem[{{Galloway-Sprietsma} {et~al.}(2025){Galloway-Sprietsma}, {Bae}, {Izquierdo}, {Stadler}, {Longarini}, {Teague}, {Andrews}, {Winter}, {Benisty}, {Facchini}, {Rosotti}, {Zawadzki}, {Pinte}, {Fasano}, {Barraza-Alfaro}, {Cataldi}, {Cuello}, {Curone}, {Czekala}, {Flock}, {Fukagawa}, {Gardner}, {Garg}, {Hall}, {Huang}, {Ilee}, {Kanagawa}, {Lesur}, {Lodato}, {Loomis}, {Menard}, {Orihara}, {Price}, {Wafflard-Fernandez}, {Wilner}, {W{\"o}lfer}, {Yen}, \& {Yoshida}}]{Galloway-SprietsmaEtal2025}
{Galloway-Sprietsma}, M., {Bae}, J., {Izquierdo}, A.~F., {et~al.} 2025, \apjl, 984, L10, \dodoi{10.3847/2041-8213/adc437}

\bibitem[{{Gangi} {et~al.}(2022){Gangi}, {Antoniucci}, {Biazzo}, {Frasca}, {Nisini}, {Alcal{\'a}}, {Giannini}, {Manara}, {Giunta}, {Harutyunyan}, {Munari}, \& {Vitali}}]{GangiEtal2022}
{Gangi}, M., {Antoniucci}, S., {Biazzo}, K., {et~al.} 2022, \aap, 667, A124, \dodoi{10.1051/0004-6361/202244042}

\bibitem[{{Garcia Lopez} {et~al.}(2006){Garcia Lopez}, {Natta}, {Testi}, \& {Habart}}]{GarciaLopezEtal2006}
{Garcia Lopez}, R., {Natta}, A., {Testi}, L., \& {Habart}, E. 2006, \aap, 459, 837, \dodoi{10.1051/0004-6361:20065575}

\bibitem[{{Garg} {et~al.}(2021){Garg}, {Pinte}, {Christiaens}, {Price}, {Lazendic}, {Boehler}, {Casassus}, {Marino}, {Perez}, \& {Zuleta}}]{GargEtal2021}
{Garg}, H., {Pinte}, C., {Christiaens}, V., {et~al.} 2021, \mnras, 504, 782, \dodoi{10.1093/mnras/stab800}

\bibitem[{{Garg} {et~al.}(2022){Garg}, {Pinte}, {Hammond}, {Teague}, {Hilder}, {Price}, {Calcino}, {Christiaens}, \& {Poblete}}]{GargEtal2022}
{Garg}, H., {Pinte}, C., {Hammond}, I., {et~al.} 2022, \mnras, \dodoi{10.1093/mnras/stac3039}

\bibitem[{{Garufi} {et~al.}(2022){Garufi}, {Podio}, {Codella}, {Segura-Cox}, {Vander Donckt}, {Mercimek}, {Bacciotti}, {Fedele}, {Kasper}, {Pineda}, {Humphreys}, \& {Testi}}]{GarufiEtal2022a}
{Garufi}, A., {Podio}, L., {Codella}, C., {et~al.} 2022, \aap, 658, A104, \dodoi{10.1051/0004-6361/202141264}

\bibitem[{{Garufi} {et~al.}(2024){Garufi}, {Ginski}, {van Holstein}, {Benisty}, {Manara}, {P{\'e}rez}, {Pinilla}, {Ribas}, {Weber}, {Williams}, {Cieza}, {Dominik}, {Facchini}, {Huang}, {Zurlo}, {Bae}, {Hagelberg}, {Henning}, {Hogerheijde}, {Janson}, {M{\'e}nard}, {Messina}, {Meyer}, {Pinte}, {Quanz}, {Rigliaco}, {Roccatagliata}, {Schmid}, {Szul{\'a}gyi}, {van Boekel}, {Wahhaj}, {Antichi}, {Baruffolo}, \& {Moulin}}]{GarufiEtal2024}
{Garufi}, A., {Ginski}, C., {van Holstein}, R.~G., {et~al.} 2024, \aap, 685, A53, \dodoi{10.1051/0004-6361/202347586}

\bibitem[{{Ginski} {et~al.}(2021){Ginski}, {Facchini}, {Huang}, {Benisty}, {Vaendel}, {Stapper}, {Dominik}, {Bae}, {M{\'e}nard}, {Muro-Arena}, {Hogerheijde}, {McClure}, {van Holstein}, {Birnstiel}, {Boehler}, {Bohn}, {Flock}, {Mamajek}, {Manara}, {Pinilla}, {Pinte}, \& {Ribas}}]{GinskiEtal2021}
{Ginski}, C., {Facchini}, S., {Huang}, J., {et~al.} 2021, \apjl, 908, L25, \dodoi{10.3847/2041-8213/abdf57}

\bibitem[{{Goldreich} \& {Tremaine}(1979)}]{GoldreichTremaine1979}
{Goldreich}, P., \& {Tremaine}, S. 1979, \apj, 233, 857, \dodoi{10.1086/157448}

\bibitem[{{Goldsmith} \& {Langer}(1999)}]{Goldsmith&Langer1999}
{Goldsmith}, P.~F., \& {Langer}, W.~D. 1999, \apj, 517, 209, \dodoi{10.1086/307195}

\bibitem[{{Gonzalez} {et~al.}(2020){Gonzalez}, {van der Plas}, {Pinte}, {Cuello}, {Nealon}, {M{\'e}nard}, {Revol}, {Rodet}, {Langlois}, \& {Maire}}]{GonzalezEtAl2020}
{Gonzalez}, J.-F., {van der Plas}, G., {Pinte}, C., {et~al.} 2020, \mnras, 499, 3837, \dodoi{10.1093/mnras/staa2938}

\bibitem[{{Grady} {et~al.}(2005){Grady}, {Woodgate}, {Bowers}, {Gull}, {Sitko}, {Carpenter}, {Lynch}, {Russell}, {Perry}, {Williger}, {Roberge}, {Bouret}, \& {Sahu}}]{GradyEtal2005}
{Grady}, C.~A., {Woodgate}, B.~E., {Bowers}, C.~W., {et~al.} 2005, \apj, 630, 958, \dodoi{10.1086/430731}

\bibitem[{{Grant} {et~al.}(2022){Grant}, {Espaillat}, {Brittain}, {Scott-Joseph}, \& {Calvet}}]{GrantEtal2022}
{Grant}, S.~L., {Espaillat}, C.~C., {Brittain}, S., {Scott-Joseph}, C., \& {Calvet}, N. 2022, \apj, 926, 229, \dodoi{10.3847/1538-4357/ac450a}

\bibitem[{{Gray} {et~al.}(2017){Gray}, {Riggs}, {Koen}, {Murphy}, {Newsome}, {Corbally}, {Cheng}, \& {Neff}}]{Gray2017}
{Gray}, R.~O., {Riggs}, Q.~S., {Koen}, C., {et~al.} 2017, \aj, 154, 31, \dodoi{10.3847/1538-3881/aa6d5e}

\bibitem[{{Gupta} {et~al.}(2023){Gupta}, {Miotello}, {Manara}, {Williams}, {Facchini}, {Beccari}, {Birnstiel}, {Ginski}, {Hacar}, {K{\"u}ffmeier}, {Testi}, {Tychoniec}, \& {Yen}}]{GuptaEtal2023}
{Gupta}, A., {Miotello}, A., {Manara}, C.~F., {et~al.} 2023, \aap, 670, L8, \dodoi{10.1051/0004-6361/202245254}

\bibitem[{{Haffert} {et~al.}(2019){Haffert}, {Bohn}, {de Boer}, {Snellen}, {Brinchmann}, {Girard}, {Keller}, \& {Bacon}}]{HaffertEtal2019}
{Haffert}, S.~Y., {Bohn}, A.~J., {de Boer}, J., {et~al.} 2019, Nature Astronomy, 3, 749, \dodoi{10.1038/s41550-019-0780-5}

\bibitem[{{Hall} {et~al.}(2020){Hall}, {Dong}, {Teague}, {Terry}, {Pinte}, {Paneque-Carre{\~n}o}, {Veronesi}, {Alexander}, \& {Lodato}}]{HallEtal2020}
{Hall}, C., {Dong}, R., {Teague}, R., {et~al.} 2020, \apj, 904, 148, \dodoi{10.3847/1538-4357/abac17}

\bibitem[{{Hammond} {et~al.}(2023){Hammond}, {Christiaens}, {Price}, {Toci}, {Pinte}, {Juillard}, \& {Garg}}]{HammondEtal2023}
{Hammond}, I., {Christiaens}, V., {Price}, D.~J., {et~al.} 2023, \mnras, 522, L51, \dodoi{10.1093/mnrasl/slad027}

\bibitem[{{Hammond} {et~al.}(2022){Hammond}, {Christiaens}, {Price}, {Ubeira-Gabellini}, {Baird}, {Calcino}, {Benisty}, {Lodato}, {Testi}, {Pinte}, {Toci}, \& {Fedele}}]{HammondEtal2022}
---. 2022, \mnras, 515, 6109, \dodoi{10.1093/mnras/stac2119}

\bibitem[{{Harris} {et~al.}(2020){Harris}, {Millman}, {van der Walt}, {Gommers}, {Virtanen}, {Cournapeau}, {Wieser}, {Taylor}, {Berg}, {Smith}, {Kern}, {Picus}, {Hoyer}, {van Kerkwijk}, {Brett}, {Haldane}, {del R{\'\i}o}, {Wiebe}, {Peterson}, {G{\'e}rard-Marchant}, {Sheppard}, {Reddy}, {Weckesser}, {Abbasi}, {Gohlke}, \& {Oliphant}}]{HarrisEtal2020}
{Harris}, C.~R., {Millman}, K.~J., {van der Walt}, S.~J., {et~al.} 2020, \nat, 585, 357, \dodoi{10.1038/s41586-020-2649-2}

\bibitem[{{Hartigan} \& {Kenyon}(2003)}]{HartiganKenyon2003}
{Hartigan}, P., \& {Kenyon}, S.~J. 2003, \apj, 583, 334, \dodoi{10.1086/345293}

\bibitem[{{Hartquist} {et~al.}(1980){Hartquist}, {Dalgarno}, \& {Oppenheimer}}]{HartquistEtal1980}
{Hartquist}, T.~W., {Dalgarno}, A., \& {Oppenheimer}, M. 1980, \apj, 236, 182, \dodoi{10.1086/157731}

\bibitem[{{Hashimoto} {et~al.}(2011){Hashimoto}, {Tamura}, {Muto}, {Kudo}, {Fukagawa}, {Fukue}, {Goto}, {Grady}, {Henning}, {Hodapp}, {Honda}, {Inutsuka}, {Kokubo}, {Knapp}, {McElwain}, {Momose}, {Ohashi}, {Okamoto}, {Takami}, {Turner}, {Wisniewski}, {Janson}, {Abe}, {Brandner}, {Carson}, {Egner}, {Feldt}, {Golota}, {Guyon}, {Hayano}, {Hayashi}, {Hayashi}, {Ishii}, {Kandori}, {Kusakabe}, {Matsuo}, {Mayama}, {Miyama}, {Morino}, {Moro-Martin}, {Nishimura}, {Pyo}, {Suto}, {Suzuki}, {Takato}, {Terada}, {Thalmann}, {Tomono}, {Watanabe}, {Yamada}, {Takami}, \& {Usuda}}]{HashimotoEtal2011}
{Hashimoto}, J., {Tamura}, M., {Muto}, T., {et~al.} 2011, \apjl, 729, L17, \dodoi{10.1088/2041-8205/729/2/L17}

\bibitem[{{Hennebelle} {et~al.}(2017){Hennebelle}, {Lesur}, \& {Fromang}}]{HennebelleEtal2017}
{Hennebelle}, P., {Lesur}, G., \& {Fromang}, S. 2017, \aap, 599, A86, \dodoi{10.1051/0004-6361/201629779}

\bibitem[{{Herczeg} \& {Hillenbrand}(2014)}]{Herczeg&Hillenbrand2014}
{Herczeg}, G.~J., \& {Hillenbrand}, L.~A. 2014, \apj, 786, 97, \dodoi{10.1088/0004-637X/786/2/97}

\bibitem[{{Herczeg} {et~al.}(2023){Herczeg}, {Chen}, {Donati}, {Dupree}, {Walter}, {Hillenbrand}, {Johns-Krull}, {Manara}, {G{\"u}nther}, {Fang}, {Schneider}, {Valenti}, {Alencar}, {Venuti}, {Alcal{\'a}}, {Frasca}, {Arulanantham}, {Linsky}, {Bouvier}, {Brickhouse}, {Calvet}, {Espaillat}, {Campbell-White}, {Carpenter}, {Chang}, {Cruz}, {Dahm}, {Eisl{\"o}ffel}, {Edwards}, {Fischer}, {Guo}, {Henning}, {Ji}, {Jose}, {Kastner}, {Launhardt}, {Principe}, {Robinson}, {Serna}, {Siwak}, {Sterzik}, \& {Takasao}}]{HerczegEtal2023}
{Herczeg}, G.~J., {Chen}, Y., {Donati}, J.-F., {et~al.} 2023, \apj, 956, 102, \dodoi{10.3847/1538-4357/acf468}

\bibitem[{{Hildebrand}(1983)}]{Hildebrand1983}
{Hildebrand}, R.~H. 1983, \qjras, 24, 267

\bibitem[{{Huang} {et~al.}(2023){Huang}, {Bergin}, {Bae}, {Benisty}, \& {Andrews}}]{HuangEtal2023}
{Huang}, J., {Bergin}, E.~A., {Bae}, J., {Benisty}, M., \& {Andrews}, S.~M. 2023, \apj, 943, 107, \dodoi{10.3847/1538-4357/aca89c}

\bibitem[{{Huang} {et~al.}(2024){Huang}, {Bergin}, {Le Gal}, {Andrews}, {Bae}, {Keyte}, \& {Sturm}}]{HuangEtal2024}
{Huang}, J., {Bergin}, E.~A., {Le Gal}, R., {et~al.} 2024, \apj, 973, 135, \dodoi{10.3847/1538-4357/ad6447}

\bibitem[{{Huang} {et~al.}(2018){Huang}, {Andrews}, {P{\'e}rez}, {Zhu}, {Dullemond}, {Isella}, {Benisty}, {Bai}, {Birnstiel}, {Carpenter}, {Guzm{\'a}n}, {Hughes}, {{\"O}berg}, {Ricci}, {Wilner}, \& {Zhang}}]{HuangEtal2018c}
{Huang}, J., {Andrews}, S.~M., {P{\'e}rez}, L.~M., {et~al.} 2018, \apjl, 869, L43, \dodoi{10.3847/2041-8213/aaf7a0}

\bibitem[{{Hunter}(2007)}]{Hunter2007}
{Hunter}, J.~D. 2007, Computing in Science and Engineering, 9, 90, \dodoi{10.1109/MCSE.2007.55}

\bibitem[{{Ilee} {et~al.}(2011){Ilee}, {Boley}, {Caselli}, {Durisen}, {Hartquist}, \& {Rawlings}}]{IleeEtal2011}
{Ilee}, J.~D., {Boley}, A.~C., {Caselli}, P., {et~al.} 2011, \mnras, 417, 2950, \dodoi{10.1111/j.1365-2966.2011.19455.x}

\bibitem[{{Isella} {et~al.}(2010){Isella}, {Carpenter}, \& {Sargent}}]{IsellaEtal2010}
{Isella}, A., {Carpenter}, J.~M., \& {Sargent}, A.~I. 2010, \apj, 714, 1746, \dodoi{10.1088/0004-637X/714/2/1746}

\bibitem[{{Izquierdo} {et~al.}(2022){Izquierdo}, {Facchini}, {Rosotti}, {van Dishoeck}, \& {Testi}}]{IzquierdoEtal2022}
{Izquierdo}, A.~F., {Facchini}, S., {Rosotti}, G.~P., {van Dishoeck}, E.~F., \& {Testi}, L. 2022, \apj, 928, 2, \dodoi{10.3847/1538-4357/ac474d}

\bibitem[{{Izquierdo} {et~al.}(2023){Izquierdo}, {Testi}, {Facchini}, {Rosotti}, {van Dishoeck}, {W{\"o}lfer}, \& {Paneque-Carre{\~n}o}}]{IzquierdoEtal2023}
{Izquierdo}, A.~F., {Testi}, L., {Facchini}, S., {et~al.} 2023, \aap, 674, A113, \dodoi{10.1051/0004-6361/202245425}

\bibitem[{{Izquierdo} {et~al.}(2025){Izquierdo}, {Stadler}, {Galloway-Sprietsma}, {Benisty}, {Pinte}, {Bae}, {Teague}, {Facchini}, {W{\"o}lfer}, {Longarini}, {Curone}, {Andrews}, {Barraza-Alfaro}, {Cataldi}, {Cuello}, {Czekala}, {Fasano}, {Flock}, {Fukagawa}, {Garg}, {Hall}, {Hammond}, {Hilder}, {Huang}, {Ilee}, {Isella}, {Kanagawa}, {Lesur}, {Lodato}, {Loomis}, {Orihara}, {Price}, {Rosotti}, {Testi}, {Yen}, {Wafflard-Fernandez}, {Wilner}, {Winter}, {Yoshida}, \& {Zawadzki}}]{IzquierdoEtal2025}
{Izquierdo}, A.~F., {Stadler}, J., {Galloway-Sprietsma}, M., {et~al.} 2025, \apjl, 984, L8, \dodoi{10.3847/2041-8213/adc439}

\bibitem[{{Jiang} {et~al.}(2023){Jiang}, {Wang}, {Ormel}, {Krijt}, \& {Dong}}]{JiangEtal2023}
{Jiang}, H., {Wang}, Y., {Ormel}, C.~W., {Krijt}, S., \& {Dong}, R. 2023, \aap, 678, A33, \dodoi{10.1051/0004-6361/202346637}

\bibitem[{{Keppler} {et~al.}(2018){Keppler}, {Benisty}, {M{\"u}ller}, {Henning}, {van Boekel}, {Cantalloube}, {Ginski}, {van Holstein}, {Maire}, {Pohl}, {Samland }, {Avenhaus}, {Baudino}, {Boccaletti}, {de Boer}, {Bonnefoy}, {Chauvin}, {Desidera}, {Langlois}, {Lazzoni}, {Marleau}, {Mordasini}, {Pawellek}, {Stolker}, {Vigan}, {Zurlo}, {Birnstiel}, {Brandner}, {Feldt}, {Flock}, {Girard}, {Gratton}, {Hagelberg}, {Isella}, {Janson}, {Juhasz}, {Kemmer}, {Kral}, {Lagrange}, {Launhardt}, {Matter}, {M{\'e}nard}, {Milli}, {Molli{\`e}re}, {Olofsson}, {P{\'e}rez}, {Pinilla}, {Pinte}, {Quanz}, {Schmidt}, {Udry}, {Wahhaj}, {Williams}, {Buenzli}, {Cudel}, {Dominik}, {Galicher}, {Kasper}, {Lannier}, {Mesa}, {Mouillet}, {Peretti}, {Perrot}, {Salter}, {Sissa}, {Wildi}, {Abe}, {Antichi}, {Augereau}, {Baruffolo}, {Baudoz}, {Bazzon}, {Beuzit}, {Blanchard}, {Brems}, {Buey}, {De Caprio}, {Carbillet}, {Carle}, {Cascone}, {Cheetham}, {Claudi}, {Costille}, {Delboulb{\'e}}, {Dohlen}, {Fantinel}, {Feautrier}, {Fusco}, {Giro}, {Gluck},
  {Gry}, {Hubin}, {Hugot}, {Jaquet}, {Le Mignant}, {Llored}, {Madec}, {Magnard}, {Martinez}, {Maurel}, {Meyer}, {M{\"o}ller-Nilsson}, {Moulin}, {Mugnier}, {Orign{\'e}}, {Pavlov}, {Perret}, {Petit}, {Pragt}, {Puget}, {Rabou}, {Ramos}, {Rigal}, {Rochat}, {Roelfsema}, {Rousset}, {Roux}, {Salasnich}, {Sauvage}, {Sevin}, {Soenke}, {Stadler}, {Suarez}, {Turatto}, \& {Weber}}]{KepplerEtal2018}
{Keppler}, M., {Benisty}, M., {M{\"u}ller}, A., {et~al.} 2018, \aap, 617, A44, \dodoi{10.1051/0004-6361/201832957}

\bibitem[{{Keyte} {et~al.}(2023){Keyte}, {Kama}, {Booth}, {Bergin}, {Cleeves}, {van Dishoeck}, {Drozdovskaya}, {Furuya}, {Rawlings}, {Shorttle}, \& {Walsh}}]{KeyteEtal2023}
{Keyte}, L., {Kama}, M., {Booth}, A.~S., {et~al.} 2023, Nature Astronomy, 7, 684, \dodoi{10.1038/s41550-023-01951-9}

\bibitem[{{Kratter} \& {Lodato}(2016)}]{KratterLodato2016}
{Kratter}, K., \& {Lodato}, G. 2016, \araa, 54, 271, \dodoi{10.1146/annurev-astro-081915-023307}

\bibitem[{{Kuffmeier} {et~al.}(2019){Kuffmeier}, {Calcutt}, \& {Kristensen}}]{KuffmeierEtal2019}
{Kuffmeier}, M., {Calcutt}, H., \& {Kristensen}, L.~E. 2019, \aap, 628, A112, \dodoi{10.1051/0004-6361/201935504}

\bibitem[{{Kuffmeier} {et~al.}(2020){Kuffmeier}, {Goicovic}, \& {Dullemond}}]{KuffmeierEtal2020}
{Kuffmeier}, M., {Goicovic}, F.~G., \& {Dullemond}, C.~P. 2020, \aap, 633, A3, \dodoi{10.1051/0004-6361/201936820}

\bibitem[{{Kuo} {et~al.}(2022){Kuo}, {Yen}, {Gu}, \& {Chang}}]{KuoEtal2022}
{Kuo}, I. H.~G., {Yen}, H.-W., {Gu}, P.-G., \& {Chang}, T.-E. 2022, \apj, 938, 50, \dodoi{10.3847/1538-4357/ac9228}

\bibitem[{{Kuznetsova} {et~al.}(2022){Kuznetsova}, {Bae}, {Hartmann}, \& {Low}}]{KuznetsovaEtal2022}
{Kuznetsova}, A., {Bae}, J., {Hartmann}, L., \& {Low}, M.-M.~M. 2022, \apj, 928, 92, \dodoi{10.3847/1538-4357/ac54a8}

\bibitem[{{Law} {et~al.}(2023){Law}, {Booth}, \& {{\"O}berg}}]{LawEtal2023}
{Law}, C.~J., {Booth}, A.~S., \& {{\"O}berg}, K.~I. 2023, \apjl, 952, L19, \dodoi{10.3847/2041-8213/acdfd0}

\bibitem[{{Law} {et~al.}(2024){Law}, {Benisty}, {Facchini}, {Teague}, {Bae}, {Isella}, {Kamp}, {{\"O}berg}, {Portilla-Revelo}, \& {Rampinelli}}]{LawEtal2024}
{Law}, C.~J., {Benisty}, M., {Facchini}, S., {et~al.} 2024, \apj, 964, 190, \dodoi{10.3847/1538-4357/ad24d2}

\bibitem[{{Le Gal} {et~al.}(2019){Le Gal}, {{\"O}berg}, {Loomis}, {Pegues}, \& {Bergner}}]{LeGalEtal2019a}
{Le Gal}, R., {{\"O}berg}, K.~I., {Loomis}, R.~A., {Pegues}, J., \& {Bergner}, J.~B. 2019, \apj, 876, 72, \dodoi{10.3847/1538-4357/ab1416}

\bibitem[{{Leemker} {et~al.}(2022){Leemker}, {Booth}, {van Dishoeck}, {P{\'e}rez-S{\'a}nchez}, {Szul{\'a}gyi}, {Bosman}, {Bruderer}, {Facchini}, {Hogerheijde}, {Paneque-Carre{\~n}o}, \& {Sturm}}]{LeemkerEtal2022}
{Leemker}, M., {Booth}, A.~S., {van Dishoeck}, E.~F., {et~al.} 2022, \aap, 663, A23, \dodoi{10.1051/0004-6361/202243229}

\bibitem[{{Lesur} {et~al.}(2015){Lesur}, {Hennebelle}, \& {Fromang}}]{LesurEtal2015}
{Lesur}, G., {Hennebelle}, P., \& {Fromang}, S. 2015, \aap, 582, L9, \dodoi{10.1051/0004-6361/201526734}

\bibitem[{{Lesur} {et~al.}(2023){Lesur}, {Flock}, {Ercolano}, {Lin}, {Yang}, {Barranco}, {Benitez-Llambay}, {Goodman}, {Johansen}, {Klahr}, {Laibe}, {Lyra}, {Marcus}, {Nelson}, {Squire}, {Simon}, {Turner}, {Umurhan}, \& {Youdin}}]{LesurEtal2023}
{Lesur}, G., {Flock}, M., {Ercolano}, B., {et~al.} 2023, in Astronomical Society of the Pacific Conference Series, Vol. 534, Astronomical Society of the Pacific Conference Series, ed. S.~{Inutsuka}, Y.~{Aikawa}, T.~{Muto}, K.~{Tomida}, \& M.~{Tamura}, 465

\bibitem[{{Long} {et~al.}(2018){Long}, {Pinilla}, {Herczeg}, {Harsono}, {Dipierro}, {Pascucci}, {Hendler}, {Tazzari}, {Ragusa}, {Salyk}, {Edwards}, {Lodato}, {van de Plas}, {Johnstone}, {Liu}, {Boehler}, {Cabrit}, {Manara}, {Menard}, {Mulders}, {Nisini}, {Fischer}, {Rigliaco}, {Banzatti}, {Avenhaus}, \& {Gully-Santiago}}]{LongEtal2018f}
{Long}, F., {Pinilla}, P., {Herczeg}, G.~J., {et~al.} 2018, \apj, 869, 17, \dodoi{10.3847/1538-4357/aae8e1}

\bibitem[{{Long} {et~al.}(2019){Long}, {Herczeg}, {Harsono}, {Pinilla}, {Tazzari}, {Manara}, {Pascucci}, {Cabrit}, {Nisini}, {Johnstone}, {Edwards}, {Salyk}, {Menard}, {Lodato}, {Boehler}, {Mace}, {Liu}, {Mulders}, {Hendler}, {Ragusa}, {Fischer}, {Banzatti}, {Rigliaco}, {van de Plas}, {Dipierro}, {Gully-Santiago}, \& {Lopez-Valdivia}}]{LongEtal2019}
{Long}, F., {Herczeg}, G.~J., {Harsono}, D., {et~al.} 2019, \apj, 882, 49, \dodoi{10.3847/1538-4357/ab2d2d}

\bibitem[{{Longarini} {et~al.}(2024){Longarini}, {Lodato}, {Clarke}, {Speedie}, {Paneque-Carre{\~n}o}, {Arrigoni}, {Curone}, {Toci}, \& {Hall}}]{LongariniEtal2024}
{Longarini}, C., {Lodato}, G., {Clarke}, C.~J., {et~al.} 2024, \aap, 686, L6, \dodoi{10.1051/0004-6361/202450187}

\bibitem[{{Loomis} {et~al.}(2025){Loomis}, {Facchini}, {Benisty}, {Curone}, {Ilee}, {Cataldi}, {Yen}, {Teague}, {Pinte}, {Huang}, {Garg}, {Orihara}, {Czekala}, {Zawadzki}, {Andrews}, {Wilner}, {Bae}, {Barraza-Alfaro}, {Fasano}, {Flock}, {Fukagawa}, {Galloway-Sprietsma}, {Izquierdo}, {Kanagawa}, {Lesur}, {Longarini}, {Menard}, {Price}, {Rosotti}, {Stadler}, {Wafflard-Fernandez}, {W{\"o}lfer}, \& {Yoshida}}]{LoomisEtal2025}
{Loomis}, R.~A., {Facchini}, S., {Benisty}, M., {et~al.} 2025, \apjl, 984, L7, \dodoi{10.3847/2041-8213/adc43a}

\bibitem[{{Luhman}(2023)}]{LuhmanEtal2023a}
{Luhman}, K.~L. 2023, \aj, 165, 37, \dodoi{10.3847/1538-3881/ac9da3}

\bibitem[{{Mac{\'\i}as} {et~al.}(2021){Mac{\'\i}as}, {Guerra-Alvarado}, {Carrasco-Gonz{\'a}lez}, {Ribas}, {Espaillat}, {Huang}, \& {Andrews}}]{MaciasEtal2021}
{Mac{\'\i}as}, E., {Guerra-Alvarado}, O., {Carrasco-Gonz{\'a}lez}, C., {et~al.} 2021, \aap, 648, A33, \dodoi{10.1051/0004-6361/202039812}

\bibitem[{{Manara} {et~al.}(2023){Manara}, {Ansdell}, {Rosotti}, {Hughes}, {Armitage}, {Lodato}, \& {Williams}}]{ManaraEtal2023}
{Manara}, C.~F., {Ansdell}, M., {Rosotti}, G.~P., {et~al.} 2023, in Astronomical Society of the Pacific Conference Series, Vol. 534, Astronomical Society of the Pacific Conference Series, ed. S.~{Inutsuka}, Y.~{Aikawa}, T.~{Muto}, K.~{Tomida}, \& M.~{Tamura}, 539, \dodoi{10.48550/arXiv.2203.09930}

\bibitem[{{Manara} {et~al.}(2019){Manara}, {Tazzari}, {Long}, {Herczeg}, {Lodato}, {Rota}, {Cazzoletti}, {van der Plas}, {Pinilla}, {Dipierro}, {Edwards}, {Harsono}, {Johnstone}, {Liu}, {Menard}, {Nisini}, {Ragusa}, {Boehler}, \& {Cabrit}}]{ManaraEtal2019}
{Manara}, C.~F., {Tazzari}, M., {Long}, F., {et~al.} 2019, \aap, 628, A95, \dodoi{10.1051/0004-6361/201935964}

\bibitem[{{Mesa} {et~al.}(2022){Mesa}, {Ginski}, {Gratton}, {Ertel}, {Wagner}, {Bonavita}, {Fedele}, {Meyer}, {Henning}, {Langlois}, {Garufi}, {Antoniucci}, {Claudi}, {Defr{\`e}re}, {Desidera}, {Janson}, {Pawellek}, {Rigliaco}, {Squicciarini}, {Zurlo}, {Boccaletti}, {Bonnefoy}, {Cantalloube}, {Chauvin}, {Feldt}, {Hagelberg}, {Hugot}, {Lagrange}, {Lazzoni}, {Maurel}, {Perrot}, {Petit}, {Rouan}, \& {Vigan}}]{MesaEtal2022}
{Mesa}, D., {Ginski}, C., {Gratton}, R., {et~al.} 2022, \aap, 658, A63, \dodoi{10.1051/0004-6361/202142219}

\bibitem[{{Montesinos} {et~al.}(2016){Montesinos}, {Perez}, {Casassus}, {Marino}, {Cuadra}, \& {Christiaens}}]{MontesinosEtal2016}
{Montesinos}, M., {Perez}, S., {Casassus}, S., {et~al.} 2016, \apjl, 823, L8, \dodoi{10.3847/2041-8205/823/1/L8}

\bibitem[{{Mooley} {et~al.}(2013){Mooley}, {Hillenbrand}, {Rebull}, {Padgett}, \& {Knapp}}]{Mooley2013}
{Mooley}, K., {Hillenbrand}, L., {Rebull}, L., {Padgett}, D., \& {Knapp}, G. 2013, \apj, 771, 110, \dodoi{10.1088/0004-637X/771/2/110}

\bibitem[{{Nealon} {et~al.}(2018){Nealon}, {Dipierro}, {Alexander}, {Martin}, \& {Nixon}}]{NealonEtal2018}
{Nealon}, R., {Dipierro}, G., {Alexander}, R., {Martin}, R.~G., \& {Nixon}, C. 2018, \mnras, 481, 20, \dodoi{10.1093/mnras/sty2267}

\bibitem[{{{\"O}berg} \& {Bergin}(2021)}]{OebergBergin2021}
{{\"O}berg}, K.~I., \& {Bergin}, E.~A. 2021, \physrep, 893, 1, \dodoi{10.1016/j.physrep.2020.09.004}

\bibitem[{{{\"O}berg} {et~al.}(2023){{\"O}berg}, {Facchini}, \& {Anderson}}]{OebergEtal2023}
{{\"O}berg}, K.~I., {Facchini}, S., \& {Anderson}, D.~E. 2023, \araa, 61, 287, \dodoi{10.1146/annurev-astro-022823-040820}

\bibitem[{{{\"O}berg} {et~al.}(2011){{\"O}berg}, {Murray-Clay}, \& {Bergin}}]{OebergEtal2011}
{{\"O}berg}, K.~I., {Murray-Clay}, R., \& {Bergin}, E.~A. 2011, \apjl, 743, L16, \dodoi{10.1088/2041-8205/743/1/L16}

\bibitem[{{{\"O}berg} {et~al.}(2021){{\"O}berg}, {Guzm{\'a}n}, {Walsh}, {Aikawa}, {Bergin}, {Law}, {Loomis}, {Alarc{\'o}n}, {Andrews}, {Bae}, {Bergner}, {Boehler}, {Booth}, {Bosman}, {Calahan}, {Cataldi}, {Cleeves}, {Czekala}, {Furuya}, {Huang}, {Ilee}, {Kurtovic}, {Le Gal}, {Liu}, {Long}, {M{\'e}nard}, {Nomura}, {P{\'e}rez}, {Qi}, {Schwarz}, {Sierra}, {Teague}, {Tsukagoshi}, {Yamato}, {van't Hoff}, {Waggoner}, {Wilner}, \& {Zhang}}]{OebergEtal2021}
{{\"O}berg}, K.~I., {Guzm{\'a}n}, V.~V., {Walsh}, C., {et~al.} 2021, \apjs, 257, 1, \dodoi{10.3847/1538-4365/ac1432}

\bibitem[{{Ohashi} {et~al.}(2020){Ohashi}, {Kataoka}, {van der Marel}, {Hull}, {Dent}, {Pohl}, {Pinilla}, {van Dishoeck}, \& {Henning}}]{OhashiEtal2020}
{Ohashi}, S., {Kataoka}, A., {van der Marel}, N., {et~al.} 2020, \apj, 900, 81, \dodoi{10.3847/1538-4357/abaab4}

\bibitem[{{Ono} {et~al.}(2025){Ono}, {Okamura}, {Okuzumi}, \& {Muto}}]{OnoEtal2025}
{Ono}, T., {Okamura}, T., {Okuzumi}, S., \& {Muto}, T. 2025, \pasj, 77, 149, \dodoi{10.1093/pasj/psae106}

\bibitem[{{Oya} {et~al.}(2025){Oya}, {Saiga}, {Miotello}, {Koutoulaki}, {Johnstone}, {Ceccarelli}, {Chandler}, {Codella}, {Sakai}, {Bianchi}, {Bouvier}, {Charnley}, {Cuello}, {De Simone}, {Francis}, {Hanawa}, {Jim{\'e}nez-Serra}, {Loinard}, {Menard}, {Sabatini}, {Vastel}, {Zhang}, {Aikawa}, {Alves}, {Balucani}, {Busquet}, {Caselli}, {Caux}, {Choudhury}, {Dulieu}, {Dur{\'a}n}, {Evans}, {Fedele}, {Feng}, {Fontani}, {Hama}, {Herbst}, {Hirano}, {Hirota}, {Isella}, {Kahane}, {Lefloch}, {Le Gal}, {Liu}, {L{\'o}pez-Sepulcre}, {Maud}, {Maureira}, {Mercimek}, {Moellenbrock}, {Mori}, {Nomura}, {Oba}, {O'Donoghue}, {Ohashi}, {Okoda}, {Ospina-Zamudio}, {Pineda}, {Podio}, {Rimola}, {Sakai}, {Segura-Cox}, {Shirley}, {Svoboda}, {Testi}, {Viti}, {Watanabe}, {Watanabe}, {Zhang}, \& {Yamamoto}}]{OyaEtal2025}
{Oya}, Y., {Saiga}, E., {Miotello}, A., {et~al.} 2025, \apj, 980, 263, \dodoi{10.3847/1538-4357/adabe7}

\bibitem[{{Pacheco-V{\'a}zquez} {et~al.}(2016){Pacheco-V{\'a}zquez}, {Fuente}, {Baruteau}, {Bern{\'e}}, {Ag{\'u}ndez}, {Neri}, {Goicoechea}, {Cernicharo}, \& {Bachiller}}]{Pacheco-VazquezEtal2016}
{Pacheco-V{\'a}zquez}, S., {Fuente}, A., {Baruteau}, C., {et~al.} 2016, \aap, 589, A60, \dodoi{10.1051/0004-6361/201527089}

\bibitem[{{Padoan} {et~al.}(2024){Padoan}, {Pan}, {Pelkonen}, {Haugboelle}, \& {Nordlund}}]{PadoanEtal2024}
{Padoan}, P., {Pan}, L., {Pelkonen}, V.-M., {Haugboelle}, T., \& {Nordlund}, A. 2024, arXiv e-prints, arXiv:2405.07334, \dodoi{10.48550/arXiv.2405.07334}

\bibitem[{{Paneque-Carre{\~n}o} {et~al.}(2021){Paneque-Carre{\~n}o}, {P{\'e}rez}, {Benisty}, {Hall}, {Veronesi}, {Lodato}, {Sierra}, {Carpenter}, {Andrews}, {Bae}, {Henning}, {Kwon}, {Linz}, {Loinard}, {Pinte}, {Ricci}, {Tazzari}, {Testi}, \& {Wilner}}]{Paneque-CarrenoEtal2021}
{Paneque-Carre{\~n}o}, T., {P{\'e}rez}, L.~M., {Benisty}, M., {et~al.} 2021, \apj, 914, 88, \dodoi{10.3847/1538-4357/abf243}

\bibitem[{{P{\'e}rez} {et~al.}(2016){P{\'e}rez}, {Carpenter}, {Andrews}, {Ricci}, {Isella}, {Linz}, {Sargent}, {Wilner}, {Henning}, {Deller}, {Chandler}, {Dullemond}, {Lazio}, {Menten}, {Corder}, {Storm}, {Testi}, {Tazzari}, {Kwon}, {Calvet}, {Greaves}, {Harris}, \& {Mundy}}]{PerezEtal2016}
{P{\'e}rez}, L.~M., {Carpenter}, J.~M., {Andrews}, S.~M., {et~al.} 2016, Science, 353, 1519, \dodoi{10.1126/science.aaf8296}

\bibitem[{{Pineda} {et~al.}(2019){Pineda}, {Szul{\'a}gyi}, {Quanz}, {van Dishoeck}, {Garufi}, {Meru}, {Mulders}, {Testi}, {Meyer}, \& {Reggiani}}]{PinedaEtal2019}
{Pineda}, J.~E., {Szul{\'a}gyi}, J., {Quanz}, S.~P., {et~al.} 2019, \apj, 871, 48, \dodoi{10.3847/1538-4357/aaf389}

\bibitem[{{Pinte} {et~al.}(2023){Pinte}, {Teague}, {Flaherty}, {Hall}, {Facchini}, \& {Casassus}}]{PinteEtal2023a}
{Pinte}, C., {Teague}, R., {Flaherty}, K., {et~al.} 2023, in Astronomical Society of the Pacific Conference Series, Vol. 534, Astronomical Society of the Pacific Conference Series, ed. S.~{Inutsuka}, Y.~{Aikawa}, T.~{Muto}, K.~{Tomida}, \& M.~{Tamura}, 645

\bibitem[{{Podio} {et~al.}(2021){Podio}, {Tabone}, {Codella}, {Gueth}, {Maury}, {Cabrit}, {Lefloch}, {Maret}, {Belloche}, {Andr{\'e}}, {Anderl}, {Gaudel}, \& {Testi}}]{PodioEtal2021}
{Podio}, L., {Tabone}, B., {Codella}, C., {et~al.} 2021, \aap, 648, A45, \dodoi{10.1051/0004-6361/202038429}

\bibitem[{{Rampinelli} {et~al.}(2024){Rampinelli}, {Facchini}, {Leemker}, {Bae}, {Benisty}, {Teague}, {Law}, {{\"O}berg}, {Portilla-Revelo}, \& {Cridland}}]{RampinelliEtal2024}
{Rampinelli}, L., {Facchini}, S., {Leemker}, M., {et~al.} 2024, \aap, 689, A65, \dodoi{10.1051/0004-6361/202449698}

\bibitem[{{Reggiani} {et~al.}(2018){Reggiani}, {Christiaens}, {Absil}, {Mawet}, {Huby}, {Choquet}, {Gomez Gonzalez}, {Ruane}, {Femenia}, {Serabyn}, {Matthews}, {Barraza}, {Carlomagno}, {Defr{\`e}re}, {Delacroix}, {Habraken}, {Jolivet}, {Karlsson}, {Orban de Xivry}, {Piron}, {Surdej}, {Vargas Catalan}, \& {Wertz}}]{ReggianiEtal2018}
{Reggiani}, M., {Christiaens}, V., {Absil}, O., {et~al.} 2018, \aap, 611, A74, \dodoi{10.1051/0004-6361/201732016}

\bibitem[{{Ren} {et~al.}(2018){Ren}, {Dong}, {Esposito}, {Pueyo}, {Debes}, {Poteet}, {Choquet}, {Benisty}, {Chiang}, {Grady}, {Hines}, {Schneider}, \& {Soummer}}]{RenEtal2018}
{Ren}, B., {Dong}, R., {Esposito}, T.~M., {et~al.} 2018, \apjl, 857, L9, \dodoi{10.3847/2041-8213/aab7f5}

\bibitem[{{Ren} {et~al.}(2020){Ren}, {Dong}, {van Holstein}, {Ruffio}, {Calvin}, {Girard}, {Benisty}, {Boccaletti}, {Esposito}, {Choquet}, {Mawet}, {Pueyo}, {Stolker}, {Chiang}, {Boer}, {Debes}, {Garufi}, {Grady}, {Hines}, {Maire}, {M{\'e}nard}, {Millar-Blanchaer}, {Perrin}, {Poteet}, \& {Schneider}}]{RenEtal2020}
{Ren}, B., {Dong}, R., {van Holstein}, R.~G., {et~al.} 2020, \apjl, 898, L38, \dodoi{10.3847/2041-8213/aba43e}

\bibitem[{{Ren} {et~al.}(2023){Ren}, {Benisty}, {Ginski}, {Tazaki}, {Wallack}, {Milli}, {Garufi}, {Bae}, {Facchini}, {M{\'e}nard}, {Pinilla}, {Swastik}, {Teague}, \& {Wahhaj}}]{RenEtal2023b}
{Ren}, B.~B., {Benisty}, M., {Ginski}, C., {et~al.} 2023, \aap, 680, A114, \dodoi{10.1051/0004-6361/202347353}

\bibitem[{{Rivi{\`e}re-Marichalar} {et~al.}(2020){Rivi{\`e}re-Marichalar}, {Fuente}, {Le Gal}, {Baruteau}, {Neri}, {Navarro-Almaida}, {Trevi{\~n}o-Morales}, {Mac{\'\i}as}, {Bachiller}, \& {Osorio}}]{Riviere-MarichalarEtal2020}
{Rivi{\`e}re-Marichalar}, P., {Fuente}, A., {Le Gal}, R., {et~al.} 2020, \aap, 642, A32, \dodoi{10.1051/0004-6361/202038549}

\bibitem[{{Safonov} {et~al.}(2022){Safonov}, {Strakhov}, {Goliguzova}, \& {Voziakova}}]{SafonovEtal2022}
{Safonov}, B.~S., {Strakhov}, I.~A., {Goliguzova}, M.~V., \& {Voziakova}, O.~V. 2022, \aj, 163, 31, \dodoi{10.3847/1538-3881/ac36cb}

\bibitem[{{Salyk} {et~al.}(2013){Salyk}, {Herczeg}, {Brown}, {Blake}, {Pontoppidan}, \& {van Dishoeck}}]{SalykEtal2013}
{Salyk}, C., {Herczeg}, G.~J., {Brown}, J.~M., {et~al.} 2013, \apj, 769, 21, \dodoi{10.1088/0004-637X/769/1/21}

\bibitem[{{Schworer} {et~al.}(2017){Schworer}, {Lacour}, {Hu{\'e}lamo}, {Pinte}, {Chauvin}, {Coud{\'e} du Foresto}, {Ehrenreich}, {Girard}, \& {Tuthill}}]{SchworerEtal2017}
{Schworer}, G., {Lacour}, S., {Hu{\'e}lamo}, N., {et~al.} 2017, \apj, 842, 77, \dodoi{10.3847/1538-4357/aa74b7}

\bibitem[{{Semenov} \& {Wiebe}(2011)}]{SemenovWiebe2011}
{Semenov}, D., \& {Wiebe}, D. 2011, \apjs, 196, 25, \dodoi{10.1088/0067-0049/196/2/25}

\bibitem[{{Soderblom} {et~al.}(2014){Soderblom}, {Hillenbrand}, {Jeffries}, {Mamajek}, \& {Naylor}}]{SoderblomEtal2014}
{Soderblom}, D.~R., {Hillenbrand}, L.~A., {Jeffries}, R.~D., {Mamajek}, E.~E., \& {Naylor}, T. 2014, in Protostars and Planets VI, ed. H.~{Beuther}, R.~S. {Klessen}, C.~P. {Dullemond}, \& T.~{Henning}, 219--241, \dodoi{10.2458/azu_uapress_9780816531240-ch010}

\bibitem[{{Speedie} {et~al.}(2024){Speedie}, {Dong}, {Hall}, {Longarini}, {Veronesi}, {Paneque-Carre{\~n}o}, {Lodato}, {Tang}, {Teague}, \& {Hashimoto}}]{SpeedieEtal2024}
{Speedie}, J., {Dong}, R., {Hall}, C., {et~al.} 2024, \nat, 633, 58, \dodoi{10.1038/s41586-024-07877-0}

\bibitem[{{Speedie} {et~al.}(2025){Speedie}, {Dong}, {Teague}, {Segura-Cox}, {Pineda}, {Calsino}, {Longarini}, {Hall}, {Tang}, {Hashimoto}, {Paneque-Carre{\~n}o}, {Lodato}, \& {Veronesi}}]{SpeedieEtal2025}
{Speedie}, J., {Dong}, R., {Teague}, R., {et~al.} 2025, arXiv e-prints, arXiv:2503.01957, \dodoi{10.48550/arXiv.2503.01957}

\bibitem[{{Stapper} {et~al.}(2022){Stapper}, {Hogerheijde}, {van Dishoeck}, \& {Mentel}}]{StapperEtal2022}
{Stapper}, L.~M., {Hogerheijde}, M.~R., {van Dishoeck}, E.~F., \& {Mentel}, R. 2022, \aap, 658, A112, \dodoi{10.1051/0004-6361/202142164}

\bibitem[{{Su} \& {Bai}(2024)}]{SuBai2024}
{Su}, Z., \& {Bai}, X.-N. 2024, \apj, 975, 126, \dodoi{10.3847/1538-4357/ad7581}

\bibitem[{{Tang} {et~al.}(2014){Tang}, {Dutrey}, {Guilloteau}, {Pi{\'e}tu}, {Di Folco}, {Beck}, {Ho}, {Boehler}, {Gueth}, {Bary}, \& {Simon}}]{TangEtal2014}
{Tang}, Y.-W., {Dutrey}, A., {Guilloteau}, S., {et~al.} 2014, \apj, 793, 10, \dodoi{10.1088/0004-637X/793/1/10}

\bibitem[{{Tang} {et~al.}(2017){Tang}, {Guilloteau}, {Dutrey}, {Muto}, {Shen}, {Gu}, {Inutsuka}, {Momose}, {Pietu}, {Fukagawa}, {Chapillon}, {Ho}, {di Folco}, {Corder}, {Ohashi}, \& {Hashimoto}}]{TangEtal2017}
{Tang}, Y.-W., {Guilloteau}, S., {Dutrey}, A., {et~al.} 2017, \apj, 840, 32, \dodoi{10.3847/1538-4357/aa6af7}

\bibitem[{{Teague}(2019)}]{Teague2019}
{Teague}, R. 2019, The Journal of Open Source Software, 4, 1632, \dodoi{10.21105/joss.01632}

\bibitem[{{Teague}(2020)}]{Teague2020}
---. 2020, {richteague/keplerian\_mask: Initial Release}, 1.0,  Zenodo, \dodoi{10.5281/zenodo.4321137}

\bibitem[{{Teague} \& {Foreman-Mackey}(2018)}]{TeagueForeman-Mackey2018}
{Teague}, R., \& {Foreman-Mackey}, D. 2018, Research Notes of the American Astronomical Society, 2, 173, \dodoi{10.3847/2515-5172/aae265}

\bibitem[{{Teague} {et~al.}(2022){Teague}, {Bae}, {Andrews}, {Benisty}, {Bergin}, {Facchini}, {Huang}, {Longarini}, \& {Wilner}}]{TeagueEtal2022}
{Teague}, R., {Bae}, J., {Andrews}, S.~M., {et~al.} 2022, \apj, 936, 163, \dodoi{10.3847/1538-4357/ac88ca}

\bibitem[{{Teague} {et~al.}(2025){Teague}, {Benisty}, {Facchini}, {Fukagawa}, {Pinte}, {Andrews}, {Bae}, {Barraza-Alfaro}, {Cataldi}, {Cuello}, {Curone}, {Czekala}, {Fasano}, {Flock}, {Galloway-Sprietsma}, {Garg}, {Hall}, {Hammond}, {Hilder}, {Huang}, {Ilee}, {Izquierdo}, {Kanagawa}, {Lesur}, {Lodato}, {Longarini}, {Loomis}, {Masset}, {Menard}, {Orihara}, {Price}, {Rosotti}, {Stadler}, {Testi}, {Yen}, {Wafflard-Fernandez}, {Wilner}, {Winter}, {W{\"o}lfer}, {Yoshida}, \& {Zawadzki}}]{TeagueEtal2025}
{Teague}, R., {Benisty}, M., {Facchini}, S., {et~al.} 2025, \apjl, 984, L6, \dodoi{10.3847/2041-8213/adc43b}

\bibitem[{{Tripathi} {et~al.}(2017){Tripathi}, {Andrews}, {Birnstiel}, \& {Wilner}}]{TripathiEtal2017}
{Tripathi}, A., {Andrews}, S.~M., {Birnstiel}, T., \& {Wilner}, D.~J. 2017, \apj, 845, 44, \dodoi{10.3847/1538-4357/aa7c62}

\bibitem[{{Ubeira Gabellini} {et~al.}(2019){Ubeira Gabellini}, {Miotello}, {Facchini}, {Ragusa}, {Lodato}, {Testi}, {Benisty}, {Bruderer}, {T. Kurtovic}, {Andrews}, {Carpenter}, {Corder}, {Dipierro}, {Ercolano}, {Fedele}, {Guidi}, {Henning}, {Isella}, {Kwon}, {Linz}, {McClure}, {Perez}, {Ricci}, {Rosotti}, {Tazzari}, \& {Wilner}}]{UbeiraGabelliniEtal2019}
{Ubeira Gabellini}, M.~G., {Miotello}, A., {Facchini}, S., {et~al.} 2019, \mnras, 486, 4638, \dodoi{10.1093/mnras/stz1138}

\bibitem[{{Uyama} {et~al.}(2020){Uyama}, {Muto}, {Mawet}, {Christiaens}, {Hashimoto}, {Kudo}, {Kuzuhara}, {Ruane}, {Beichman}, {Absil}, {Akiyama}, {Bae}, {Bottom}, {Choquet}, {Currie}, {Dong}, {Follette}, {Fukagawa}, {Guidi}, {Huby}, {Kwon}, {Mayama}, {Meshkat}, {Reggiani}, {Ricci}, {Serabyn}, {Tamura}, {Testi}, {Wallack}, {Williams}, \& {Zhu}}]{UyamaEtal2020b}
{Uyama}, T., {Muto}, T., {Mawet}, D., {et~al.} 2020, \aj, 159, 118, \dodoi{10.3847/1538-3881/ab7006}

\bibitem[{{van Boekel} {et~al.}(2017){van Boekel}, {Henning}, {Menu}, {de Boer}, {Langlois}, {M{\"u}ller}, {Avenhaus}, {Boccaletti}, {Schmid}, {Thalmann}, {Benisty}, {Dominik}, {Ginski}, {Girard}, {Gisler}, {Lobo Gomes}, {Menard}, {Min}, {Pavlov}, {Pohl}, {Quanz}, {Rabou}, {Roelfsema}, {Sauvage}, {Teague}, {Wildi}, \& {Zurlo}}]{vanBoekelEtal2017}
{van Boekel}, R., {Henning}, T., {Menu}, J., {et~al.} 2017, \apj, 837, 132, \dodoi{10.3847/1538-4357/aa5d68}

\bibitem[{{van der Marel} {et~al.}(2021{\natexlab{a}}){van der Marel}, {Booth}, {Leemker}, {van Dishoeck}, \& {Ohashi}}]{vanderMarelEtal2021b}
{van der Marel}, N., {Booth}, A.~S., {Leemker}, M., {van Dishoeck}, E.~F., \& {Ohashi}, S. 2021{\natexlab{a}}, \aap, 651, L5, \dodoi{10.1051/0004-6361/202141051}

\bibitem[{{van der Marel} {et~al.}(2021{\natexlab{b}}){van der Marel}, {Birnstiel}, {Garufi}, {Ragusa}, {Christiaens}, {Price}, {Sallum}, {Muley}, {Francis}, \& {Dong}}]{vanderMarelEtal2021a}
{van der Marel}, N., {Birnstiel}, T., {Garufi}, A., {et~al.} 2021{\natexlab{b}}, \aj, 161, 33, \dodoi{10.3847/1538-3881/abc3ba}

\bibitem[{{van Gelder} {et~al.}(2021){van Gelder}, {Tabone}, {van Dishoeck}, \& {Godard}}]{vanGelderEtal2021}
{van Gelder}, M.~L., {Tabone}, B., {van Dishoeck}, E.~F., \& {Godard}, B. 2021, \aap, 653, A159, \dodoi{10.1051/0004-6361/202141591}

\bibitem[{{van't Hoff} {et~al.}(2023){van't Hoff}, {Tobin}, {Li}, {Ohashi}, {J{\o}rgensen}, {Lin}, {Aikawa}, {Aso}, {de Gregorio-Monsalvo}, {Gavino}, {Han}, {Koch}, {Kwon}, {Lee}, {Lee}, {Looney}, {Narayanan}, {Plunkett}, {Sai}, {Santamar{\'\i}a-Miranda}, {Sharma}, {Sheehan}, {Takakuwa}, {Thieme}, {Williams}, {Lai}, {Phuong}, \& {Yen}}]{van'tHoffEtal2023}
{van't Hoff}, M. L.~R., {Tobin}, J.~J., {Li}, Z.-Y., {et~al.} 2023, \apj, 951, 10, \dodoi{10.3847/1538-4357/accf87}

\bibitem[{{Veronesi} {et~al.}(2021){Veronesi}, {Paneque-Carre{\~n}o}, {Lodato}, {Testi}, {P{\'e}rez}, {Bertin}, \& {Hall}}]{VeronesiEtal2021}
{Veronesi}, B., {Paneque-Carre{\~n}o}, T., {Lodato}, G., {et~al.} 2021, \apjl, 914, L27, \dodoi{10.3847/2041-8213/abfe6a}

\bibitem[{{Vidal} {et~al.}(2017){Vidal}, {Loison}, {Jaziri}, {Ruaud}, {Gratier}, \& {Wakelam}}]{VidalEtal2017}
{Vidal}, T. H.~G., {Loison}, J.-C., {Jaziri}, A.~Y., {et~al.} 2017, \mnras, 469, 435, \dodoi{10.1093/mnras/stx828}

\bibitem[{{Vieira} {et~al.}(2003){Vieira}, {Corradi}, {Alencar}, {Mendes}, {Torres}, {Quast}, {Guimar{\~a}es}, \& {da Silva}}]{VieiraEtal2003}
{Vieira}, S.~L.~A., {Corradi}, W.~J.~B., {Alencar}, S.~H.~P., {et~al.} 2003, \aj, 126, 2971, \dodoi{10.1086/379553}

\bibitem[{{Vioque} {et~al.}(2018){Vioque}, {Oudmaijer}, {Baines}, {Mendigut{\'\i}a}, \& {P{\'e}rez-Mart{\'\i}nez}}]{VioqueEtal2018}
{Vioque}, M., {Oudmaijer}, R.~D., {Baines}, D., {Mendigut{\'\i}a}, I., \& {P{\'e}rez-Mart{\'\i}nez}, R. 2018, \aap, 620, A128, \dodoi{10.1051/0004-6361/201832870}

\bibitem[{{Virtanen} {et~al.}(2020){Virtanen}, {Gommers}, {Oliphant}, {Haberland}, {Reddy}, {Cournapeau}, {Burovski}, {Peterson}, {Weckesser}, {Bright}, {van der Walt}, {Brett}, {Wilson}, {Millman}, {Mayorov}, {Nelson}, {Jones}, {Kern}, {Larson}, {Carey}, {Polat}, {Feng}, {Moore}, {VanderPlas}, {Laxalde}, {Perktold}, {Cimrman}, {Henriksen}, {Quintero}, {Harris}, {Archibald}, {Ribeiro}, {Pedregosa}, {van Mulbregt}, \& {SciPy 1. 0 Contributors}}]{VirtanenEtal2020}
{Virtanen}, P., {Gommers}, R., {Oliphant}, T.~E., {et~al.} 2020, Nature Methods, 17, 261, \dodoi{10.1038/s41592-019-0686-2}

\bibitem[{{Wagner} {et~al.}(2019){Wagner}, {Stone}, {Spalding}, {Apai}, {Dong}, {Ertel}, {Leisenring}, \& {Webster}}]{WagnerEtal2019b}
{Wagner}, K., {Stone}, J.~M., {Spalding}, E., {et~al.} 2019, \apj, 882, 20, \dodoi{10.3847/1538-4357/ab32ea}

\bibitem[{{Wagner} {et~al.}(2023){Wagner}, {Stone}, {Skemer}, {Ertel}, {Dong}, {Apai}, {Spalding}, {Leisenring}, {Sitko}, {Kratter}, {Barman}, {Marley}, {Miles}, {Boccaletti}, {Assani}, {Bayyari}, {Uyama}, {Woodward}, {Hinz}, {Briesemeister}, {Lawson}, {M{\'e}nard}, {Pantin}, {Russell}, {Skrutskie}, \& {Wisniewski}}]{WagnerEtal2023}
{Wagner}, K., {Stone}, J., {Skemer}, A., {et~al.} 2023, Nature Astronomy, 7, 1208, \dodoi{10.1038/s41550-023-02028-3}

\bibitem[{{Wagner} {et~al.}(2024){Wagner}, {Leisenring}, {Cugno}, {Mullin}, {Dong}, {Wolff}, {Greene}, {Johnstone}, {Meyer}, {Beichman}, {Boyer}, {Horner}, {Hodapp}, {Kelly}, {McCarthy}, {Roellig}, {Rieke}, {Rieke}, {Sitko}, {Stansberry}, \& {Young}}]{WagnerEtal2024}
{Wagner}, K., {Leisenring}, J., {Cugno}, G., {et~al.} 2024, \aj, 167, 181, \dodoi{10.3847/1538-3881/ad11d5}

\bibitem[{{Wichittanakom} {et~al.}(2020){Wichittanakom}, {Oudmaijer}, {Fairlamb}, {Mendigut{\'\i}a}, {Vioque}, \& {Ababakr}}]{WichittanakomEtal2020}
{Wichittanakom}, C., {Oudmaijer}, R.~D., {Fairlamb}, J.~R., {et~al.} 2020, \mnras, 493, 234, \dodoi{10.1093/mnras/staa169}

\bibitem[{{Winter} {et~al.}(2024){Winter}, {Benisty}, \& {Andrews}}]{WinterEtal2024}
{Winter}, A.~J., {Benisty}, M., \& {Andrews}, S.~M. 2024, \apjl, 972, L9, \dodoi{10.3847/2041-8213/ad6d5d}

\bibitem[{{W{\"o}lfer} {et~al.}(2021){W{\"o}lfer}, {Facchini}, {Kurtovic}, {Teague}, {van Dishoeck}, {Benisty}, {Ercolano}, {Lodato}, {Miotello}, {Rosotti}, {Testi}, \& {Ubeira Gabellini}}]{WoelferEtal2021}
{W{\"o}lfer}, L., {Facchini}, S., {Kurtovic}, N.~T., {et~al.} 2021, \aap, 648, A19, \dodoi{10.1051/0004-6361/202039469}

\bibitem[{{W{\"o}lfer} {et~al.}(2023){W{\"o}lfer}, {Facchini}, {van der Marel}, {van Dishoeck}, {Benisty}, {Bohn}, {Francis}, {Izquierdo}, \& {Teague}}]{WoelferEtal2023}
{W{\"o}lfer}, L., {Facchini}, S., {van der Marel}, N., {et~al.} 2023, \aap, 670, A154, \dodoi{10.1051/0004-6361/202243601}

\bibitem[{{W{\"o}lfer} {et~al.}(2025){W{\"o}lfer}, {Barraza-Alfaro}, {Teague}, {Curone}, {Benisty}, {Fukagawa}, {Bae}, {Cataldi}, {Czekala}, {Facchini}, {Fasano}, {Flock}, {Galloway-Sprietsma}, {Garg}, {Hall}, {Huang}, {Ilee}, {Izquierdo}, {Kanagawa}, {Lesur}, {Longarini}, {Loomis}, {Menard}, {Nath}, {Orihara}, {Pinte}, {Price}, {Rosotti}, {Stadler}, {Wafflard-Fernandez}, {Winter}, {Yen}, {Yoshida}, \& {Zawadzki}}]{WoelferEtal2025}
{W{\"o}lfer}, L., {Barraza-Alfaro}, M., {Teague}, R., {et~al.} 2025, \apjl, 984, L22, \dodoi{10.3847/2041-8213/adc42c}

\bibitem[{{Yamato} {et~al.}(2023){Yamato}, {Aikawa}, {Ohashi}, {Tobin}, {J{\o}rgensen}, {Takakuwa}, {Aso}, {Sai}, {Flores}, {de Gregorio-Monsalvo}, {Hirano}, {Han}, {Kido}, {Koch}, {Kwon}, {Lai}, {Lee}, {Lee}, {Li}, {Lin}, {Looney}, {Mori}, {Narayanan}, {Phuong}, {Saigo}, {Santamar{\'\i}a-Miranda}, {Sharma}, {Thieme}, {Tomida}, {van't Hoff}, \& {Yen}}]{YamatoEtal2023}
{Yamato}, Y., {Aikawa}, Y., {Ohashi}, N., {et~al.} 2023, \apj, 951, 11, \dodoi{10.3847/1538-4357/accd71}

\bibitem[{{Yang} {et~al.}(2023){Yang}, {Fern{\'a}ndez-L{\'o}pez}, {Li}, {Stephens}, {Looney}, {Lin}, \& {Harrison}}]{YangEtal2023h}
{Yang}, H., {Fern{\'a}ndez-L{\'o}pez}, M., {Li}, Z.-Y., {et~al.} 2023, \apjl, 948, L2, \dodoi{10.3847/2041-8213/acccf8}

\bibitem[{{Yen} {et~al.}(2016){Yen}, {Koch}, {Liu}, {Puspitaningrum}, {Hirano}, {Lee}, \& {Takakuwa}}]{YenEtal2016}
{Yen}, H.-W., {Koch}, P.~M., {Liu}, H.~B., {et~al.} 2016, \apj, 832, 204, \dodoi{10.3847/0004-637X/832/2/204}

\bibitem[{{Yoneda} {et~al.}(2016){Yoneda}, {Tsukamoto}, {Furuya}, \& {Aikawa}}]{YonedaEtal2016}
{Yoneda}, H., {Tsukamoto}, Y., {Furuya}, K., \& {Aikawa}, Y. 2016, \apj, 833, 105, \dodoi{10.3847/1538-4357/833/1/105}

\bibitem[{{Yoshida} {et~al.}(2024){Yoshida}, {Nomura}, {Law}, {Teague}, {Shibaike}, {Furuya}, \& {Tsukagoshi}}]{YoshidaEtal2024}
{Yoshida}, T.~C., {Nomura}, H., {Law}, C.~J., {et~al.} 2024, \apjl, 971, L15, \dodoi{10.3847/2041-8213/ad654c}

\bibitem[{{Zhang} {et~al.}(2024){Zhang}, {Sakai}, {Ohashi}, {Murillo}, {Chandler}, {Svoboda}, {Ceccarelli}, {Codella}, {Cacciapuoti}, {O'Donoghue}, {Viti}, {Aikawa}, {Bianchi}, {Caselli}, {Charnley}, {Hanawa}, {J{\'\i}menez-Serra}, {Liu}, {Loinard}, {Oya}, {Podio}, {Sabatini}, {Vastel}, \& {Yamamoto}}]{ZhangEtal2024_SO}
{Zhang}, Z.~E., {Sakai}, N., {Ohashi}, S., {et~al.} 2024, \apj, 966, 207, \dodoi{10.3847/1538-4357/ad3921}

\bibitem[{{Zhu} {et~al.}(2015){Zhu}, {Dong}, {Stone}, \& {Rafikov}}]{ZhuEtal2015b}
{Zhu}, Z., {Dong}, R., {Stone}, J.~M., \& {Rafikov}, R.~R. 2015, \apj, 813, 88, \dodoi{10.1088/0004-637X/813/2/88}

\bibitem[{{Zhu} {et~al.}(2025){Zhu}, {Zhang}, \& {Johnson}}]{ZhuEtal2025}
{Zhu}, Z., {Zhang}, S., \& {Johnson}, T.~M. 2025, \apj, 980, 259, \dodoi{10.3847/1538-4357/adae0d}

\bibitem[{{Ziampras} {et~al.}(2025){Ziampras}, {Dullemond}, {Birnstiel}, {Benisty}, \& {Nelson}}]{ZiamprasEtal2025}
{Ziampras}, A., {Dullemond}, C.~P., {Birnstiel}, T., {Benisty}, M., \& {Nelson}, R.~P. 2025, \mnras, \dodoi{10.1093/mnras/staf785}

\end{thebibliography}
\bibliographystyle{aasjournal}



\end{document}